\documentclass[a4paper,12pt]{article}
\pdfoutput=1
\usepackage{graphicx,subfigure,amsmath,amssymb,multirow}
\usepackage{cite}
\usepackage{longtable}
\usepackage{color}

\renewcommand{\baselinestretch}{1.5}
\newlength{\dinwidth}
\newlength{\dinmargin}
\def\be{\begin{equation}}
\def\ee{\end{equation}}
\def\ba{\begin{eqnarray}}
\def\ea{\end{eqnarray}}
 \def\la{ \langle}
  \def\ra{ \rangle}
     \def\e{ \epsilon}
      \def\r{ \gamma}
       \def\lbd{\lambda}
        \def \d {{\rm d}}
           \def\w{\omega}
            \def\u{\mu}
              \def\a{\alpha}
  \def\b{\beta}
\def\v{\nu}
     \def\ve{ \varepsilon}
          \def\qb{{ \bf q}_\bot}
           \def\pb{{ \bf p}_\bot}
           \def\kb{{ \bf k}_\bot}

\def\tb{\textcolor{black}}

\allowdisplaybreaks

\begin{document}
\title{\bf  Form factors of $P\to T$ transition within the light-front quark models}
{\author{ Lili Chen$^{a}$, Yu-Wan Ren$^{a}$,   Li-Ting Wang$^{a}$ and Qin Chang$^{a}$\footnote{changqin@htu.edu.cn}\\
{ $^a$\small Institute of Particle and Nuclear Physics, Henan Normal University, Henan 453007, China}
}
\date{}

\maketitle

\begin{abstract}
In this paper, we  calculate the vector, axial-vector and tensor form factors of $P\to T$ transition within the standard light-front~(SLF) and covariant light-front~(CLF) quark models~(QMs). The self-consistency and Lorentz covariance of CLF QM with two types of correspondence schemes are investigated. The zero-mode effects and the spurious $\w$-dependent contributions to the form factors of $P\to T$ transition are analyzed.   Employing a self-consistent CLF QM, we present our numerical predictions for the vector, axial-vector and tensor form factors of $c\to (q,s)$~($q=u,d$) induced $D \to (a_2,K^*_2)$, $D_s \to (K^*_2,f'_{2})$, $\eta_c(1S) \to (D^*_2,D^*_{s2})$, $ B_c \to (B^*_2,B^*_{s2})$  transitions and $b\to (q,s,c)$  induced $B \to (a_2,K^*_2,D^*_2)$, $B_s \to (K^*_2,f'_2,D^*_{s2})$, $B_c \to (D^*_2,D^*_{s2},\chi_{c2}(1P))$, $\eta_b(1S) \to (B^*_2,B^*_{s2})$ transitions. Finally, in order to test the obtained form factors, the semileptonic  $B\to \bar{D}_2^*(2460)\ell^+\nu_\ell$~($\ell=e,\u$) and $\bar{D}_2^*(2460)\tau^+\nu_{\tau}$ decays are studied. It is expected that  our results for the form factors of $P\to T$ transition can be applied further to the relevant phenomenological studies of meson decays.


\end{abstract}

\newpage
\section{Introduction }
In the quark model, mesons are bound states of quark $q$ and antiquark $\bar{q}'$, and thus the spin-parity quantum number $J^P$ of mesons are consequently fixed by the constituent quark pair, for instances $J^P= 0^-$ for pseudoscalar~(P) meson and  $J^P= 2^+$ for p-wave tensor~(T) meson. Many tensor mesons have been well established in various processes~\cite{Zyla:2020zbs}. Following the flavor $SU(3)$ symmetry, nine possible tensor $(q\bar{q}')$ states containing the light $u$, $d$, and $s$ quarks, isovector mesons $a_2(1320)$, isodoulet states $K^*_2(1430)$ and two isosinglet mesons $f_2(1270)$, $f'_2(1525)$ form the $1^3P_2$ nonet~\cite{Zyla:2020zbs}. The known open charm and bottom p-wave tensor states include $D_{2}^*(2460)$, $D_{s2}^*(2573)$, $B_2^*(5747)$ and $B_{s2}^*(5840)$, the charmonium and bottomonium tensor mesons are $\chi_{c2}(1P )$ and $\chi_{b2}(1P )$, respectively.

The  $B$ and $D$ meson decays induced by  heavy-to-light transition provide a fertile ground for testing the Standard Model~(SM) and searching for new physics~(NP).  Some discrepancies between the experimental data and the SM predictions have been found, for instance,  the SM predictions for $R_{D^{(*)}}$ deviate from data by more than $3\sigma$ errors~\cite{HFLAV:2019otj}. If these tensions are confirmed by the forthcoming experiments, the discrepancies should also be seen in $B$ transitions to tensor mesons in addition to $B$ decays to pseudoscalar  or vector mesons. The decay modes  involving tensor final states are of great interest because the tensor meson  has additional polarization states compared with the (pseudo)scalar and (axial)vector  mesons and thus the relevant decays may have more  kinematical quantities related to  the underlying helicity structure. Some theoretical studies on these decays have been made in, for instance, Refs.~\cite{Cheng:2010yd,Zou:2020ool,Li:2019jlp,Li:2018lbd,Liu:2017cwl,Morales:2016pcq,Qin:2014xta,Zou:2012sy,Zou:2012sx,Zou:2012td,Li:2010ra,Mohapatra:2021izl,Zuo:2021kui,Rajeev:2020aut,Das:2018orb,Ju:2014oha,Ahmed:2012zzc,Aliev:2011gc,Junaid:2011egj}. In the calculation of the amplitudes of semi-leptonic, non-leptonic and radiative $B$ and $D$ meson decays, the form factors serve as the basic and important input parameters.

There are many  approaches for evaluating  the form factors,  for instance, Wirbel-Stech-Bauer model~\cite{Wirbel:1985ji}, lattice QCD~(LQCD)~\cite{Daniel:1990ah}, perturbative QCD~(PQCD)  with some nonperturbative inputs~\cite{Lepage:1980fj,Li:1992nu}, QCD sum rules~(QCD SR)~\cite{Shifman:1978bx,Shifman:1978by}, light-front quark models~(LF QMs)~\cite{Terentev:1976jk,Berestetsky:1977zk,Jaus:1989au,Jaus:1989av,Cheng:1997au,Carbonell:1998rj,Jaus:1999zv}, etc. Some $B\to T$ transition form factors have be evaluated by employing the light-cone sum rules~(LCSRs) approach~\cite{Yang:2010qd,Aliev:2019ojc,Wang:2010tz}, QCD SR~\cite{Khosravi:2015jfa},  the large energy effective theory (LEET)~\cite{Ebert:2001pc,Hatanaka:2009gb} and PQCD approach~\cite{Wang:2010ni}. In this paper, we will evaluate the form factors of  $c\to (q,s)$~($q=u,d$) induced $D \to (a_2,K^*_2)$, $D_s \to (K^*_2,f'_{2})$, $\eta_c(1S) \to (D^*_2,D^*_{s2})$, $ B_c \to (B^*_2,B^*_{s2})$  transitions and $b\to (q,s,c)$  induced $B \to (a_2,K^*_2,D^*_2)$, $B_s \to (K^*_2,f'_2,D^*_{s2})$, $B_c \to (D^*_2,D^*_{s2},\chi_{c2}(1P))$, $\eta_b(1S) \to (B^*_2,B^*_{s2})$ transitions within the standard and the covariant light-front quark models.


The standard light-front quark model~(SLF QM) ~\cite{Terentev:1976jk,Berestetsky:1977zk,Jaus:1989au,Jaus:1989av} is based on the LF formalism~\cite{Dirac:1949cp} and LF quantization of QCD~\cite{Brodsky:1997de}, and provides a conceptually simple and  phenomenologically feasible framework for evaluating nonperturbative quantities. However,  in this approach, the matrix element lacks manifest Lorentz covariance and the zero-mode contributions can not be determined explicitly.  Therefore, a basically different technique is developed by Jaus to deal with the covariance and the zero-mode problems with the help of a manifestly covariant Bethe-Saltpeter~(BS) approach as a guide to the calculation~\cite{Jaus:1999zv}. In such a covariant light-front quark model~(CLF QM), the zero-mode contributions can be well determined, and the  matrix element is expected to be covariant because  the $\w$-dependent spurious contributions, where $\w$ is the light-like four-vector used to define  light-front by $\w\cdot x=0$, can be eliminated  by the inclusion of zero-mode contributions~\cite{Jaus:1999zv}.

Unfortunately, the covariance of matrix element in fact can not be fully recovered in such traditional CLF QM because there are still  some residual  $\w$-dependent spurious contributions associated with $B$ functions~\cite{Jaus:1999zv,Chang:2018zjq}. In addition, the traditional CLF QM suffers from a self-consistence problem for a long time, for instance, it has been found that  the CLF results for $f_V$ obtained respectively via  longitudinal ($\lbd=0$) and  transverse ($\lbd=\pm$) polarization states are  inconsistent with each other~\cite{Cheng:2003sm},  $[f_V]_{\rm CLF}^{\lbd=0}\neq [f_V]_{\rm CLF}^{\lbd=\pm}$,  because the former receives an additional contribution characterized by the $B_1^{(2)}$ function. Another self-consistence problem has also been found in Ref.~\cite{Chang:2020wvs}.

 In order to deal with the self-consistence problem, Choi and Ji present a modified correspondence scheme between the  covariant BS  approach and the LF approach~(named as type-II scheme)~\cite{Choi:2013mda}, which requires an additional $M\to M_0$ replacement relative to the traditional  type-I correspondence scheme.  By using such improved self-consistent CLF QM, one can obtain the self-consistent results~\cite{Choi:2013mda}, and moreover, the covariance of matrix element can be fully recovered~\cite{Chang:2018zjq,Chang:2019mmh,Chang:2019obq,Chang:2020wvs}. In this paper, we would like to further test the self-consistence and covariance of the self-consistent CLF QM via the form factors of $P\to T$ transition.

Our paper is organized as follows. In sections 2, the SLF and the CLF QMs are review briefly, and then our theoretical results for the vector, axial-vector and tensor form  factors of  $P\to T $ transition are  presented. In section 3, the self-consistency and covariance of CLF results are discussed, and  the zero-mode and the valence contributions are analyzed. After that, our numerical results for the $c\to (q,s)$~($q=u,d$) induced $D\to (a_2,K^*_2)$, $D_s\to (K^*_2, f'_2)$, $\eta_c(1S)\to (D^*_2,D^*_{s2})$, $B_c\to (B^*_2,B^*_{s2} )$ transitions and the $b\to (q,s,c)$  induced $B\to (a_2,K^*_2, D^*_2), B_s\to (K^*_2, f'_2, D^*_{s2}), B_c\to (D^*_2, D^*_{s2}, X_{c2}(1P))$, $\eta_b(1S)\to (B^*_2,B^*_{s2})$  transitions are given. Finally, a summary is given in section 4.

\section{ Theoretical framework and results }
\subsection{ Definitions of form factors}
\label{sec:2}
The matrix elements  of $P\to T$ transition with vector, axial-vector and tensor currents are commonly factorized in terms of form factors as ~\cite{Cheng:2010yd,Cheng:2004yj}
\begin{align}
\la T(\epsilon^{\mu\nu},p'')|\bar q_2\gamma_\mu q_1|P(p')\ra
=&i \varepsilon_{\mu\nu\alpha\beta}e^{*\nu}P^\alpha q^\beta\frac{V(q^2)}{M'+M''}\,,
\label{eq:depvt}\\
\la T(\epsilon^{\mu\nu},p'')|\bar q_2\gamma_\mu\gamma_5 q_1|P(p')\ra
=&-2M''\frac{e^*\cdot q}{q^2} q_\mu A_0(q^2)-\Big(M'+M''\Big)\Big(e^*_\mu-\frac{e^*\cdot q}{q^2} q_\mu \Big)A_1(q^2) \nonumber\\
&+\frac{e^*\cdot q}{M'+M''}\left(P _\mu-\frac{M'^2-M''^2}{q^2}q_\mu\right)A_2 (q^2)\,,
\label{eq:depat}\\
\la T(\epsilon^{\mu\nu},p'')|\bar q_2\sigma_{\mu\nu}q^\nu(1+\gamma_5)q_1|P(p')\ra
=& - \varepsilon_{\mu\nu\alpha\beta}e^{*\nu} P^\alpha q^\beta \,T_1(q^2)\nonumber\\ &+i\Big[(M'^2-M''^2)e^*_{\mu}-(e^*\cdot q)P_{\mu})\Big] T_2(q^2)\nonumber\\
&+i\Big(e^*\cdot q\Big)\Big[q_\mu-\frac{q^2}{M'^2-M''^2}P_\mu\Big]T_3(q
^2)\,,
\label{eq:deptt}
\end{align}
where, $\ve_{0123}=-1$, $P=p'+p''$, $q=p'-p''$, $e^{*\nu}\equiv\frac{\epsilon^{*\mu\nu}\cdot p'_{\mu}}{M'}$, $M^{'('')}$ is the mass of initial (final) state,  $\epsilon^{\mu\nu}$ is the  polarization tensor of tensor  meson and  satisfies $\epsilon^{\mu\nu} p''_{\nu}=0$.  The polarization tensors $\epsilon^{\mu\nu}(\lambda)$ with $\lambda$ helicity~($\lambda=0,\pm1,\pm2$ ) can be constructed in terms of the polarization vectors of vector state $\epsilon^{\mu}{(\lambda')}$, and explicitly written as
\begin{align}
\epsilon^{\mu\nu}{(0)}&=\sqrt{\frac{1}{6}}\left[\epsilon^\mu(+1)\epsilon^\nu(-1)+\epsilon^\mu(-1)\epsilon^\nu(+1)\right]+\sqrt{\frac{2}{3}}\epsilon^\mu(0)\epsilon^\nu(0)\,,\\
\epsilon^{\mu\nu}{(\pm1)}&=\sqrt{\frac{1}{2}}\left[\epsilon^\mu(\pm1)\epsilon^\nu(0)+\epsilon^\mu(0)\epsilon^\nu(\pm1)\right]\,,\\
\epsilon^{\mu\nu}{(\pm2)}&=\epsilon^\mu(\pm1)\epsilon^\nu(\pm1)\,,
\end{align}
where,
\begin{align}
\epsilon^\mu(0)&=\frac{1}{M}(p^+,\frac{-M^2+{ \pb^2}}{p^+},-\pb)\,,\nonumber\\
 \epsilon^\mu(\pm)&=(0,-\frac{2}{p^+} {\boldsymbol{\epsilon}_\perp\cdot \pb},{ \boldsymbol{\epsilon}_\perp})\,,\quad \boldsymbol{\epsilon}_\perp\equiv\mp\frac{(1,\pm i)}{\sqrt 2}\,.
\end{align}

Using the identity $2\sigma^{\mu\nu}\r_5=i\ve^{\u\v\a\b}\sigma_{\a\b} $, one can rewrite the definition of tensor form factors $T_{1,2,3}$, Eq.~\eqref{eq:deptt}, as
\begin{align}
\la T(\epsilon^{\mu\nu},p'')|\bar q_2\sigma_{\mu\nu}q_1|P(p')\ra =&  \varepsilon_{\mu\nu\alpha\beta}\bigg\{-e^{*\alpha} P^\beta\, T_1(q^2)+\frac{M'^2-M''^2}{q^2} e^{*\alpha} q^\beta\left[T_1(q^2)-T_2(q^2)\right]\nonumber\\
&-\frac{e^*\cdot q}{q^2}P^\alpha q^\beta\Big[T_1(q^2)-T_2(q^2)-\frac{q^2}{(M'^2-M''^2)}T_3(q^2)\Big]\bigg\}\,,
\label{eq:deptts}
\end{align}
which is much more convenient for extracting the form factors within the SLF QM.
\subsection{Theoretical results in the SLF QM }

In order to clarify the convention and notation used in this paper,  we would like to review briefly the framework of SLF QM.  One  may refer to, for instance,  Refs.~\cite{Terentev:1976jk,Berestetsky:1977zk,Jaus:1989au,Jaus:1989av} for details.

The main work of LF approach is to evaluate the current matrix element,
\begin{align}\label{eq:amp1}
{\cal B} \equiv \la  T(\e^{\u\v*},p'') | \bar{q}''_1 (k_1'')\Gamma q'_1(k_1') |P(p') \ra \,,\qquad \Gamma=\r_\u\,,\r_\u\r_5\,, \sigma_{\mu\nu}q^\nu(1+\r_5)\,,\sigma_{\mu\nu}\,,
\end{align}
which will be further used to extract the form factors. The meson bound-state ($q_1\bar q_2$) with total momentum $p$ and spin $J$ can be written as
\begin{align}
|M(p,L,J)\ra = & \sum_{h_1,h_2} \int \frac{\d^3 {k}_1}{(2\pi)^32\sqrt{k_1^+}} \frac{\d^3 {k}_2}{(2\pi)^32\sqrt{k_2^+}} (2\pi)^3 \delta^3 ({{p}-{k}_1-{k}_2}) \nonumber\\
&\Psi_{LS}^{JJ_z}({k}_1,h_1,{k}_2,h_2)|q_1({k}_1,h_1)\ra|\bar{q}_2({k}_2,h_2)\ra\,,
\label{eq:Fockexp}
\end{align}
where $k_1$ and $k_2$ are the on-mass-shell light-front momenta and can be written in terms of the internal relative momentum variables $(x,{\mathbf{ k}_{\bot}})$ as
\begin{align}\label{eq:momk1}
k_1^+=xp^+\,,\quad\, \mathbf{k}_{1\bot}=x\mathbf{p}_{\bot}+\mathbf{k}_{\bot} \,,\qquad  k_2^+=\bar{x}p^+ \,,\quad\, \mathbf{k}_{2\bot}=\bar{x}\mathbf{p}_{\bot}-\mathbf{k}_{\bot}\,,
\end{align}
with $\bar{x}\equiv1-x$.

The momentum-space wavefunction~(WF) for a $^{2S+1}L_J$ meson, $\Psi_{LS}^{JJ_z}({k}_1,h_1,{k}_2,h_2)$, in Eq.~\eqref{eq:Fockexp} satisfies the normalization condition and can be expressed as
\begin{align}
\label{eq:LFWFP2}
\Psi_{LS}^{JJ_z}({k}_1,h_1,{k}_2,h_2)=S_{h_1,h_2}(x,\mathbf{k}_{\bot}) \psi(x,\mathbf{k}_{\bot}) \,,
\end{align}
where, $\psi(x,\mathbf{k}_{\bot})$ is the radial WF and responsible for describing the momentum distribution of the constituent quarks in the bound-state; $S_{h_1,h_2}(x,\mathbf{k}_{\bot})$  is the spin-orbital WF and responsible for constructing a state of definite spin $(S,S_z)$ out of the LF helicity $(h_1,h_2)$ eigenstates.  For the former, we adopt the Gaussian type WF
\begin{align}
\label{eq:RWFs}
\psi_{1s}(x,\mathbf{k}_{\bot}) =&4\frac{\pi^{\frac{3}{4}}}{\beta^{\frac{3}{2}}} \sqrt{ \frac{\partial k_z}{\partial x}}\exp\left[ -\frac{k_z^2+\mathbf{k}_\bot^2}{2\beta^2}\right]\,,\nonumber\\
\psi_{1p}(x,\mathbf{k}_{\bot})=&\frac{\sqrt 2}{\beta} \psi_{1s}(x,\mathbf{k}_{\bot})\,,
\end{align}
where, $k_z$ is the relative momentum in $z$-direction, $k_z=(x-\frac{1}{2})M_0+\frac{m_2^2-m_1^2}{2 M_0}$, with the invariant mass $M_0^2=\frac{m_1^2+\mathbf{k}_{\bot}^2}{x}+\frac{m_2^2+\mathbf{k}_{\bot}^2}{\bar{x}}$.
The spin-orbital WF, $S_{h_1,h_2}(x,\mathbf{k}_{\bot}) $, can be obtained by the interaction-independent Melosh transformation.  It is convenient to use its covariant form, which can be further reduced by using the equation of motion on spinors and finally written as~\cite{Jaus:1989av,Cheng:2003sm}
\begin{align}\label{eq:defS2}
S_{h_1,h_2}=\frac{\bar{u}(k_1,h_1)\Gamma' v(k_2,h_2)}{\sqrt{2} \hat{M}_0}\,,
\end{align}
where, $\hat{M}_0^2=M_0^2-(m_1-m_2)^2$. For $P$ and $T$ mesons, the vertices $\Gamma'$ are written as
\begin{align}
\Gamma'_P&=\gamma_5\,,\\
\Gamma'_T&=-\frac{1}{2}\hat\epsilon^{\mu\nu}\left[\gamma_\mu-\frac{(k_1-k_2)_\mu}{D_{T,LF}}\right](k_1-k_2)_\nu\,,\quad D_{T,{\rm LF}}=M_0+m_1+m_2\,,
\label{eq:vSLF}
\end{align}
where $\hat\epsilon^{\mu\nu}=\epsilon^{\mu\nu}(M\to M_0)$.

Using the formulae given above, the matrix element of $M'\to M''$ transition can be written as
\begin{eqnarray}
{\cal B}_{\rm  SLF}=\sum_{h'_1,h''_1,h_2} \int  \frac{\d x \,\d^2{ \mathbf{k}_\bot'}}{(2\pi)^3\,2x}  {\psi''}^{*}(x,\mathbf{k}_{\bot}''){\psi'}(x,\mathbf{k}_{\bot}')
S''^{\dagger}_{h''_1,h_2}(x,\mathbf{k}_{\bot}'')\,C_{h''_1,h'_1}(x,\mathbf{k}_{\bot}',\mathbf{k}_{\bot}'')\,S'_{h'_1,h_2}(x,\mathbf{k}_{\bot}')\,,
\label{eq:B}
\end{eqnarray}
where $S'^{(\prime\prime)}$ and $\psi'^{(\prime\prime)}$ are the WFs of initial~(final) state; $C_{h''_1,h'_1}(x,\mathbf{k}_{\bot}',\mathbf{k}_{\bot}'') \equiv  \bar{u}_{h''_1}(x,\mathbf{k}_{\bot}'')  \Gamma   u_{h'_1}(x,\mathbf{k}_{\bot}')$ with $\Gamma$ given by Eq.~\eqref{eq:amp1} for $P\to T$ transition. In the further calculation, it is covariant to use the Drell-Yan-West frame, $q^+=0$, and take a Lorentz frame, $\mathbf{p}_{\bot}'=0$. In this frame, the momenta of constituent quarks in initial and final states are written as
\begin{align}\label{eq:momk2}
{k}_1'=(xp'^+,\mathbf{k}_{\bot}')  \,,\quad  {k}_1''=(xp'^+, x\mathbf{p}_{\bot}'' +\mathbf{k}_{\bot}'') \,,\quad   {k}_2=(\bar{x}p'^+ ,-\mathbf{k}_{\bot}' ) =(\bar{x}p'^+ ,\bar{x}\mathbf{p}_{\bot}'' -\mathbf{k}_{\bot}'' )\,,
\end{align}
where $\mathbf{p}_{\bot}''=\mathbf{p}_{\bot}'-\mathbf{q}_{\bot}=-\mathbf{q}_{\bot}$  and $\mathbf{k}_{\bot}''=\mathbf{k}_{\bot}'-\bar{x}\mathbf{q}_{\bot}$.

In the SLF QM, in order to extract the form factors of $P\to T$ transition defined by Eqs.~(\ref{eq:depvt}-\ref{eq:deptt}), one has to take explicit values of $\mu$ and/or $\lambda$. In this work, we take the following strategy\footnote{ \tb{This strategy is not unique. It is possible to extract the form factor by choosing the other values of  $\mu$ or $\lambda$, but the other strategies may make the calculation complicated relative to our strategy. In addition, different strategy may result in different expression for the form factor, which can be easily understood because the manifest covariance is violated in the SLF QM,  this is exactly why the CLF QM is needed. A simple example for this issue is $f_V$~(one may refer to Ref.~\cite{Chang:2018zjq} for details). } }:
 \begin{itemize}
 \item For the vector form factor $V$, we take $\lbd''=+2$ and multiply  both sides of Eq.~\eqref{eq:depvt} by $\e^{\u}$.
 \item For the axial-vector form factors, we take $\u=+$, and then use  ${\cal B}_{\rm  SLF}^+$ with $\lbd''=+2$ and $+1$ to extract $A_{2}$ and $A_{1}$, respectively; we take $\lbd''=+2$, and multiply  both sides of Eq.~\eqref{eq:depat} by $q^{\u}$  to extract  $A_{0}$.
\item For the tensor form factors, we take $\lbd''=+2$, and multiply  both sides of Eq.~\eqref{eq:deptts} by $\e^{\u}q^{\v}$, $\e^{\u}P^{\v}$ and $\e^{\u}\e^{*\v}$ to extract $T_1$, $T_2$ and $T_3$, respectively.
\end{itemize}

After some derivations and simplifications,   we  finally obtain the SLF results for the form factors of $P\to T$ transition written as
\begin{equation}\label{eq:FSLF}
[{\cal F}(q^2)]_{\rm  SLF}=\int\frac{\d x\,\d^2{\bf k_\bot'}}{(2\pi)^3\,2x}\frac{{\psi''_T}^*(x,{\bf k''_\bot})\,{\psi'_P}(x,{\bf k_\bot'})}{2\hat {M}'_0\hat {M}''_0}\,{\cal \widetilde{F}}^{\rm SLF}(x,{\bf k}_\bot',q^2)\,,
\end{equation}
where,  the integrands are
\begin{align}
\widetilde{V}^{\rm SLF}=&\frac{M'(M'+M'')}{x\bar xq^2(M'^2-M''^2+\qb^2)}\bigg\{2\kb''\cdot\qb\Big[(2x-1)\kb'^2(m'_1-m''_1)+2\bar x\kb'\cdot\qb\big((1-2x)m'_1\nonumber\\
&+xm''_1+xm_2\big)+(\bar xm'_1+xm_2)\big( (m'_1-m''_1)(\bar x m''_1+xm_2)+\bar x(2x-1)\qb^2\big)\Big]
\nonumber\\
&-\frac{2}{D''_{T,{\rm LF}}}\Big[(2x-1)\kb'^4\qb^2-2x\kb'\cdot\kb''(\kb'\cdot\qb)^2+2\bar x(1+2x)\kb'\cdot\kb''\kb'\cdot\qb\qb^2\nonumber\\
&+2x\bar x(\kb'\cdot\qb)^3-2x\bar x\kb'^2\kb''\cdot\qb\qb^2+\bar x(1-6x)\kb'^2\kb'\cdot\qb\qb^2-\bar x^2\kb'\cdot\kb''\qb^4\nonumber\\
&-\bar x(\kb'^2\kb''\cdot\qb-\bar x\kb'\cdot\kb''\qb^2)(m'^2_1-m''^2_1)+(\kb'^2\qb^2-2\kb'\cdot\qb\kb''\cdot\qb\nonumber\\
&-\bar x^2\qb^4)(x^2m^2_2-\bar x^2m'^2_1)
\Big]
\bigg\}\,,\\
\widetilde{A}_0^{\rm{ SLF}}=&\frac{M'}{M''q^4}\bigg\{(\kb'\cdot\kb''\qb^2-2\kb'\cdot\qb\kb''\cdot\qb)\Big[
4\kb'\cdot\qb(m_2-m'_1)+(M'^2-M''^2)(2xm'_1\nonumber\\
&-m'_1-m''_1-2xm_2)+\big((3-2x)m'_1+m''_1-2\bar xm_2\big)\qb^2\Big]-\frac{\kb''\cdot\qb\qb^2}{x\bar x}\Big[\kb'^2(m'_1+m''_1)\nonumber\\
&-2\bar x\kb'\cdot\qb \big((1-2x+2x^2)m'_1+xm''_1+x(1-2x)m_2\big)+\Big(xm_2+\bar xm'_1\Big)\Big((xm_2\nonumber\\
&+\bar xm''_1)(m'_1+m''_1)-x\bar x(2x-1)(M'^2-M''^2)-\bar x^2(2x-1)\qb^2\Big)\Big]\nonumber\\
&-\frac{2}{x\bar xD''_{T,{\rm LF}}}\Big(2\kb'\cdot\qb\kb''\cdot\qb-\kb'^2\qb^2+\bar x^2\qb^4\Big)\Big[2\kb'\cdot\kb''\kb'\cdot\qb+\bar x^2\kb'\cdot\qb\qb^2\nonumber\\
&+(\kb'^2+x^2m^2_2)[x(M''^2-M'^2)-\bar x\qb^2]+\kb'\cdot\qb\big(2(xm_2+\bar xm'_1)(xm_2-\bar xm''_1)\nonumber\\
&+\bar x(m'_1+m''_1)^2+x\bar x(M'^2-M''^2)\big)+x\bar x(M''^2-M'^2)(xm_2m'_1-xm''_1m_2-\bar xm'_1m''_1)\nonumber\\
&+\bar xx^2m_2(m'_1-m''_1)\qb^2-\bar xm'_1(\bar xm'_1+x\bar xm''_1+2xm_2)\qb^2
\Big]
\bigg\}\,,\\
\widetilde{A}_1^{\textup{\scriptsize SLF}}=&\frac{M'M''}{x\bar xM''_0(M'+M'')\qb^2}\bigg\{\kb'^2\kb''\cdot\qb\Big[(1-8x\bar x)m'_1-4xm''_1+2x(3-4x)m_2\Big]\nonumber\\
&+(1-2x)\Big[\kb''^2\kb'\cdot\qb m''_1+x\kb'\cdot\kb''\qb^2 m_2\Big]-2\bar x\kb'\cdot\qb\kb''\cdot\qb\Big[(1-8x\bar x)m'_1\nonumber\\
&-2xm''_1+4x(1-2x)m_2\Big]+(1-8x\bar x)\kb''\cdot\qb(xM'''^2+\bar x\qb^2)(\bar xm'_1+xm_2)\nonumber\\
&-4x\kb''\cdot\qb m''_1m_2(\bar xm'_1+xm_2)
+\kb'\cdot\qb\big[(1-2x)m'_1+m''_1+2xm_2\big]\big(\bar xm''^2_1-xm^2_2\big)
\nonumber\\
&+x(1-2x)\kb'\cdot\qb M''^2(
\bar xm''_1+xm_2)+\qb^2\big(\bar xm'_1+xm_2\big)\big[(1-2x)(xm^2_2-\bar x m''^2_1)\nonumber\\
&+4x\bar xm''_1m_2\big]
-\frac{4}{\bar x D''_{T,{\rm LF}}}
\kb''\cdot\qb\big[\kb'\cdot\kb''+(\bar xm'_1+xm_2)(xm_2-\bar xm''_1)
\big]\big[(2x-1)\kb''^2\nonumber\\
&-\bar xm''^2_1+xm^2_2+x\bar x(2x-1)M''^2
\big]\bigg\}+\frac{2(M'^2-M''^2+\qb^2)}{(M'+M'')^2}\widetilde{A}_2^{\rm{ SLF}}(x,{\bf k'_\bot},q^2)\,,\\
\widetilde{A}_2^{\rm SLF}=&\frac{M'(M'+M'')}{q^4}\bigg\{-2(2\kb'\cdot\qb\kb''\cdot\qb-\kb'\cdot\kb''\qb^2)[(1-2x)m'_1+m''_1+2xm_2]\nonumber\\
&+2(1-2x)\kb''\cdot\qb\qb^2(\bar xm'_1+xm_2)+\frac{2}{\bar xD''_{T,{\rm LF}}}\Big[2\kb'^4\qb^2+2\bar x\kb'^2\kb'\cdot\qb\qb^2\nonumber\\
&-4\kb'^2(\kb'\cdot\qb)^2+4\bar x\kb''\cdot\qb(\kb'\cdot\qb)^2-2\bar x^2\kb'\cdot\kb''\qb^4-2(2\kb'\cdot\qb\kb'\cdot\qb\nonumber\\
&-\kb'^2\qb^2+\bar x^2\qb^4)(\bar xm'_1+xm_2)(xm_2-\bar xm''_1)
\Big]
\bigg\}\,,\\
\widetilde{T}_{1}^{\rm{ SLF}}=&\frac{M'}{x\bar xq^2(M'^2-M''^2+\qb^2)}\bigg\{(2x-1)[(3-x)\kb'^2\kb''\cdot\qb\qb^2-4\kb'\cdot\kb''(\kb'\cdot\qb)^2\nonumber\\
&+3\bar x^2\kb'\cdot\kb''\qb^4-4\bar x^2\kb'\cdot\qb\kb''\cdot\qb\qb^2]-x\bar x\kb''\cdot\qb\qb^2[2x(M'^2-M''^2)+\qb^2]\nonumber\\
&+\kb'^2\kb''\cdot\qb[2x^2(M'^2-M''^2)+(-4x^2+7x-2)\qb^2]+\kb''\cdot\qb[(2\bar xxm'_1-\bar xm'_1\nonumber\\
&+xm_2)\qb^2+2x(\bar xm''_1+xm_2)(M'^2-M''^2)](\bar xm'_1+xm_2)-(2\kb'\cdot\qb\kb''\cdot\qb\nonumber\\
&-\kb'\cdot\kb''\qb^2)[\bar x(2x-1)m'^2_1-\bar xm''^2_1+2x^2m^2_2]
\nonumber\\
&+\frac{2}{D''_{T,{\rm LF}}}\Big[\bar x\big(3\kb'^2\kb''\cdot\qb\qb^2-4\kb''\cdot\qb(\kb'\cdot\qb)^2+\bar x\kb'\cdot\kb''\qb^4\big)\big(2xm'_1-m'_1+xm''_1\big)\nonumber\\
&+x\bar xm_2(4\kb''\cdot\qb(\kb'\cdot\qb)^2-2\kb'^2\kb'\cdot\qb\qb^2+2\bar x^2\kb'\cdot\qb\qb^4)+\bar x\big(\kb'^2\kb''\cdot\qb\nonumber\\
&-\bar x\kb'\cdot\kb''\qb^2\big)\big(x(m'_1+m''_1)(M'^2-M''^2)+\bar xm'_1\qb^2\big)+\big(2\kb'\cdot\qb\kb''\cdot\qb-\kb'\cdot\kb''\qb^2\nonumber\\
&-\bar x\kb''\cdot\qb\qb^2\big)\big((2x-1)\kb'^2(m'_1+m''_1)+2x\bar x\kb'\cdot\qb m_2+x\bar xm_2(m'^2_1-m''^2_1)+(m'_1\nonumber\\
&+m''_1)(x^2m^2_2-\bar x^2m'_1m''_1)+x\bar x(M''^2-M'^2)(\bar xm'_1+xm_2)-x\bar x^2(m'_1+m_2)\qb^2\big)
\Big]
\bigg\}\,,\\
\widetilde{T}_{2}^{\rm{ SLF}}=&\frac{M'}{x^2\bar x(M'^2-M''^2)(M'^2-M''^2+\qb^2)}\bigg\{x\bar x(2x-3)[4\kb''\cdot\qb(\kb'\cdot\qb)^2-2\kb'^2\kb''\cdot\qb\qb^2\nonumber\\
&-\bar x\kb'\cdot\kb''\kb'\cdot\qb\qb^2]-4\bar x^2\kb'^2\kb'\cdot\kb''\qb^2+(2\kb'\cdot\qb\kb''\cdot\qb-\kb'\cdot\kb''\qb^2)
[2(5x-2\nonumber\\
&-2x^2)\kb'^2+\bar x(2x^2+3x-4)m'^2_1+3x\bar xm''^2_1+2x^2m^2_2+4x\bar x(m'_1-m_2)(\bar xm'_1-xm''_1)]\nonumber\\
&+\bar x\kb'\cdot\kb''\qb^2[4(\bar xm'_1+xm_2)(xm''_1-\bar x m'_1)-4xm'_1m''_1+2x^3(M'^2+M''^2)+x^2\qb^2]\nonumber\\
&+\kb''\cdot\qb\big[4\bar x\kb'^4+\kb'^2\qb^2(4-14x+13x^3-4x^3)+\kb'^2\big(4m'_1(\bar x^2m'_1+x^2m''_1+x\bar xm_2)\nonumber\\
&+4\bar xm''_1(m''_1+xm_2)-2x^3(M'^2+M''^2)\big)-\big(\bar xm'_1+xm_2\big)\big(4m'_1m''_1(xm_2-\bar xm''_1)\nonumber\\
&+2x^2(M'^2+M''^2)(\bar xm''_1+xm_2)+\qb^2(2\bar xx^2m''_1-m'_1(4-11x+9x^2-2x^3)+x^2m_2)\big)\big]
\nonumber\\
&+\frac{2x}{D''_{T,{\rm LF}}}\Big[\bar x\big(4\kb''\cdot\qb(\kb'\cdot\qb)^2-3\kb'^2\kb''\cdot\qb\qb^2-\bar x\kb'\cdot\kb''\qb^4\big)\big(m'_1-xm''_1-xm_2\big)\nonumber\\
&-2\bar x\big(\kb'^4\qb^2-2\bar x\kb'\cdot\kb''\kb'\cdot\qb\qb^2-\bar x^2\kb'^2\qb^4\big)\big(m'_1-m_2\big)+\big(2\kb'\cdot\qb\kb''\cdot\qb-\kb'^2\qb^2\nonumber\\
&+\bar x^2\qb^4\big)\big(\kb'^2(2x-1)(m'_1+m''_1)-2\kb'^2m'_1+(\bar xm'_1+xm_2)((m'_1+m''_1)(xm_2-\bar xm''_1)\nonumber\\
&-2m'_1m_2+x\bar x(M'^2+M''^2)-\bar x^2\qb^2\big)
+\big(\kb'^2\kb''\cdot\qb-\bar x\kb'\cdot\kb''\qb^2\big)\big((2\kb'^2+2\bar x(m'_1\nonumber\\
&-m_2)(m''_1+m_2)+2m^2_2-x\bar x(M'^2+M''^2))(m'_1+m''_1)+\bar x(3\bar xm'_1+3xm_2-2m_2)\qb^2\big)
\Big]
\bigg\}\nonumber\\
&-\widetilde{T}_{1}^{\rm{ SLF}}(x,{\bf k'_\bot},q^2)\,,\\
\widetilde{T}_{3}^{\rm{ SLF}}=&\frac{2M'(M'^2-M''^2)}{x\bar x\qb^4(M'^2-M''^2+\qb^2)}\bigg\{(4x-3)\kb'^2\kb''\cdot\qb\qb^2+4\bar x\kb''\cdot\qb(\kb'\cdot\qb)^2-\bar x\kb'\cdot\kb''\qb^4\nonumber\\
&+2(2\kb'\cdot\qb\kb''\cdot\qb-\kb'^2\qb^2+\bar x^2\qb^4)(2x\kb'\cdot\kb''-\kb'^2)+[\kb''\cdot\qb\qb^2+4\kb'\cdot\qb\kb''\cdot\qb\nonumber\\
&-2\kb'^2\qb^2+2\bar x^2\qb^4](x^2m^2_2-\bar x^2m'^2_1)+\bar x(2\kb'\cdot\qb\kb''\cdot\qb-\kb'\cdot\kb''\qb^2)(m'^2_1-m''^2_1)\nonumber\\
&-\frac{2}{D''_{T,{\rm LF}}}\Big(2\kb'\cdot\qb\kb''\cdot\qb-\kb'^2\qb^2+\bar x^2\qb^4\Big)\Big[(2x\kb'\cdot\kb''-\kb'^2)(m'_1+m''_1)-\big(\bar xm'_1\nonumber\\
&+xm_2\big)\left(2\bar x\kb'\cdot\qb-\bar x(1-2x)\qb^2+(m'_1+m''_1)(xm_2-\bar xm''_1)\right)
\Big]
\bigg\}\nonumber\\
&-\frac{M'^2-M''^2}{q^2}\Big[\widetilde{T}_{1}^{\rm{ SLF}}(x,{\bf k'_\bot},q^2)-\widetilde{T}_{2}^{\rm{ SLF}}(x,{\bf k'_\bot},q^2)\Big]\,.\label{eq:A3SLF1}
\end{align}

\subsection{Theoretical results in the  CLF QM}
\begin{figure}[t]
\caption{The Feynman diagram for the matrix element within the CLF QM.}
\begin{center}
\includegraphics[scale=0.3]{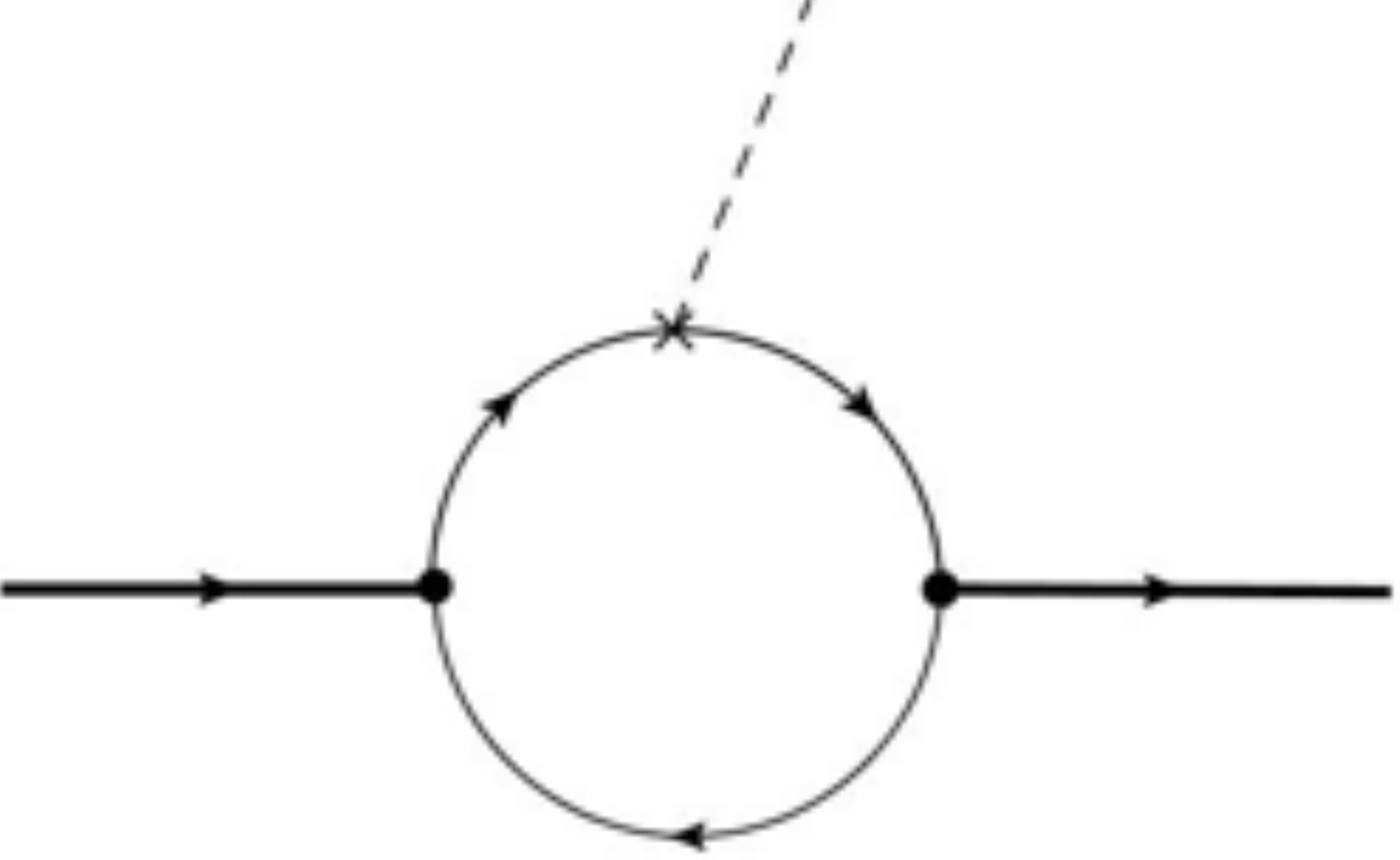}
\end{center}
\label{fig:fayn}
\end{figure}
In order to treat the complete Lorentz covariance of matrix elements and investigate the effect of zero-mode contribution,  a theoretical framework of CLF approach is developed by Jaus with the help of a manifestly covariant BS approach as a guide to the calculation. One may refer to Refs.~\cite{Jaus:1999zv,Jaus:2002sv,Cheng:2003sm} for detail. In the CLF QM, the matrix element is obtained by calculating the Feynman diagram shown by Fig.~\ref{fig:fayn}. For the $P\to T$ transition, the matrix element  can be expressed as
\begin{eqnarray}\label{eq:Bclf1}
{\cal B}=N_c \int \frac{\d^4 k_1'}{(2\pi)^4} \frac{H_{P}H_{T}}{N_1'\,N_1''\,N_2}iS_{\cal B}\cdot \epsilon^{*}\,,
\end{eqnarray}
where $\d^4 k_1'=\frac{1}{2} \d k_1'^- \d k_1'^+ \d^2 \mathbf{k}_{\bot}'$, the denominators $N_{1}^{(\prime,\prime\prime)}=k_{1}^{(\prime,\prime\prime)2}-m_1^{(\prime,\prime\prime)2}+i\varepsilon$ and $N_{2}=k_{2}^{2}-m_2^{2}+i\varepsilon$  come from the fermion propagators,  $H_{P, T}$ are  vertex functions, and $\epsilon$ is the  polarization tensor of tensor  meson. The trace term  $S_{\cal B}$ is associated with the fermion loop and is written as
\begin{eqnarray}
S_{\cal B}={\rm Tr}\left[\Gamma\, (\not\!k'_1+m'_1)\,(i\Gamma_{P})\,(-\!\not\!k_2+m_2)\,(i\r^0{\Gamma}_{T}^{\dag}\r^0) (\not\!k_1''+m_1'')\right]\,,
\end{eqnarray}
where $\Gamma_{P,T}$ are the vertex operators written as~\cite{Cheng:2003sm}
\begin{eqnarray}
i\Gamma_P&=&-i\gamma_5\,,\\
i\Gamma_T&=&i\frac{1}{2}\left[\gamma_\mu-\frac{(k''_1-k_2)_\mu}{D_{T,{\rm con}}}\right]\left(k''_1-k_2\right)_\nu\,,\quad D_{T,{\rm con}}=M+m''_1+m_2\,.
\end{eqnarray}

As has been stressed in Ref.~\cite{Jaus:1999zv},  the covariant calculation and the calculation of the light-front formalism give identical results at the one-loop level if the vertex functions $H_{P, T}$  are  analytic in the upper complex $k_1'^-$ plane. By closing the contour in the upper complex $k_1'^-$ plane and assuming that $H_{P, T}$ are analytic within the contour, the integration picks up a residue at $k_2^2=\hat{k}_2^2=m_2^2$ corresponding to putting the spectator antiquark on its mass-shell.  After integrating the minus component of loop momentum, the covariant calculation becomes the LF one. Such manipulation ask for the following replacements~\cite{Jaus:1999zv,Cheng:2003sm}
\begin{eqnarray}
N_1^{\prime(\prime\prime)} \to \hat{N}_1^{\prime(\prime\prime)}=x \left( M^{\prime(\prime\prime)2}-M_0^{\prime(\prime\prime)2}\right)
\end{eqnarray}
and
\begin{eqnarray}\label{eq:type1}
\chi_{P(T)} = H_{P(T)}/N^{\prime(\prime\prime)}\to h_{P(T)}/\hat{N}^{\prime(\prime\prime)}\,,\qquad  D_{T,{\rm  con}} \to D_{T,{\rm  LF}}\,,\qquad \text{(type-I)}
\end{eqnarray}
where the LF form of vertex function, $h_{P(T)}$,  is given by
\begin{eqnarray}
h_{P(T)}/\hat{N}^{\prime(\prime\prime)}&=&\frac{1}{\sqrt{2N_c}}\sqrt{\frac{\bar{x}}{x}}\frac{\psi_{1s(1p)}}{\hat{M}_0^{\prime(\prime\prime)}}\,,
\label{eq:vPV}
\end{eqnarray}
The Eq.~\eqref{eq:type1} shows the correspondence between manifestly covariant  and LF approaches~\cite{Jaus:1999zv,Cheng:2003sm},  the correspondence between $\chi$ and $h$  can be clearly derived by matching the CLF expressions to the SLF ones via some zero-mode independent  quantities~\cite{Jaus:1999zv,Cheng:2003sm}. However, the validity of the correspondence for the $D$ factor appearing in the vertex operator cannot been clarified explicitly~\cite{Choi:2013mda}. In fact, the traditional type-I correspondence may give self-contradictory results for some quantities. An obvious example is  $f_{V}$ noted by the authors of Ref.~\cite{Cheng:2003sm}. In order to get self-consistent results for $f_{V}$, a much more generalized correspondence scheme is proposed~\cite{Choi:2013mda},
\begin{eqnarray}\label{eq:type2}
\chi_{P(T)} = H_{P(T)}/N^{\prime(\prime\prime)}\to h_{P(T)}^{\prime(\prime\prime)}/\hat{N}^{\prime(\prime\prime)}\,,\qquad  M\to  M_0\,.\qquad \text{(type-II)}
\end{eqnarray}
Within this updated self-consistent scheme, the CLF QM can give a self-consistent result for $f_{V,A}$ and form factors of $P\to(V,A)$ and $V'\to V''$ transitions~\cite{Choi:2013mda,Chang:2018zjq,Chang:2019mmh,Chang:2019obq}, while the self-consistency of the form factors of $P\to T$ transition remains to be tested.

Using above formulas and  integrating out $k_1'^-$, the matrix element, Eq.~\eqref{eq:Bclf1}, can be reduced to the LF form
\begin{eqnarray}
\label{eq:Bclf2}
\hat{{\cal B}}=N_c \int \frac{\d x \d^2 \mathbf{k}_{\bot}'}{2(2\pi)^3}\frac{h_{P}h_{T}}{\bar{x} \hat{N}_1'\,\hat{N}_1''\,}\hat{S}_{\cal B}\cdot \epsilon^{*}\,.
\end{eqnarray}
However, the matrix element obtained in this way contains spurious $\w^\mu$-dependent contributions, which violate the covariance of matrix element. It should be noted that the contribution of the zero-mode from the $k_1'^+=0$ is not taken into account in the contour integration. It is interesting that the spurious $\w^\mu$-dependent terms in $\hat{{\cal B}}$ can be eliminated by the zero-mode contributions~\cite{Jaus:1999zv}. The inclusion of  zero-mode contributions in practice amount to the following replacements
\begin{align}
 \label{eq:repFF1}
\hat{k}_1'^{\mu} \to& P^\u A_1^{(1)}+q^\u A_2^{(1)} \,,\\
\hat{k}_1'^{\mu}\hat{k}_1'^{\nu} \to &g^{\u\v}A_1^{(2)}+P^\u P^\v A_2^{(2)}+(P^\u q^\v+q^\u P^\v)A_3^{(2)}+q^\u q^\v A_4^{(2)}\nonumber\\
&+\frac{P^\u\omega^\v+\omega^\u P^\v}{\omega\cdot P}B_1^{(2)}\,,\label{eq:repk2}\\
\hat k_1'^{\mu}\hat k_1'^{\nu}\hat k_1'^{\alpha}\to&\left(g^{\mu \nu}P^\alpha+g^{\mu \alpha}P^\nu+g^{\nu\alpha}P^\mu\right)A_1^{(3)}+\left(g^{\mu \nu}q^\alpha+g^{\mu \alpha}q^\nu+g^{\nu\alpha}q^\mu\right)A_2^{(3)}\nonumber\\
&+P^\mu P^\nu P^\alpha A_3^{(3)}+\left(P^\mu P^\nu q^\alpha+P^\mu q^\nu P^\alpha+q^\mu P^\nu P^\alpha\right)A_4^{(3)}\nonumber\\
&+\left(q^\mu q^\nu P^\alpha+q^\mu P^\nu q^\alpha+P^\mu q^\nu q^\alpha\right)A_5^{(3)}+q_\mu q_\nu q_\alpha A_6^{(3)}\nonumber\\
&+\frac{1}{\omega \cdot P}\left(P^\mu P^\nu\omega^\alpha+P^\mu \omega^\nu P^\alpha+\omega^\mu P^\nu P^\alpha\right)B_1^{(3)}\nonumber\\
&+\frac{1}{\omega\cdot P}\left[\left(P^\mu q^\nu+q^\mu P^\nu\right)\omega^\alpha+\left(P^\mu q^\alpha+q^\mu P^\alpha\right)\omega^\nu+\left(P^\alpha q^\nu+q^\alpha P^\nu\right)\omega^\mu\right]B_2^{(3)}\,,\label{eq:repk3}\\
k_1'^{\mu}\hat{N}_2\to& q^\u\left(A_2^{(1)}Z_2+\frac{q\cdot P}{q^2}A_1^{(2)} \right) \,,\\
\hat{k}_1'^{\mu}\hat{k}_1'^{\nu}\hat{N}_2\to &g^{\u\v}A_1^{(2)}Z_2+q^\u q^\v\left( A_4^{(2)}Z_2+2\frac{q\cdot P}{q^2}A_2^{(1)}A_1^{(2)}\right)+\frac{P^\u\omega^\v+\omega^\u P^\v}{\omega\cdot P}B_3^{(3)}\,,\label{eq:repk2N}\\
\hat k_1'^{\mu}\hat k_1'^{\nu}\hat k_1'^{\alpha}\hat N_2\to &\left(g^{\mu \nu}q^\alpha+g^{\mu \alpha}q^\nu+g^{\nu \alpha}q^\mu\right)\left(A_2^{(3)}Z_2+\frac{q\cdot P}{q^2}A_1^{(4)}\right)+q^\mu q^\nu q^\alpha \left(A_6^{(3)}Z_2+3\frac{q\cdot P}{q^2}A_4^{(4)}\right)\nonumber\\
&+\frac{1}{\omega\cdot P}\left[\left(P^\mu q^\nu+q^\mu P^\nu\right)\omega^\alpha+\left(P^\mu q^\alpha+q^\mu P^\alpha\right)\omega^\nu+\left(P^\alpha q^\nu+q^\alpha P^\nu\right)\omega^\mu\right]B_5^{(4)}\,,
 \label{eq:repFF}
\end{align}
where $A$ and $ B$ functions are given by
\begin{align}
A_1^{(1)}&=  \frac{x}{2}\,,\qquad
A_2^{(1)}=\frac{x}{2} -\frac{\kb' \cdot \qb }{q^2}\,;\\
A_1^{(2)}&=-\kb'^2 -\frac{(\kb' \cdot \qb)^2}{q^2}\,,\qquad
A_2^{(2)}=(A_1^{(1)})^2\,,\qquad
A_3^{(2)}=A_1^{(1)}A_2^{(1)}\,,\nonumber\\
A_4^{(2)}&=(A_2^{(1)})^2-\frac{1}{q^2}A_1^{(2)}\,;\\
A_1^{(3)}&=A_1^{(1)}A_1^{(2)}\,,\qquad
A_2^{(3)}=A_2^{(1)}A_1^{(2)}\,,\qquad
A_3^{(3)}=A_1^{(1)}A_2^{(2)}\,,\qquad
A_4^{(3)}=A_2^{(1)}A_2^{(2)}\,,\nonumber\\
A_5^{(3)}&=A_1^{(1)}A_4^{(2)}\,,\qquad
A_6^{(3)}=A_2^{(1)}A_4^{(2)}-\frac{2}{q^2}A_2^{(1)}A_1^{(2)}\,;\\
A_3^{(4)}&=A_1^{(1)}A_2^{(3)}\,;\\
B_1^{(2)}&=\frac{x}{2}Z_2-A_1^{(2)}\,;\\
B_2^{(3)}&=\frac{x}{2}Z_2A_2^{(1)}+A_1^{(1)}A_1^{(2)} \frac{P\cdot q}{q^2}-A_2^{(1)}A_1^{(2)},
~~B_3^{(3)}=B_1^{(2)}Z_2+\left(P^2-\frac{(q\cdot P)^2}{q^2}\right)A_1^{(1)}A_1^{(2)};\\
 B_1^{(4)}&=\frac{x}{2}Z_2A_1^{(2)}-A_1^{(4)}\,,\qquad
B_5^{(4)}=B_2^{(3)}Z_2+\frac{q\cdot P}{q^2}B_1^{(4)}+\left[P^2-\frac{\left(q\cdot P\right)^2}{q^2}\right]A_3^{(4)} \,;\\
Z_2&=\hat{N}_1'+m_1'^2-m_2^2+(\bar{x}-x)M'^2+(q^2+q\cdot P)\frac{\kb' \cdot \qb}{q^2}\,.
\end{align}
The $\w$-dependent terms associated with the  $C$ functions are not shown because they can be eliminated exactly by the inclusion of the zero-mode contributions~\cite{Jaus:1999zv}. However, there are still some residual $\w$-dependences associated with the  $B$ functions, which are irrelevant zero-mode contribution~\cite{Jaus:1999zv}, and possibly result in the inconsistence and covariance problems~\cite{Choi:2013mda,Chang:2018zjq,Chang:2019mmh,Chang:2019obq}. 

Using the formulae given above, one can obtain CLF results for the matrix elements of $P\to T$ transition,  ${\cal B}_{\rm  CLF}^\u$.  Then, matching obtained ${\cal B}_{\rm  CLF}^\u(\Gamma=\r_\u)$ , ${\cal B}_{\rm  CLF}^\u(\Gamma=\r_\u\r_5)$ and ${\cal B}_{\rm  CLF}^\u(\Gamma=\sigma_{\u\v}(1+\r_5)q^\v)$  to the definitions of  form factors given by Eqs.~(\ref{eq:depvt}-\ref{eq:deptt}), the CLF results for the form factors of $P\to T$ transition can be extracted directly. They can be written as
\begin{eqnarray}\label{eq:FCLF}
[\mathcal F(q^2)]_{\rm CLF}=N_c\int\frac{\d x\d^2{\bf k'_\bot}}{2(2\pi)^3}\frac{\chi_P'\chi_T''}{\bar x}{\cal \widetilde{\cal F}}^{\rm CLF}(x,{\bf k'_\bot},q^2)\,,
\end{eqnarray}
where the integrands are
\begin{align}
\label{eq:VCLF}
\widetilde{V}^{\rm{CLF}}=&2M'(M'+M'')\Big[m'_1(1-2A^{(1)}_1-2A^{(1)}_2+A^{(2)}_2+2A^{(2)}_3+A^{(2)}_4)\nonumber\\
&+m''_1(-A^{(1)}_1+A^{(1)}_2+A^{(2)}_2-A^{(2)}_4)+2m_2(A^{(1)}_1-A^{(2)}_2-A^{(2)}_3)\nonumber\\
&+\frac{4}{D''_{T,{\rm con}}}(-A^{(2)}_1+A^{(3)}_1+A^{(3)}_2)
\Big]\,,
\\
\widetilde{A}_0^{\rm CLF}=&\frac{M'q^2}{M''}\bigg\{m'_1(3-6A^{(1)}_1-10A^{(1)}_2+3A^{(2)}_2+14A^{(2)}_3+11A^{(2)}_4-4A^{(3)}_4-8A^{(3)}_5-4A^{(3)}_6)\nonumber\\
&+m''_1(-A^{(1)}_1+A^{(1)}_2+A^{(2)}_2-A^{(2)}_4)
+2m_2(A^{(1)}_1+2A^{(1)}_2-A^{(2)}_2-5A^{(2)}_3-4A^{(2)}_4\nonumber\\
&+2A^{(3)}_4+4A^{(3)}_5+2A^{(3)}_6)
+\frac{1}{D''_{T,{\rm con}}}\Big[\big(-2M'^2+2(m'_1-m_2)^2-(m'_1+m''_1)^2+\hat N'_1\nonumber\\
&-\hat N''_1-\qb^2\big)(1-2A^{(1)}_1-2A^{(1)}_2+A^{(2)}_2+2A^{(2)}_3+A^{(2)}_4)+2\big(M'^2+M''^2\nonumber\\
&+2(m'_1-m_2)(m''_1+m_2)+\qb^2\big)(A^{(1)}_2-2A^{(2)}_3-2A^{(2)}_4+A^{(3)}_4+2A^{(3)}_5+A^{(3)}_6)\nonumber\\
&+2Z_2(1-4A^{(1)}_2+5A^{(2)}_4-2A^{(3)}_6)+\frac{4(M'^2-M''^2)}{q^2}(-2A^{(2)}_1+5A^{(1)}_2A^{(2)}_1-3A^{(4)}_4)
\Big]\bigg\}\nonumber\\
&+\frac{M'+M''}{2M''}\widetilde{A}_1^{\rm CLF}(x,{\bf k'_\bot},q^2)-\frac{M'-M''}{2M''}\widetilde{A}_2^{\rm CLF}(x,{\bf k'_\bot},q^2)\,,
\\
\widetilde{A}_1^{\rm CLF}=&\frac{2M'}{M'+M''}\bigg\{m'_1(M''^2-m''^2_1-m^2_2-\hat N''_1-Z_2)+m''_1(M'^2-m'^2_1-m^2_2-\hat N'_1-Z_2)\nonumber\\
&+m_2[(m'_1+m''_1)^2+\hat N'_1+\hat N''_1+\qb^2]-\Big[m'_1(M''^2-m''^2_1-m^2_2-\hat N''_1)+m''_1(M'^2\nonumber\\
&-m'^2_1-m^2_2-\hat N'_1)+m_2((m'_1+m''_1)^2+\hat N'_1+\hat N''_1+\qb^2)
\Big]\Big[A^{(1)}_1+A^{(1)}_2\Big]\nonumber\\
&+(m'_1+m''_1)\left[Z_2A^{(1)}_2+\frac{M'^2-M''^2}{q^2}A^{(2)}_1\right]+2(4m'_1-m''_1-3m_2)A^{(2)}_1\nonumber\\
&-8(m'_1-m_2)(A^{(3)}_1+A^{(3)}_2)-\frac{4}{D''_{T,{\rm con}}}\Big[\big(M'^2+M''^2+\qb^2+2(m'_1-m_2)(m''_1\nonumber\\
&+m_2)\big)(A^{(2)}_1-A^{(3)}_1-A^{(3)}_2)+2\left(-Z_2A^{(2)}_1+Z_2A^{(3)}_2+\frac{M'^2-M''^2}{q^2}A^{(4)}_1\right)
\Big]\bigg\}\,,
\\
\widetilde{A}_2^{\rm CLF}=&2M'(M'+M'')\bigg\{-m'_1+(6m'_1-m''_1-2m_2)A^{(1)}_1+(2m'_1+m''_1)A^{(1)}_2-(m'_1+m''_1)A^{(2)}_4\nonumber\\
&-(9m'_1-m''_1-6m_2)A^{(2)}_2-(10m'_1-6m_2)A^{(2)}_3+4(m'_1-m_2)(A^{(3)}_3+2A^{(3)}_4+A^{(3)}_5)\nonumber\\
&-\frac{1}{D''_{T,{\rm con}}}\Big[\big((m'_1+m''_1)^2+\hat N'_1+\hat N''_1+\qb^2\big)\big(-1+2A^{(1)}_1+2A^{(1)}_2-A^{(2)}_2-2A^{(2)}_3-A^{(2)}_4
\big)\nonumber\\
&+2\big(M'^2+M''^2+\qb^2+2(m'_1-m_2)(m''_1+m_2)\big)(A^{(1)}_1-2A^{(2)}_2-2A^{(2)}_3+A^{(3)}_3+2A^{(3)}_4\nonumber\\
&+A^{(3)}_5)
\Big]\bigg\}\,,\\
\widetilde{T}_1^{\rm{ CLF}}=&M'\bigg\{4A^{(3)}_1+12A^{(3)}_2-10A^{(2)}_1+4(M'^2-M''^2)(A^{(2)}_2-A^{(2)}_3-A^{(3)}_3+A^{(3)}_5)\nonumber\\
&+4q^2(A^{(1)}_2-A^{(1)}_1+A^{(2)}_2+A^{(2)}_3-2A^{(2)}_4-A^{(3)}_4+A^{(3)}_6)+\big[(m'_1+m''_1)^2+\hat N'_1+\hat N''_1\nonumber\\
&+\qb^2\big](1-A^{(1)}_1-A^{(1)}_2)+2\big[M'^2-M''^2-\qb^2+2(m'_1+m''_1)(m_2-m'_1)-2\hat N'_1\big](A^{(1)}_1\nonumber\\
&-A^{(2)}_2-A^{(2)}_3)+\frac{8(m'_1+m''_1)}{D''_{T,{\rm con}}}(A^{(3)}_1+A^{(3)}_2-A^{(2)}_1)
\bigg\}\,,
\\
\widetilde{T}_2^{\rm{CLF}}=&M'\bigg\{
\Big[(m'_1-m''_1)^2+\hat N'_1+\hat N''_1+\qb^2
-\frac{q^2}{M'^2-M''^2}\big(2(m'_1-m_2)^2-(m'_1-m''_1)^2\nonumber\\
&-2M'^2+\hat N'_1-\hat N''_1-\qb^2\big)\Big](A^{(1)}_1+A^{(1)}_2-1)+\frac{2Z_2q^2}{M'^2-M''^2}(1-3A^{(1)}_2+2A^{(2)}_4)\nonumber\\
&+2\Big[M'^2+M''^2-2(m'_1-m_2)(m''_1-m_2)+\qb^2\Big]\Big[A^{(1)}_1-A^{(2)}_2-A^{(2)}_3+\frac{q^2}{M'^2-M''^2}(A^{(1)}_2\nonumber\\
&-A^{(2)}_3-A^{(2)}_4)\Big]+8(A^{(1)}_2A^{(2)}_1-A^{(3)}_1-A^{(3)}_2)+\frac{2q^2}{M'^2-M''^2}(5A^{(2)}_1-4A^{(3)}_1-4A^{(3)}_2)\nonumber\\
&+\frac{8}{D''_{T,{\rm con}}}\Big[m'_1+m''_1-\frac{q^2}{M'^2-M''^2}(m'_1-m''_1-2m_2)
\Big]\Big[A^{(3)}_1+A^{(3)}_2-A^{(2)}_1\Big]\bigg\}\,,
\\
\widetilde{T}_3^{\rm{CLF}}=&M'\bigg\{[2M'^2+(m'_1-m''_1)^2-2(m'_1-m_2)^2-\hat N'_1+\hat N''_1+\qb^2](1-A^{(1)}_1-A^{(1)}_2)\nonumber\\
&+2[M'^2+M''^2+2(m'_1-m_2)(m_2-m''_1)+\qb^2](A^{(2)}_3+A^{(2)}_4-A^{(1)}_2)+2(4A^{(3)}_1+4A^{(3)}_2\nonumber\\
&-5A^{(2)}_1)+2(M'^2-M''^2)\Big[2(A^{(1)}_1-A^{(1)}_2-2A^{(2)}_2+2A^{(2)}_4+A^{(3)}_3+A^{(3)}_4-A^{(3)}_5-A^{(3)}_6)\nonumber\\
&+\frac{1}{q^2}(3A^{(2)}_1-4A^{(1)}_2A^{(2)}_1)\Big]+2Z_2(-1+3A^{(1)}_2-2A^{(2)}_4)\nonumber\\
&+\frac{4}{D''_{T,{\rm con}}}\Big[m'_1(M'^2-M''^2)(3A^{(1)}_1+3A^{(1)}_2-3A^{(2)}_2-6A^{(2)}_3-3A^{(2)}_4+A^{(3)}_3+3A^{(3)}_4+3A^{(3)}_5\nonumber\\
&+A^{(3)}_6-1)+2m'_1(-A^{(2)}_1+A^{(3)}_1+A^{(3)}_2)+m''_1(M'^2-M''^2)(A^{(1)}_2-A^{(1)}_1+2A^{(2)}_2-2A^{(2)}_4\nonumber\\
&-A^{(3)}_3-A^{(3)}_4+A^{(3)}_5+A^{(3)}_6)+2m''_1(A^{(2)}_1-A^{(3)}_1-A^{(3)}_2)+2m_2(M'^2-M''^2)(2A^{(2)}_2\nonumber\\
&-A^{(1)}_1+2A^{(2)}_3-A^{(3)}_3-2A^{(3)}_4-A^{(3)}_5)+4m_2(A^{(2)}_1-A^{(3)}_1-A^{(3)}_2)
\Big]\bigg\}\,.
\label{eq:T3CLF}
\end{align}
\tb{The CLF results given above are independent of $\u$ and $\lbd$, which implies that the CLF contributions are irrelevant to the self-consistence and covariance problems. This is a significant advantage of  CLF QM compared the SLF QM.}  However, it should be noted that the contributions associated with $B$ functions are not included in above formulae. These contributions may result in the self-consistence and covariance problems \tb {of CLF QM except they are equal to zero numerically}, and will be analyzed   in the next section.

In our following discussions, we also need the valence contributions, which can be obtained by assuming $k_2^+\neq 0$~(or $p^+\neq 0$). This assumption ensures the pole of $N_2$ is safely located in the contour of  $k_1'^-$ integral~( the pole of $N_2$ is finite) and implies that the zero-mode contributions are not taken into account. At this moment, the replacements for $\hat{k}_1'^{\u}$  given above have to be disregarded. Instead, one just need to directly use the on-mass-shell condition of spectator antiquark, $k_2^2=m_2^2$, and the conservation of four-momentum at each vertex. The valence contributions to the form factors can also be expressed as Eq.~\eqref{eq:FCLF} with the integrands  written as
\begin{eqnarray}
\widetilde{V}^{\rm val.}&=&\frac{2M^{\prime}(M^{\prime}+M^{\prime\prime})}{\bar xq^2(M^{\prime2}-M^{\prime\prime2}+\textbf{q}_\bot^2)}\bigg\{2\bar x\textbf{k}^{\prime}_\bot\cdot\textbf{q}_\bot\textbf{k}^{\prime\prime}_\bot\cdot\textbf{q}_\bot(m^{\prime}_1
-m^{\prime\prime}_1)+\textbf{k}^{\prime\prime}_\bot\cdot\textbf{q}_\bot\big[(\textbf{k}_\bot^{\prime2}+m_2^2)(m''_1-m'_1
)\nonumber\\
&&+\bar x^2((m^{\prime\prime}_1-m_2)M^{\prime2}
-(m^{\prime}_1-m_2)(M^{\prime\prime2}-\textbf{q}_\bot^2))+\bar x m_2(M^{\prime2}-M^{\prime\prime2}+\textbf{q}_\bot^2)\big]
\nonumber\\
&&-\frac{1}{D''_{T,{\rm con}}}\Big[(\textbf{k}^{\prime2}_\bot\textbf{q}_\bot^2-2\bar x\textbf{k}^{\prime}_\bot\cdot\textbf{q}_\bot\textbf{k}^{\prime\prime}_\bot\cdot\textbf{q}_\bot-\bar x^2\textbf{q}_\bot^4)(\textbf{k}_\bot^{\prime2}-\bar x^2M^{\prime2}+m_2^2)
+(\bar x\textbf{k}^{\prime2}_\bot\textbf{k}^{\prime\prime}_\bot\cdot\textbf{q}_\bot
\nonumber\\
&&+\bar x^2\textbf{q}_\bot^2\textbf{k}^{\prime}_\bot\cdot\textbf{k}^{\prime\prime}_\bot)(M^{\prime2}-M^{\prime\prime2})
+\bar x\big(4(\textbf{k}^{\prime}_\bot\cdot\textbf{q}_\bot)^2\textbf{k}^{\prime\prime}_\bot\cdot\textbf{q}_\bot-2\textbf{k}_\bot^{\prime2}\textbf{q}_\bot^2
\textbf{k}^{\prime\prime}_\bot\cdot\textbf{q}_\bot-\textbf{k}^{\prime2}_\bot\textbf{k}^{\prime}_\bot\cdot\textbf{q}_\bot\textbf{q}_\bot^2
\nonumber\\&&
+\bar x^2\textbf{k}^{\prime}_\bot\cdot\textbf{q}_\bot\textbf{q}_\bot^4\big)\Big]\bigg\}\,,
\\
\widetilde{A}_0^{\rm val.}&=&\frac{M^{\prime}}{\bar xM^{\prime\prime}\textbf{q}_\bot^4}\bigg\{\bar x\Big[8(\textbf{k}^{\prime}_\bot\cdot\textbf{q}_\bot)^2\textbf{k}^{\prime\prime}_\bot\cdot\textbf{q}_\bot-6\textbf{k}^{\prime2}_\bot\textbf{k}^{\prime\prime}_\bot\cdot\textbf{q}_\bot
\textbf{q}_\bot^2
-4\bar x\textbf{k}_\bot^{\prime2}\textbf{q}_\bot^4
+4\bar x(\textbf{k}^{\prime}_\bot\cdot\textbf{q}_\bot)^2\textbf{q}_\bot^2\Big]
(m^{\prime}_1-m_2)\nonumber\\
&&+\bar x\Big[2\textbf{k}^{\prime}_\bot\cdot\textbf{q}_\bot\textbf{k}^{\prime\prime}_\bot\cdot\textbf{q}_\bot-
\textbf{k}^{\prime}_\bot\cdot\textbf{k}^{\prime\prime}_\bot\textbf{q}_\bot^2\Big]\Big[\big((1-2x)m^{\prime}_1+m^{\prime\prime}_1
+2xm_2\big)(M^{\prime2}-M^{\prime\prime2})\nonumber\\&&
-2\textbf{q}_\bot^2\big(2\bar xm^{\prime}_1-(1-2x)m_2\big)\Big]
+\bar x\textbf{k}^{\prime}_\bot\cdot\textbf{k}^{\prime\prime}_\bot\textbf{q}_\bot^4\Big[(2x-1)m^{\prime}_1
+m^{\prime\prime}_1-2xm_2\Big]\nonumber\\
&&+\textbf{k}^{\prime\prime}_\bot\cdot\textbf{q}_\bot\textbf{q}_\bot^2\Big[\bar x^2M^{\prime2}\big((2x-1)m^{\prime}_1-m^{\prime\prime}_1-2xm_2\big)+2(xm_2+\bar xm^{\prime}_1)(\bar x^2\textbf{q}_\bot^2
-x\bar xM^{\prime\prime2})\nonumber\\
&&+\textbf{k}_\bot^{\prime2}\big((1-2x)m^{\prime}_1-m^{\prime\prime}_1+2xm_2\big)
-m_2^2(m^{\prime}_1+m^{\prime\prime}_1-2m_2)+2\bar xm_2(m^{\prime}_1
-m_2)(m_2-m^{\prime\prime}_1)\Big]
\nonumber\\
&&-\frac{2}{D''_{T,{\rm con}}}\Big[2\bar x\big(2(\textbf{k}^{\prime}_\bot\cdot\textbf{q}_\bot)^2\textbf{k}^{\prime\prime}_\bot\cdot\textbf{q}_\bot
-\textbf{k}^{\prime}_\bot\cdot\textbf{q}_\bot\textbf{k}^{\prime2}_\bot\textbf{q}_\bot^2+\bar x^2\textbf{k}^{\prime}_\bot\cdot\textbf{q}_\bot
\textbf{q}_\bot^4\big)\big(M^{\prime2}
+(m^{\prime}_1-m_2)(m^{\prime\prime}_1+m_2)\big)\nonumber\\
&&+\Big(
\textbf{k}^{\prime2}_\bot\textbf{q}_\bot^2-2\textbf{k}^{\prime}_\bot\cdot\textbf{q}_\bot\textbf{k}^{\prime\prime}_\bot\cdot\textbf{q}_\bot-\bar x^2\textbf{q}_\bot^4\Big)\Big(\bar x\textbf{q}_\bot^2\big(
\bar xM^{\prime2}-(x-2)m^{\prime}_1m_2+m^{\prime\prime}_1(\bar xm^{\prime}_1+xm_2)-\bar xm_2^2\big)\nonumber\\
&&+\big(\textbf{k}_\bot^{\prime2}+m_2^2
+\bar xm_2(m'_1-m''_1-2m_2)+\bar x^2(m_2-m^{\prime}_1)(m^{\prime\prime}_1+m_2)\big)\big(M^{\prime2}-M^{\prime\prime2}\big)\Big)\Big]\bigg\}\,,
\\
\widetilde{A}_1^{\rm val.}&=&\frac{M^{\prime}}{\bar x(M^{\prime}+M^{\prime\prime})\textbf{q}_\bot^2}\bigg\{-\bar x\Big[2(\textbf{k}^{\prime}_\bot\cdot\textbf{q}_\bot)^2-\textbf{k}_\bot^{\prime2}\textbf{q}_\bot^2\Big]\Big[-2
m^{\prime\prime}_1+(1-4x)m_2+(4x-3)m^{\prime}_1\Big]
\nonumber\\
&&-(\textbf{k}^{\prime}_\bot\cdot\textbf{q}_\bot)\Big[\textbf{k}_\bot^{\prime2}\big((3-4x)m^{\prime}_1
+3m^{\prime\prime}_1+(4x-2)m_2\big)+\bar x\textbf{q}_\bot^2\big(9\bar x^2m^{\prime}_1-3x\bar xm'_1
+3\bar xm^{\prime\prime}_1
\nonumber\\
&&+12x\bar xm_2-2m_2\big)
+\bar x^2M^{\prime\prime2}\big(-2xm_2-m^{\prime\prime}_1
+(4x-1)m^{\prime}_1\big)+2xm_2^3
+(1-2x)m^{\prime}_1m_2^2\nonumber\\
&&+(1+2x)m^{\prime\prime}_1m_2^2+2\bar xm^{\prime}_1m^{\prime\prime}_1m_2\Big]-\textbf{q}_\bot^2\Big[\textbf{k}_\bot^{\prime2}\big(\bar x(8x-6)m^{\prime}_1-4\bar xm^{\prime\prime}_1
+(8x^2-10x+3)m_2\big)\nonumber\\
&&+(1-4x)(M^{\prime\prime2}-\qb^2)(x\bar x^2m_2+\bar x^3m^{\prime}_1)-2\textbf{q}_\bot^2(x\bar x^2m_2+\bar x^3m^{\prime}_1)
-2\bar x^2m^{\prime}_1m^{\prime\prime}_1m_2\nonumber\\
&&+(2x-1)m_2^2(xm_2
+\bar xm^{\prime}_1)-2x\bar xm^{\prime\prime}_1m_2^2\Big]
+\frac{2}{\bar xD''_{T,{\rm con}}}\Big[2\bar x\textbf{k}^{\prime}_\bot\cdot\textbf{q}_\bot\textbf{k}^{\prime\prime}_\bot\cdot\textbf{q}_\bot\big(3\textbf{k}_\bot^{\prime2}+m_2^2
\nonumber\\
&&+2(\bar xm'_1+xm_2)(xm_2-\bar xm''_1)
-\bar x^2(M^{\prime\prime2}-\textbf{q}_\bot^2)\big)
-4\bar x^2(\textbf{k}^{\prime}_\bot\cdot\textbf{q}_\bot)^2\textbf{k}^{\prime\prime}_\bot\cdot\textbf{q}_\bot\nonumber\\
&&-2\textbf{k}^{\prime\prime}_\bot\cdot\textbf{q}_\bot\big(\textbf{k}_\bot^{\prime2}
+(\bar xm^{\prime}_1+xm_2)(xm_2-\bar xm''_1)\big)\big(\textbf{k}_\bot'^2+m_2^2
-\bar x^2(M^{\prime\prime2}-\textbf{q}_\bot^2)\big)\Big]\bigg\}
\nonumber\\
&&
+\frac{2(M^{\prime2}-M^{\prime\prime2}+\textbf{q}_\bot^2)}{(M^{\prime}+M^{\prime\prime})^2}\widetilde{A}_2^{\rm val.}(x,\textbf{k}^{\prime}_\bot,q^2)\,,
\\
\widetilde{A}_2^{\rm val.}&=&\frac{2M^{\prime}(M^{\prime}+M^{\prime\prime})}{q^4}\bigg\{(\textbf{k}^{\prime}_\bot\cdot\textbf{k}^{\prime\prime}_\bot
\textbf{q}_\bot^2-2\textbf{k}^{\prime}_\bot
\cdot\textbf{q}_\bot\textbf{k}^{\prime\prime}_\bot\cdot\textbf{q}_\bot)\big[(1-2x)m^{\prime}_1+m^{\prime\prime}_1+2xm_2\big]\nonumber\\
&&+\textbf{k}^{\prime\prime}_\bot\cdot\textbf{q}_\bot\textbf{q}_\bot^2(1-2x)(\bar xm^{\prime}_1+xm_2)
+\frac{2}{\bar xD''_{T,{\rm con}}}\Big[\big(\textbf{k}^{\prime2}_\bot\textbf{q}_\bot^2-2\textbf{k}^{\prime}_\bot\cdot\textbf{q}_\bot
\textbf{k}^{\prime\prime}_\bot\cdot\textbf{q}_\bot-\bar x^2\textbf{q}_\bot^4\big)\big(\textbf{k}_\bot^{\prime2}\nonumber\\
&&+(\bar xm'_1+xm_2)(xm_2-\bar xm''_1)\big)
+2\bar x(\textbf{k}^{\prime}_\bot\cdot\textbf{q}_\bot)^2\textbf{k}^{\prime\prime}_\bot\cdot\textbf{q}_\bot
-\bar x\textbf{k}^{\prime2}_\bot\textbf{k}^{\prime}_\bot\cdot\textbf{q}_\bot\textbf{q}_\bot^2+\bar x^3\textbf{k}^{\prime}_\bot\cdot\textbf{q}_\bot\textbf{q}_\bot^4\Big]\bigg\}\,,
\\
\widetilde{T}_1^{\rm val.}&=&\frac{M^{\prime}}{\bar xq^2(M^{\prime2}-M^{\prime\prime2}+\textbf{q}_\bot^2)}\bigg\{
\bar x\Big[8(\textbf{k}^{\prime}_\bot\cdot\textbf{q}_\bot)^2\textbf{k}^{\prime\prime}_\bot\cdot\textbf{q}_\bot-6\textbf{k}^{\prime2}_\bot
\textbf{q}_\bot^2\textbf{k}^{\prime\prime}_\bot\cdot\textbf{q}_\bot+4\bar x(\textbf{k}^{\prime}_\bot\cdot\textbf{q}_\bot)^2\textbf{q}_\bot^2-4\bar x\textbf{k}^{\prime2}_\bot\textbf{q}_\bot^4\Big]
\nonumber\\
&&+\Big(\kb'\cdot\kb''\qb^2-2\textbf{k}^{\prime}_\bot\cdot\textbf{q}_\bot\textbf{k}^{\prime\prime}_\bot\cdot\textbf{q}_\bot\Big)\Big[2\textbf{k}_\bot^{\prime2}+\bar x((2x-1)M^{\prime2}-M^{\prime\prime2}+2\bar x\textbf{q}_\bot^2)+2m_2^2\Big]
\nonumber\\
&&-\bar x\textbf{k}^{\prime\prime}_\bot\cdot\textbf{q}_\bot\textbf{q}_\bot^2[2xM^{\prime2}-2xM^{\prime\prime2}+\textbf{q}_\bot^2]
+\textbf{k}^{\prime\prime}_\bot\cdot\textbf{q}_\bot\Big[2\bar x\textbf{q}_\bot^2\big(\textbf{k}_\bot^{\prime2}+m_2(m^{\prime}_1+m^{\prime\prime}_1-m_2)\big)\nonumber\\
&&-2\bar x\big(\textbf{k}_\bot^{\prime2}
-m_2(m^{\prime}_1+m^{\prime\prime}_1-2m_2)(M^{\prime2}-M^{\prime\prime2})\big)
+(\textbf{k}_\bot^{\prime2}+m_2^2)(2M^{\prime2}-2M^{\prime\prime2}+\textbf{q}_\bot^2)
+\nonumber\\
&&\bar x^2(2x-1)M'^2\qb^2+2\bar x^2(m^{\prime}_1-m_2)(m^{\prime\prime}_1-m_2)(M'^2-M''^2+\textbf{q}_\bot^2)
\Big]\nonumber\\
&&-\frac{2}{D''_{T,{\rm con}}}\Big(m^{\prime}_1+m^{\prime\prime}_1\Big)\Big[4\bar x(\textbf{k}^{\prime}_\bot\cdot\textbf{q}_\bot)^2\textbf{k}^{\prime\prime}_\bot\cdot\textbf{q}_\bot-\bar x\textbf{k}^{\prime2}_\bot\textbf{q}_\bot^2\textbf{k}^{\prime\prime}_\bot
\cdot\textbf{q}_\bot-\bar x^2\textbf{k}^{\prime}_\bot\cdot\textbf{k}_\bot^{\prime\prime}\textbf{q}_\bot^4
\nonumber\\
&&+(\textbf{k}^{\prime2}_\bot\textbf{q}_\bot^2
-2\textbf{k}^{\prime}_\bot\cdot\textbf{q}_\bot\textbf{k}^{\prime\prime}_\bot\cdot\textbf{q}_\bot
-\bar x^2\textbf{q}_\bot^4)(\textbf{k}_\bot^{\prime2}-\bar x^2M^{\prime2}+m_2^2)\nonumber\\
&&-\bar x(\kb'^2\kb''\cdot\qb-\bar x\kb'\cdot\kb''\qb^2)(M^{\prime2}
-M^{\prime\prime2})\Big]\bigg\}\,,
\\
\widetilde{T}_2^{\rm val.}&=&\frac{M^{\prime}}{\bar x(M^{\prime2}-M^{\prime\prime2})(M^{\prime2}-M^{\prime\prime2}+\textbf{q}_\bot^2)}\bigg\{\bar x\Big[8(\textbf{k}^{\prime}_\bot\cdot\textbf{q}_\bot)^2\textbf{k}^{\prime\prime}_\bot\cdot\textbf{q}_\bot-6\textbf{k}^{\prime2}_\bot\textbf{q}_\bot^2
\textbf{k}^{\prime\prime}_\bot\cdot\textbf{q}_\bot-4\bar x\textbf{k}^{\prime2}_\bot\textbf{q}_\bot^4
\nonumber\\
&&+4\bar x(\textbf{k}^{\prime}_\bot\cdot\textbf{q}_\bot)^2\textbf{q}_\bot^2\Big]-\Big(2\textbf{k}^{\prime\prime}_\bot\cdot\textbf{q}_\bot\textbf{k}^{\prime}_\bot\cdot\textbf{q}_\bot
-\textbf{q}_\bot^2\textbf{k}^{\prime}_\bot\cdot\textbf{k}^{\prime\prime}_\bot\Big)\Big[2\textbf{k}_\bot^{\prime2}+2m_2^2-\bar x\big(M^{\prime2}+3M^{\prime\prime2}
\nonumber\\
&&+4(m^{\prime}_1-m_2)(m_2
-m^{\prime\prime}_1)\big)-2\bar x^2(M^{\prime2}-\textbf{q}_\bot^2)\Big]+\bar x\textbf{k}^{\prime}_\bot\cdot\textbf{k}^{\prime\prime}_\bot
\textbf{q}_\bot^2\Big[4(m^{\prime}_1
-m_2)(m_2-m^{\prime\prime}_1)\nonumber\\
&&+\textbf{q}_\bot^2+2(2-x)M^{\prime2}+2xM^{\prime\prime2}
\Big]
-\textbf{k}^{\prime\prime}_\bot\cdot\textbf{q}_\bot\Big[2\big(\textbf{k}_\bot^{\prime2}+m_2^2
+\bar x^2(m_2-m^{\prime}_1)(m_2
-m^{\prime\prime}_1)\big)\big(M^{\prime2}\nonumber\\
&&+M^{\prime\prime2}\big)+2\bar x\big(\textbf{k}_\bot^{\prime2}+m_2(m^{\prime}_1-m^{\prime\prime}_1)\big)\big(M^{\prime2}-M^{\prime\prime2}\big)
+4(m^{\prime}_1-m_2)(m_2
-m^{\prime\prime}_1)(\textbf{k}_\bot^{\prime2}+m_2^2)\nonumber\\
&&-4\bar x^2M^{\prime2}M^{\prime\prime2}+\textbf{q}_\bot^2\big((2x-1)\textbf{k}_\bot^{\prime2}-\bar x^2(2x-3)M^{\prime2}-2(\bar xm'_1+xm_2)(\bar x m''_1+xm_2)\nonumber\\
&&+4\bar xm'_1m_2+3(x-\bar x)m^2_2\big)\Big]
-\frac{2}{D''_{T,{\rm con}}}\Big(-m^{\prime}_1+m^{\prime\prime}_1+2m_2\Big)\Big[\bar x(M^{\prime2}-M^{\prime\prime2})(2\bar x\textbf{k}_\bot^{\prime2}\textbf{q}_\bot^2\nonumber\\
&&-\textbf{k}_\bot^{\prime2}\textbf{k}^{\prime}_\bot\cdot\textbf{q}_\bot
-\bar x^2\textbf{k}^{\prime}_\bot\cdot\textbf{q}_\bot\textbf{q}_\bot^2)
+4\bar x(\textbf{k}^{\prime}_\bot\cdot\textbf{q}_\bot)^2\textbf{k}^{\prime\prime}_\bot\cdot\textbf{q}_\bot
-3\bar x\textbf{k}^{\prime2}_\bot\textbf{q}_\bot^2\textbf{k}^{\prime\prime}_\bot\cdot\textbf{q}_\bot-\bar x^2\textbf{q}_\bot^4\textbf{k}^{\prime}_\bot\cdot\textbf{k}^{\prime\prime}_\bot
\nonumber\\
&&+(\textbf{k}^{\prime2}_\bot\textbf{q}_\bot^2-2\textbf{k}^{\prime}_\bot\cdot\textbf{q}_\bot\textbf{k}^{\prime\prime}_\bot\cdot\textbf{q}_\bot
-\bar x^2\textbf{q}_\bot^4)(\textbf{k}_\bot^{\prime2}+m_2^2-\bar x^2M^{\prime2})\Big]\bigg\}+\widetilde{T}_1^{\rm val.}(x,\textbf{k}^{\prime}_\bot,q^2)\,,
\\
\widetilde{T}_3^{\rm val.}&=&\frac{2M^{\prime}(M^{\prime2}-M^{\prime\prime2})}{\bar xq^4(M^{\prime2}-M^{\prime\prime2}+\textbf{q}_\bot^2)}\bigg\{\bar x\Big[6\textbf{k}^{\prime2}_\bot\textbf{k}^{\prime\prime}_\bot\cdot\textbf{q}_\bot\textbf{q}_\bot^2-8(\textbf{k}^{\prime}_\bot\cdot\textbf{q}_\bot)^2\textbf{k}^{\prime\prime}_\bot\cdot\textbf{q}_\bot
+4\bar x\textbf{k}^{\prime2}_\bot\textbf{q}_\bot^4
\nonumber\\
&&-4\bar x(\textbf{k}^{\prime}_\bot\cdot\textbf{q}_\bot)^2\textbf{q}_\bot^2\Big]
-\textbf{k}^{\prime\prime}_\bot\cdot\textbf{q}_\bot\textbf{q}_\bot^2\Big[(3-4x)\textbf{k}_\bot^{\prime2}+(1-2x)(m_2^2
-\bar x^2M^{\prime2})\Big]\nonumber\\
&&-\Big[\textbf{k}^{\prime}_\bot
\cdot\textbf{k}^{\prime\prime}_\bot\textbf{q}_\bot^2-2\textbf{k}^{\prime}_\bot\cdot\textbf{q}_\bot\textbf{k}^{\prime\prime}_\bot\cdot\textbf{q}_\bot\Big]\Big[2\textbf{k}_\bot^{\prime2}+2m_2^2+\bar x((2x-1)M^{\prime2}
-M^{\prime\prime2}+2\bar x\textbf{q}_\bot^2)\Big]\nonumber\\
&&+\bar x(1
-2x)\textbf{k}^{\prime}_\bot\cdot\textbf{k}^{\prime\prime}_\bot\textbf{q}_\bot^4
+\frac{2}{D''_{T,{\rm con}}}\Big[\bar x(m^{\prime}_1+m^{\prime\prime}_1)(4(\textbf{k}^{\prime}_\bot\cdot\textbf{q}_\bot)^2\textbf{k}^{\prime\prime}_\bot\cdot\textbf{q}_\bot
-2\textbf{k}^{\prime2}_\bot\textbf{k}^{\prime}_\bot\cdot\textbf{q}_\bot\textbf{q}_\bot^2
\nonumber\\
&&+2\bar x\textbf{k}^{\prime}_\bot\cdot\textbf{q}_\bot\textbf{q}_\bot^4)
+\big(\textbf{k}^{\prime2}_\bot\textbf{q}_\bot^2-2\textbf{k}^{\prime}_\bot\cdot\textbf{q}_\bot
\textbf{k}^{\prime\prime}_\bot\cdot\textbf{q}_\bot-\bar x^2\textbf{q}_\bot^4\big)\big((m^{\prime}_1+m^{\prime\prime}_1)(\textbf{k}_\bot^{\prime2}+
m_2^2)\nonumber\\&&
+x\bar xm_2
(M^{\prime2}-M^{\prime\prime2}+\textbf{q}_\bot^2)-\bar x^2m^{\prime\prime}_1M^{\prime2}-\bar x^2m^{\prime}_1(M^{\prime\prime2}-\textbf{q}_\bot^2)\big)\Big]\bigg\}
\nonumber\\
&&+\frac{M^{\prime2}-M^{\prime\prime2}}{q^2}\Big[\widetilde{T}_1^{\rm val.}(x,\textbf{k}^{\prime}_\bot,q^2)-\widetilde{T}_2^{\rm val.}(x,\textbf{k}^{\prime}_\bot,q^2)\Big]\,.
\end{eqnarray}


\section{Numerical results and discussion}
With the theoretical results given above and the values of input parameters collected in appendix A, we then present our numerical results and discussions in this section.
As has been mentioned in the last section, the contributions associated with $B$ functions are not included in the CLF results, Eqs.~(\ref{eq:VCLF}-\ref{eq:T3CLF}). These contributions to the matrix elements of $P \to T$ transition can be written as
\begin{eqnarray}\label{eq:Bcontr}
[\mathcal B]_{\rm B}=N_c\int\frac{\d x\d^2{\bf k'_\bot}}{2(2\pi)^3}\frac{\chi_P'\chi_T''}{\bar x}{\cal \widetilde{\cal B}}_{\rm B}\,,
\end{eqnarray}
where, the integrands are 
\begin{align}\label{eq:pvtB}
{\cal \widetilde{B}}_{\rm B}^\u(\Gamma=\r_\u)=&2i\varepsilon_{\mu\lambda\rho \a }\epsilon^{*\lambda \delta}q_{\delta}P^{\a}\omega^{\rho}\frac{B_1^{(2)}}{\omega\cdot P}\left(-m_1'+m_1''\right)
\nonumber\\
&+2i\varepsilon_{\mu\lambda\rho \a }\epsilon^{*\lambda \delta}q_{\delta}q^{\a}\omega^{\rho}\frac{B_1^{(2)}}{\omega\cdot P}\left(m_1'+m_1''-2m_2\right)
\nonumber\\
&+4i\varepsilon_{\mu\rho \a \b}\epsilon^{*\lambda \delta}q_{\lambda}q_{\delta}P^{\a}q^\b\omega^{\rho}\frac{B_1^{(3)}+2B_2^{(3)}-2B_1^{(2)}}{\omega\cdot P}\cdot\frac{1}{D_{T,{\rm con}}}\,,
\\
\label{eq:patB}
{\cal \widetilde{B}}_{\rm B}^\u(\Gamma=\r_\u\r_5)=&4P_\mu\epsilon^{*\lambda \delta} q_\lambda\omega_\delta \bigg\{\frac{B_1^{(2)}}{\omega\cdot P}\Big[-5m'_1+3m_2+\frac{1}{D_{T,{\rm con}}}\big(2M'^2+2M''^2+\hat N'_1+\hat N''_1\nonumber\\
&+(m'_1+m''_1)^2+4(m'_1-m_2)(m''_1+m_2)-3q^2
\big)\Big]+\frac{2(B_1^{(3)}+B_2^{(3)})}{\omega\cdot P}\Big[2m'_1-2m_2\nonumber\\
&-\frac{1}{D_{T,{\rm con}}}\big(M'^2+M''^2+2(m'_1-m_2)(m''_1+m_2)-q^2
\big)\Big]-\frac{4(B_3^{(3)}-B_5^{(4)})}{\omega\cdot P}\cdot \frac{1}{D_{T,{\rm con}}}
\bigg\}\,\nonumber\\
&+4q_\mu\epsilon^{*\lambda \delta} q_\lambda\omega_\delta \bigg\{\frac{B_1^{(2)}}{\omega\cdot P}\Big[-3m'_1-m''_1-2m_2+\frac{1}{D_{T,{\rm con}}}\big(2M'^2-\hat N'_1+\hat N''_1\nonumber\\
&-2(m'_1-m_2)^2+(m'_1+m''_1)^2-q^2
\big)\Big]+\frac{2B_2^{(3)}}{\omega\cdot P}\Big[2m'_1-2m_2-\frac{1}{D_{T,{\rm con}}}\big(M'^2\nonumber\\
&+M''^2-q^2+2(m'_1-m_2)(m''_1+m_2)
\big)\Big]-\frac{2(B_3^{(3)}-B_5^{(4)})}{\omega\cdot P}\cdot \frac{1}{D_{T,{\rm con}}}
\bigg\}
\nonumber\\
&+4\omega_\mu\epsilon^{*\lambda \delta} q_\lambda q_\delta \bigg\{\frac{B_1^{(2)}}{\omega\cdot P}\Big[-4m'_1+m''_1+3m_2+\frac{2}{D_{T,{\rm con}}}\big(M'^2+M''^2+2(m'_1\nonumber\\
&-m_2)(m''_1+m_2)-q^2
\big)\Big]+\frac{B_1^{(3)}+2B_2^{(3)}}{\omega\cdot P}\Big[2m'_1-2m_2-\frac{1}{D_{T,{\rm con}}}\big(M'^2+M''^2\nonumber\\
&+2(m'_1-m_2)(m''_1+m_2)-q^2
\big)\Big]-\frac{4(B_3^{(3)}-B_5^{(4)})}{\omega\cdot P}\cdot \frac{1}{D_{T,{\rm con}}}
\bigg\}\,,
\\
\label{eq:ptt1B}
{\cal \widetilde{B}}_{\rm B}(\Gamma=\sigma_{\mu\nu}q^\v)=&
4\varepsilon_{\nu\lambda\alpha\beta}P^{\a}q^{\v}\omega^{\b}\epsilon^{*\lambda \delta}q_{\delta}P_{\mu}\frac{B^{(3)}_1+B^{(3)}_2-B^{(2)}_1}{\omega\cdot P}
\nonumber\\
&+2\varepsilon_{\nu\lambda\alpha\beta}P^{\a}q^{\v}\omega^{\b}\epsilon^{*\lambda \delta}
q_{\delta}q_{\mu}\frac{2B^{(3)}_2-B^{(2)}_1}{\omega\cdot P}
\nonumber\\
&+2\varepsilon_{\mu\lambda\alpha\beta}P^{\a}\omega^{\b}\epsilon^{*\lambda \delta}q_{\delta}
\left[2(M'^2-M''^2)\frac{B^{(2)}_1-B^{(3)}_1-B^{(3)}_2}{\omega\cdot P}+q^2\cdot\frac{B^{(2)}_1-2B^{(3)}_2}{\omega\cdot P}\right]\nonumber\\
&+2\varepsilon_{\mu\lambda\alpha\beta}q^{\a}\omega^{\b}\epsilon^{*\lambda \delta}q_{\delta}
\bigg\{\Big[(m'_1+m''_1)(m'_1-m_2)+\hat N'_1\Big]\frac{2B^{(2)}_1}{\omega\cdot P}\nonumber\\
&+2(M'^2-M''^2)\frac{2B^{(2)}_1-B^{(3)}_1-B^{(3)}_2}{\omega\cdot P}+q^2\cdot\frac{B^{(2)}_1-2B^{(3)}_2}{\omega\cdot P}\bigg\}\nonumber\\
&+2\varepsilon_{\mu\lambda\alpha\beta}P^{\a}q^{\b}\epsilon^{*\lambda \delta}\w_{\delta}
\bigg\{\Big[(m'_1+m''_1)(m_2-m'_1)-\hat N'_1\Big]\frac{2B^{(2)}_1}{\omega\cdot P}\nonumber\\
&+\Big(M'^2-M''^2\Big)\frac{B^{(2)}_1+2B^{(3)}_1-2B^{(3)}_2}{\omega\cdot P}+q^2\cdot\frac{3B^{(2)}_1+2B^{(3)}_2}{\omega\cdot P}\bigg\}\nonumber\\
&+4\varepsilon_{\mu\nu\alpha\beta}P^{\nu}q^{\a}\w^{\b}\epsilon^{*\lambda \delta}q_{\lambda}q_{\delta}
\left(\frac{m'_1+m''_1}{D_{T,con}}\cdot\frac{2B^{(2)}_1-B^{(3)}_1-2B^{(3)}_2}{\omega\cdot P}\right)\,,
\\
\label{eq:ptt5B}
{\cal \widetilde{B}}_{\rm B}(\Gamma=\sigma_{\mu\nu}\gamma_5q^\v)=&4iP_{\mu}\epsilon^{*\lambda \delta}q_{\lambda}\omega_{\delta}\bigg\{\frac{q^2B^{(2)}_1}{\omega\cdot P}\left[1+\frac{2(2m'_1-m''_1-2m_2)}{D_{T,{\rm con}}}\right]+\frac{B^{(3)}_2}{\omega\cdot P}\frac{4q^2(m_2-m^{\prime}_1)}{D_{T,{\rm con}}}
\nonumber\\
&-\frac{2q^2B^{(3)}_1}{\omega\cdot P}\left[1+\frac{m^{\prime}_1-m^{\prime\prime}_1-2m_2}{D_{T,{\rm con}}}\right]-\frac{B^{(3)}_3}{\omega\cdot P}\bigg\}
\nonumber\\
&+2iq_{\mu}\epsilon^{*\lambda \delta}q_{\lambda}\omega_{\delta}\Big(M'^2-M''^2\Big)\bigg\{\frac{B^{(2)}_1}{\omega\cdot P}
\left[3+\frac{4(m^{\prime\prime}_1+2m_2-2m^{\prime}_1)}{D_{T,{\rm con}}}\right]\nonumber\\
&+\frac{4B^{(3)}_1}{\omega\cdot P}\Big(1+\frac{m^{\prime}_1-m''_1-2m_2}{D_{T,{\rm con}}}\Big)
+\frac{8B^{(3)}_2}{\omega\cdot P}\cdot\frac{m^{\prime}_1-m_2}{D_{T,{\rm con}}}
\bigg\}
\nonumber\\
&+4i\omega_{\mu}\epsilon^{*\lambda\delta}q_{\lambda}q_{\delta}\bigg\{
\frac{B^{(2)}_1}{\omega\cdot P}\Big[2M^{\prime2}-M^{\prime\prime2}
+(m'_1-m_2)(m_2-m''_1)+q^2\nonumber\\
&-\frac{2}{{D_{T,{\rm con}}}}\left((M^{\prime2}-M^{\prime\prime2})(m^{\prime}_1
+m^{\prime\prime}_1)-q^2(m^{\prime}_1-m^{\prime\prime}_1-2m_2)\right)\Big]
\nonumber\\
&+\frac{B^{(3)}_1+2B^{(3)}_2}{\omega\cdot P}\Big[-M^{\prime2}+M^{\prime\prime2}-q^2+\frac{1}{{D_{T,{\rm con}}}}\big((M^{\prime2}-M^{\prime\prime2})(m^{\prime}_1
+m^{\prime\prime}_1)\nonumber\\
&-q^2(m^{\prime}_1-m^{\prime\prime}_1-2m_2)\big)\Big]
-\frac{B^{(3)}_3}{\omega\cdot P}\bigg\}
\nonumber\\
&+2i\epsilon^{*\mu\delta}\omega_{\delta}\left({M^{\prime2}-M^{\prime\prime2}}\right)\bigg\{
\frac{B^{(2)}_1}{\omega\cdot P}\Big[-(M^{\prime2}+M^{\prime\prime2})+2(m_2-m'_1)(m_2-m''_1)\nonumber\\
&-q^2\Big]+\frac{2B^{(3)}_3}{\omega\cdot P}\bigg\}\,.
\end{align}
After extracting their contributions to the form factors, $[{\cal F}]^{\rm B}$,  one can obtain the full result of form factor in the CLF QM, which can be expressed as
\begin{align}\label{eq:full}
[{\cal F}]^{\rm full}=[{\cal F}]^{\rm CLF}+[{\cal F}]^{\rm B}\,.
\end{align}

\begin{figure}[t]
\caption{The dependences of $[\Delta_B]_{T_{3}}(x)$ on $x$ for  $D\to K^*_2$ transition at ${\bf q}_{\bot}^2=(0,0.2)\,{\rm GeV^2}$ and  for $B_c\to D^*_2$ transition at ${\bf q}_{\bot}^2=(0,9)\,{\rm GeV^2}$. }
\begin{center}
\subfigure[]{\includegraphics[scale=0.5]{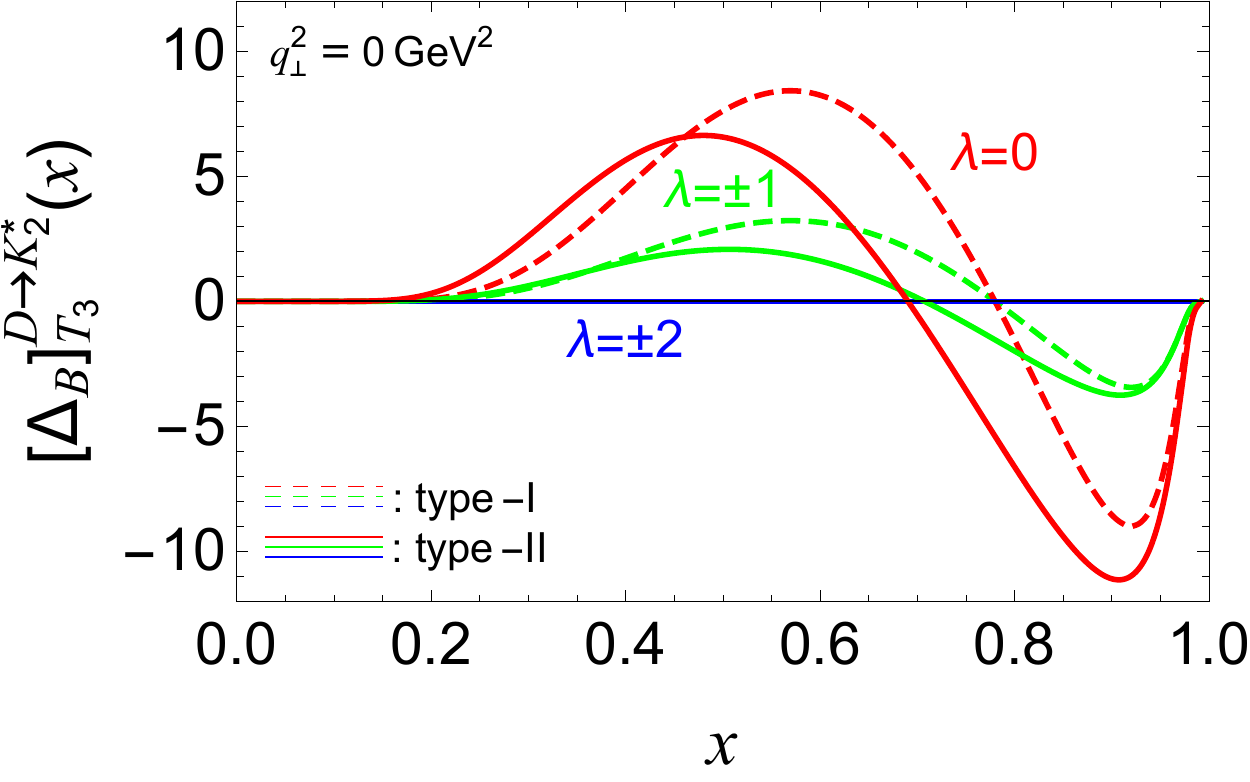}}\qquad
\subfigure[]{\includegraphics[scale=0.5]{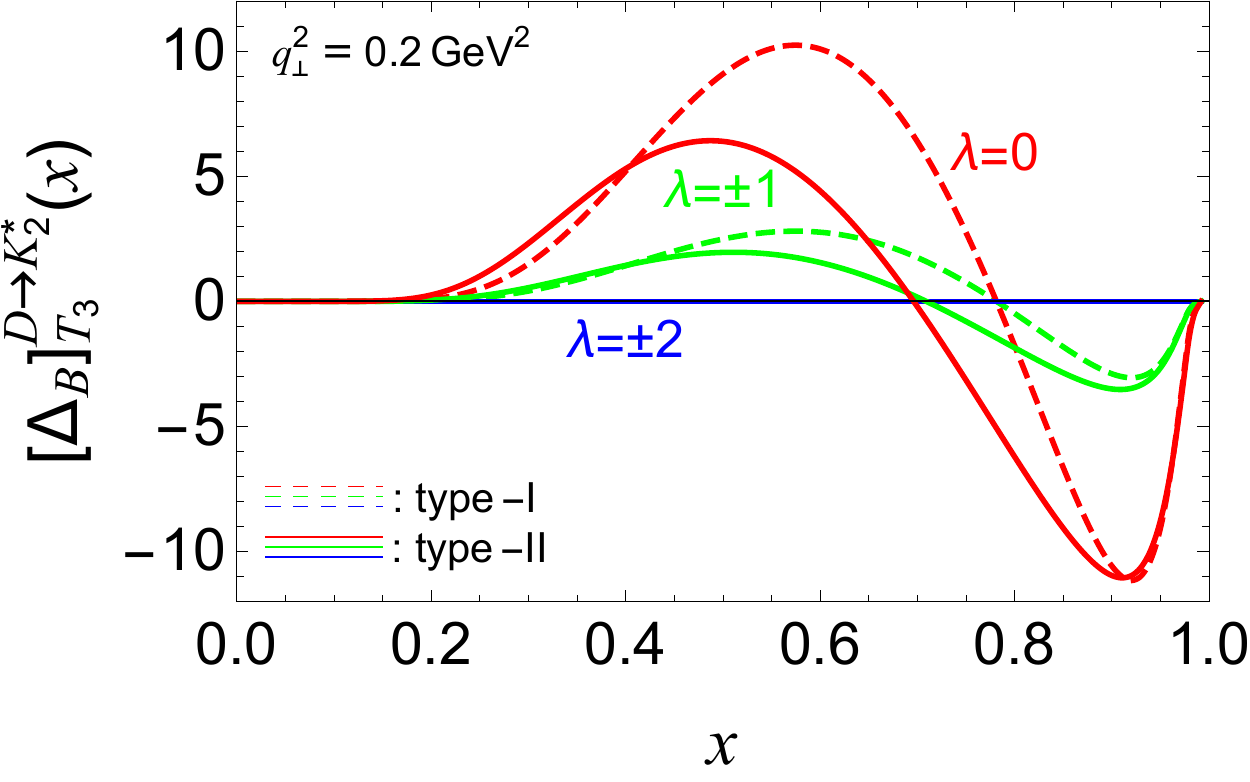}}\\
\subfigure[]{\includegraphics[scale=0.5]{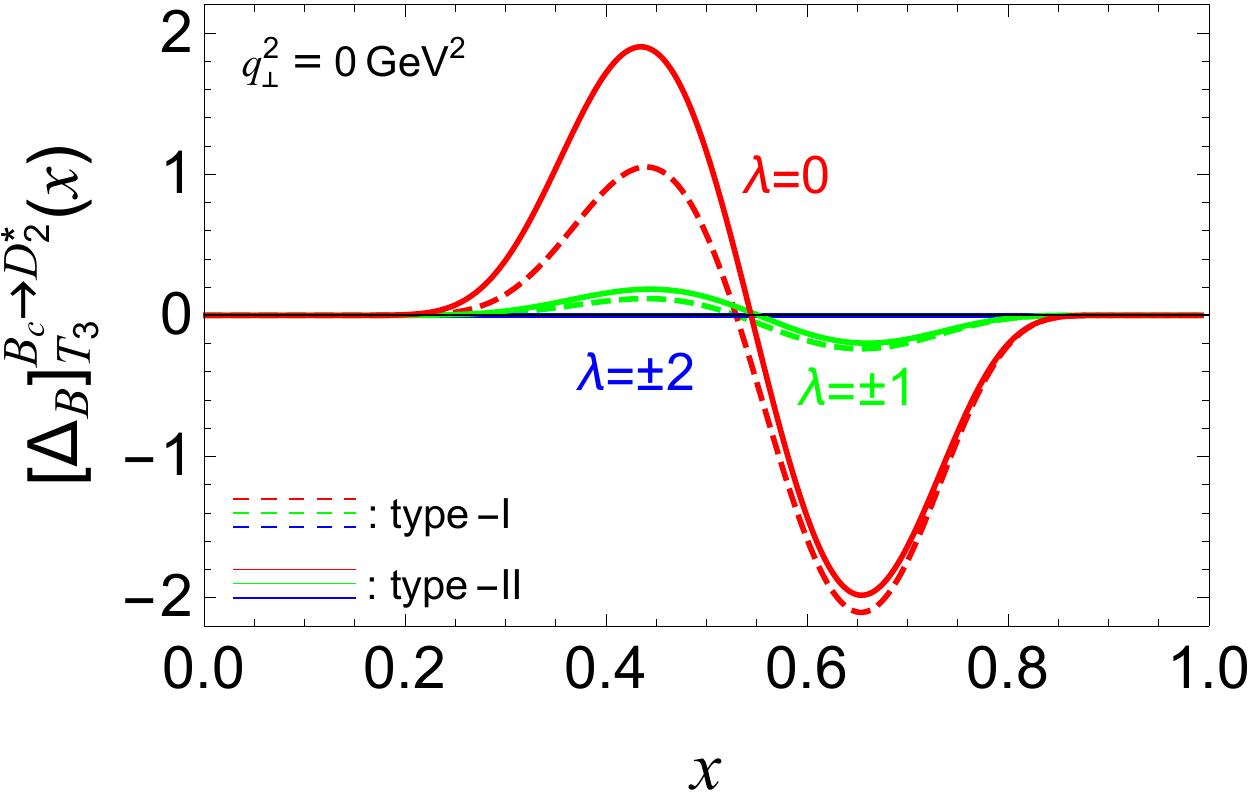}}\qquad
\subfigure[]{\includegraphics[scale=0.5]{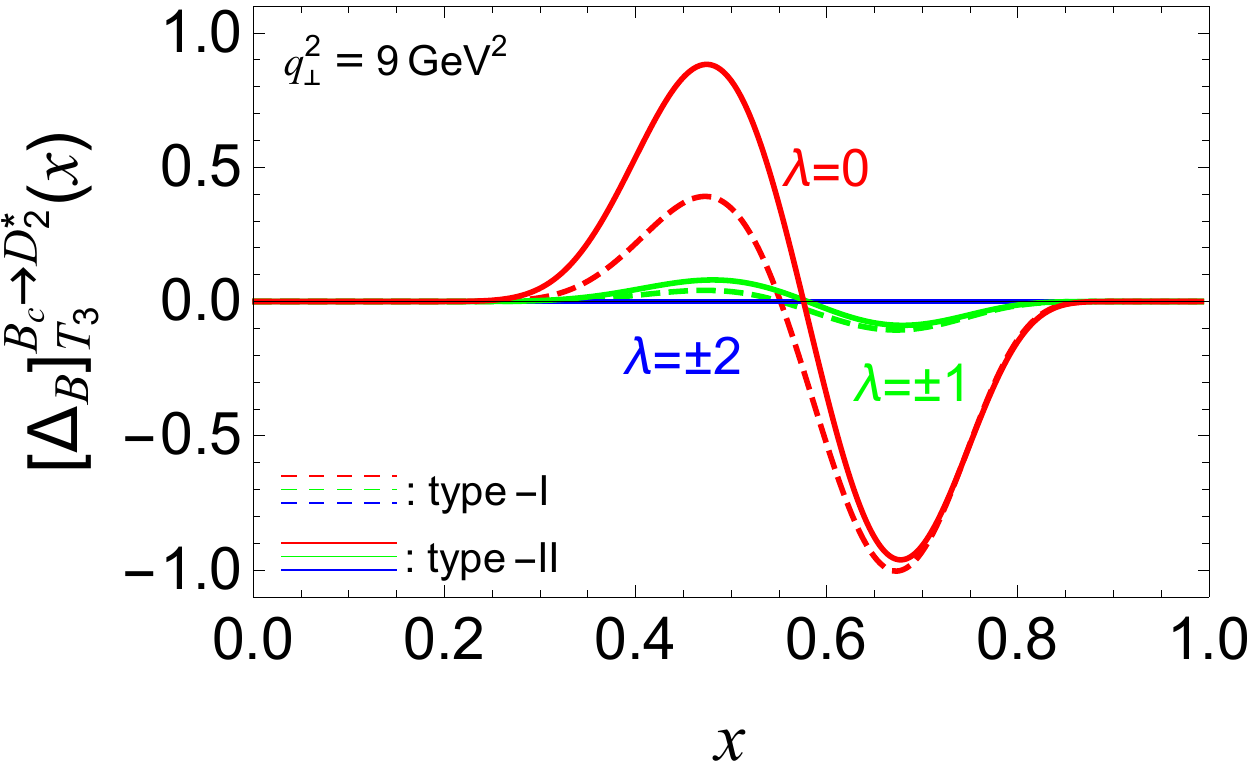}}\\
\end{center}
\label{fig:T13}
\end{figure}

\begin{table}[t]
\begin{center}
\caption{\label{tab:T1} \small Numerical results of form factor $T_{3}({\bf{q}}_\perp^2)$  at ${\bf q}_{\bot}^2=(0,1,4,9) \,{\rm GeV^2}$ for $B_c\to D^*_2$ transition.  }
\vspace{0.2cm}
\let\oldarraystretch=\arraystretch
\renewcommand*{\arraystretch}{1}
\setlength{\tabcolsep}{4pt}
\begin{tabular}{l|cccccccccccc}
\hline\hline
 $B_c\to D^*_2$ & &$ [T_3]^{\text{SLF}}_{\lambda''=\pm2}$
                 &$[T_3]^{\rm full}_{\lambda''=0}$
                 &$[T_3]^{\rm full}_{\lambda''=\pm1}$
                 &$[T_3]^{\rm full}_{\lambda''=\pm2}$
                &$[T_3]^{\rm val.}$
               &$[T_3]^{\rm CLF}$\\\hline
\multirow{2}{*}{${\bf{q}}_\perp^2=0$}
&type-I    &$0.03$&$-0.14\quad$&$-0.04\quad$&$0.05$&$0.10$&$0.05$
              \\
&type-II    &$0.08$&$0.08$&$0.08$&$0.08$&$0.08$&$0.08$
\\\hline
\multirow{2}{*}{${\bf{q}}_\perp^2=1$}
&type-I     &$0.03$&$-0.13\quad$&$-0.04\quad$&$0.05$&$0.09$&$0.05$
              \\
&type-II    &$0.07$&$0.07$&$0.07$&$0.07$&$0.07$&$0.07$
\\\hline
\multirow{2}{*}{${\bf{q}}_\perp^2=4$}
&type-I     &$0.02$&$-0.11\quad$&$-0.04\quad$&$0.03$&$0.07$&$0.03$
              \\
&type-II    &$0.05$&$0.05$&$0.05$&$0.05$&$0.05$&$0.05$
\\\hline
\multirow{2}{*}{${\bf{q}}_\perp^2=9$}
&type-I     &$0.02$&$-0.08\quad$&$-0.03\quad$&$0.02$&$0.05$&$0.02$
              \\
&type-II    &$0.04$&$0.04$&$0.04$&$0.04$&$0.04$&$0.04$
\\\hline\hline
\end{tabular}
\end{center}
\end{table}

Based on these results, we have following discussions and findings:
\begin{itemize}
\item Here, we take ${\cal \widetilde{B}}_{\rm B}(\Gamma=\sigma_{\mu\nu}\r_5q^\v)$ given by Eq.~\eqref{eq:ptt5B} as an example. From this equation, it can be found that the third term is proportional to $\omega_\u$. This  spurious $\omega_\u$-dependent contribution  corresponds to an unphysical form factor and may violate the covariance of matrix element if it is non-zero. The other  terms  would present contributions to the tensor form factors, $T_2$ and $T_3$. For convenience of discussion, we take $T_3$ as an example,  which could receive the contribution from the second term  written as
\begin{align}
{\widetilde{ T}_3}^{\rm B}=&\frac{2(M'^2-M''^2)}{ e^{*}\cdot q}\,\frac{\epsilon^{\lambda\delta*}q_\lambda\omega_\delta }{\omega\cdot P}\bigg\{B^{(2)}_1
\left[3+\frac{4(m^{\prime\prime}_1+2m_2-2m^{\prime\prime}_1)}{D_{T,{\rm con}}}\right]\nonumber\\
&+{4B^{(3)}_1}\left(1+\frac{m^{\prime}_1-m''_1-2m_2}{D_{T,{\rm con}}}\right)+\frac{8(m^{\prime}_1-m_2){B^{(3)}_2}}{D_{T,{\rm con}}}
\bigg\}\,,
\end{align}
which  is obviously dependent on the choice of $\lambda''$. For different values of $\lambda''$, ${\widetilde{ T}_3}^{\rm B}$ can be explicitly written as
\begin{equation}
{ \widetilde{T}_3}^{\rm B}=\left\{
             \begin{array}{lr}
           \frac{2M'(M'^2-M''^2)(M'^2-M''^2+q_\bot^2)}{(M'^2-M''^2)^2+2(M'^2-2M''^2)\qb^2+\qb^4}\bigg\{B^{(2)}_1
           \left[3+\frac{4(m^{\prime\prime}_1+2m_2-2m^{\prime}_1)}{D_{T,{\rm con}}}\right]\\
           +{4B^{(3)}_1}\left(1+\frac{m^{\prime}_1-m''_1-2m_2}{D_{T,{\rm con}}}\right)
           +{B^{(3)}_2}\frac{8(m^{\prime}_1-m_2)}{D_{T,{\rm con}}}
           \bigg\}\,,  &\lambda''=0 \\
           \frac{M'(M'^2-M''^2)}{M'^2-M''^2+q_\bot^2}\bigg\{B^{(2)}_1
           \left[3+\frac{4(m^{\prime\prime}_1+2m_2-2m^{\prime}_1)}{D_{T,{\rm con}}}\right]
          \\
         +{4B^{(3)}_1}\left(1+\frac{m^{\prime}_1-m''_1-2m_2}{D_{T,{\rm con}}}\right) +{B^{(3)}_2}\frac{8(m^{\prime}_1-m_2)}{D_{T,{\rm con}}}
           \bigg\}\,, &\lambda''=\pm1 \\
             0\,. &\lambda''=\pm2  \\
             \end{array}
\right.
\end{equation}
Further considering the fact that $[{\cal F}]^{\rm CLF}$ is independent of $\lambda''$, it is clearly seen that $T_3$ possibly suffers from a self-consistence problem ({\it i.e.} $[T_{3}]^{\rm full}_{\lambda=0}\neq[T_{3}]^{\rm full}_{\lambda=\pm1}\neq[T_{3}]^{\rm full}_{\lambda=\pm2}$)  caused by the $B$ function contributions in the CLF QM. Comparing with the other transitions, the $P \to T$ transition is associated with much more $B$ functions, and thus presents a tougher challenge to the CLF QM.

\item  In order to clearly show the possible self-consistence problem caused by $B$ functions within type-I and type-II schemes, we define the  contributions of $B$ functions $\Delta_{\rm B}(x)$ as
\begin{align}
\Delta_{\rm B}(x) \equiv \frac{\d [{\cal F}^{\rm B}]_{\lbd''}}{ \d x}\,,
\end{align}
which is equal to $N_c\int\frac{\d^2{\bf k'_\bot}}{2(2\pi)^3}\frac{\chi_P'\chi_T''}{\bar x} {{\widetilde {T}}}^{\rm B}_{3}$ for the case of $T_3$. Taking $D\to K^*_2$  and  $B_c\to D^*_2$ transitions as examples,  the dependence of  $\Delta_{\rm B}(x)$ on $x$ are shown in Fig.~\ref{fig:T13}. It can be seen that the self-consistence  is violated within the type-I scheme, but it can be satisfied within the type-II scheme due to $\int_0^1\d x\,[\Delta_{\rm B}]_{T_3}(x)=0$ for any values of $\lambda''$.

In order to further confirm such finding, we list the numerical results of $[T_{3}]^{\rm full}$ for $B_c\to D^*_2 $ transition at $\qb^2=(0,1,4,9)\,{\rm GeV^2}$ with $\lambda''=(0,\pm1,\pm2)$ in Table~\ref{tab:T1}, in which the SLF, valence and CLF results are also given for comparison. From these numerical results, it is  found that $[T_{3}]^{\rm full}_{\lambda=0}\neq [T_{3}]^{\rm full}_{\lambda=\pm1}\neq [T_{3}]^{\rm full}_{\lambda=\pm2}$ within the  traditional type-I  scheme, while
\begin{align}
[T_{3}]^{\rm full}_{\lambda=0}\doteq[T_{3}]^{\rm full}_{\lambda=\pm1}\doteq[T_{3}]^{\rm full}_{\lambda=\pm2}\,, \qquad{ (\text{type-II})}
\end{align}
which \tb{confirms again that the contribution associated with $B$ functions vanishes  numerically within the type-II scheme, even though it exists formally in the expression of form factor given by Eq.~\eqref{eq:full}.} In addition, one can also find from Table~\ref{tab:T1} that the SLF results are also consistent with the CLF ones, $[T_{3}]^{\rm full}\dot{=}[T_{3}]^{\rm SLF}$, in type-II scheme. These results confirm again the finding obtained from  Fig.~\ref{fig:T13}.  Moreover, we have checked  the contributions of $B$ functions to the other form factors, and obtain the same conclusion. Therefore, it can be concluded that the CLF results for the form factors of $P\to T$ transitions have a self-consistence problem caused by $B$ function contributions within the  type-I  scheme, but the type-II  scheme can give self-consistent results.

\item Besides of the self-consistence, the contributions of $B$ functions also possibly result in a covariance problem  because some terms in ${\cal \widetilde{B}}_\u$ are associated with $\w_{\u}$. Taking the third term in ${\cal \widetilde{B}}_{\rm B}(\Gamma=\sigma_{\mu\nu}\r_5q^\v)$ given by Eq.~\eqref{eq:ptt5B} as an example, this  spurious $\w_\u$-dependent contribution  corresponds to an unphysical form factor and would violate the covariance of matrix element except it vanishes numerically.  Similar to the case of  $[T_{3}]^{\rm B}$ discussed above, it is found that this spurious $\w_\u$-dependent contribution is (non)zero within the type-II~(I) scheme, which implies that the Lorentz covariance of ${\cal {B}}^\u$ is violated within the type-I correspondence scheme, but such problem can be avoided by employing  the type-II scheme.

\begin{figure}[t]
\caption{\label{fig:T13ZM} The dependences of $\d [{ T_3}]^{\rm z.m.}/\d x$ on $x$ for  $D\to K^*_2$ transition at ${\bf q}_{\bot}^2=(0,0.2)\,{\rm GeV^2}$ and  for $B_c\to D^*_2$ transition at ${\bf q}_{\bot}^2=(0,9)\,{\rm GeV^2}$. }
\begin{center}
\subfigure[]{\includegraphics[scale=0.5]{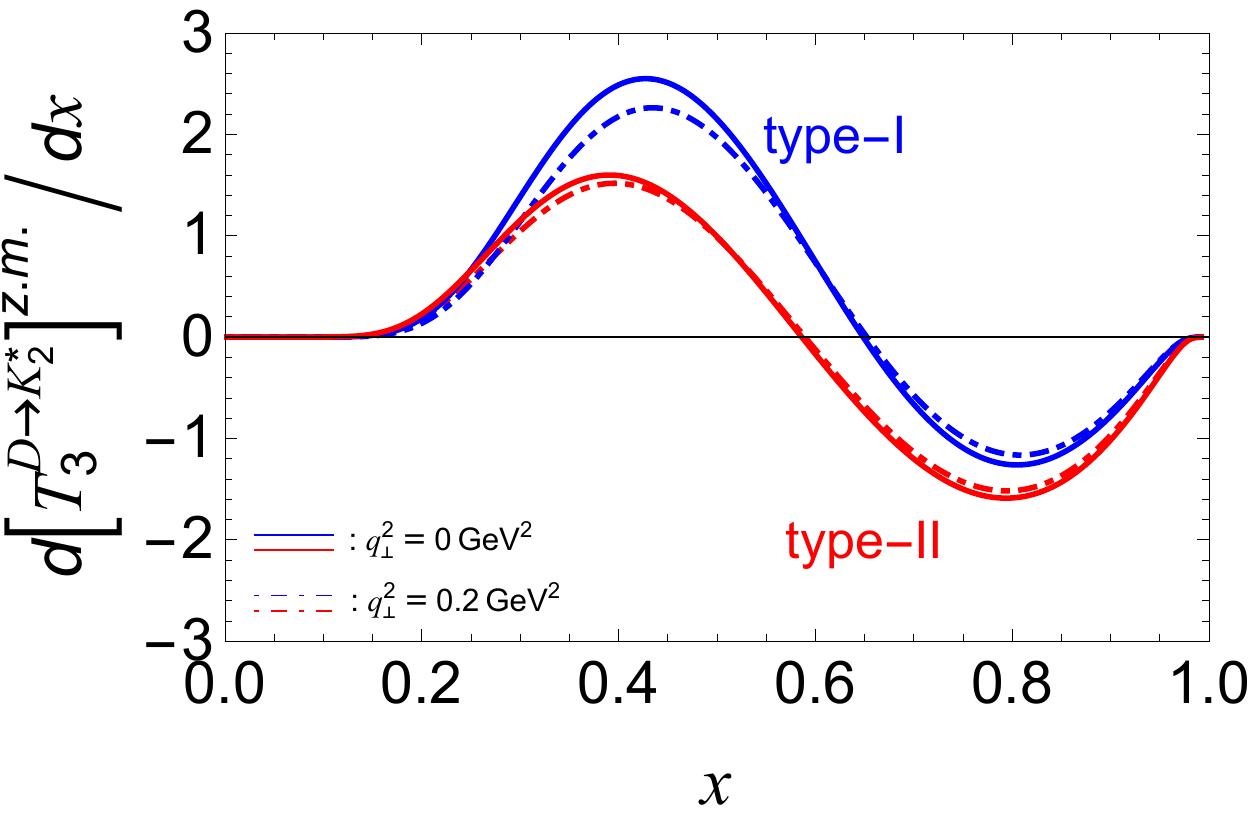}}\qquad
\subfigure[]{\includegraphics[scale=0.5]{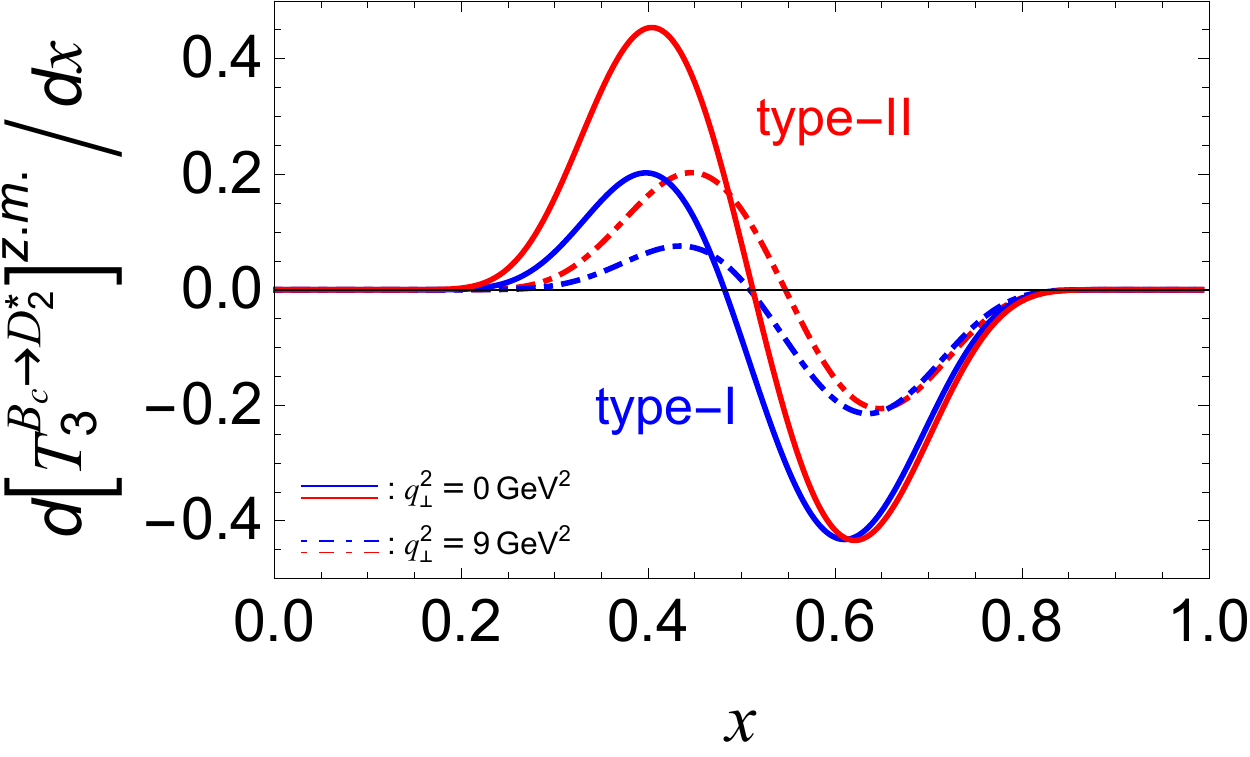}}\\
\end{center}
\end{figure}

 \item As has been mentioned above, the spurious $\w_\u$-dependent contributions associated with $C$ functions can be canceled by the zero-mode contributions~\cite{Jaus:1999zv}. The residual zero-mode contributions  to form factors can be obtained via $[{\cal F}]^{\rm CLF}=[{\cal F}]^{\rm val.}+[{\cal F}]^{\rm z.m.}$. In order to clearly show the effect of zero-mode contribution, we take $T_{3}^{B\to K^*_2\,,D\to K^*_2}$ as examples and plot the dependence of $\d [{\cal F}]^{\rm z.m.}/\d x$ on $x$ in Fig.~\ref{fig:T13ZM}.  It can be found that  zero-mode presents nonzero contributions within the traditional type-I correspondence scheme; while, these contributions, although existing formally, vanish numerically in the  type-II correspondence scheme, {\it i.e.},  $[T_{3}(q^2)]_{\rm z.m.}\dot{=} 0$~(type-II), because the contribution with small $x$ and the one with large $x$  cancel each other out exactly at each ${\qb^2}$ point. This can also be found from the numerical results given by Table~\ref{tab:T1}.
 \end{itemize}

\begin{figure}[t]
\caption{The dependences of $\Delta_{T_{2,3}}^{\rm CLF}$ on $x$ for  $D\to K^*_2$ transition at ${\bf q}_{\bot}^2=(0,0.2)\,{\rm GeV^2}$ and  for $B_c\to D^*_2$ transition at ${\bf q}_{\bot}^2=(0,9)\,{\rm GeV^2}$. }
\begin{center}
\subfigure[]{\includegraphics[scale=0.4]{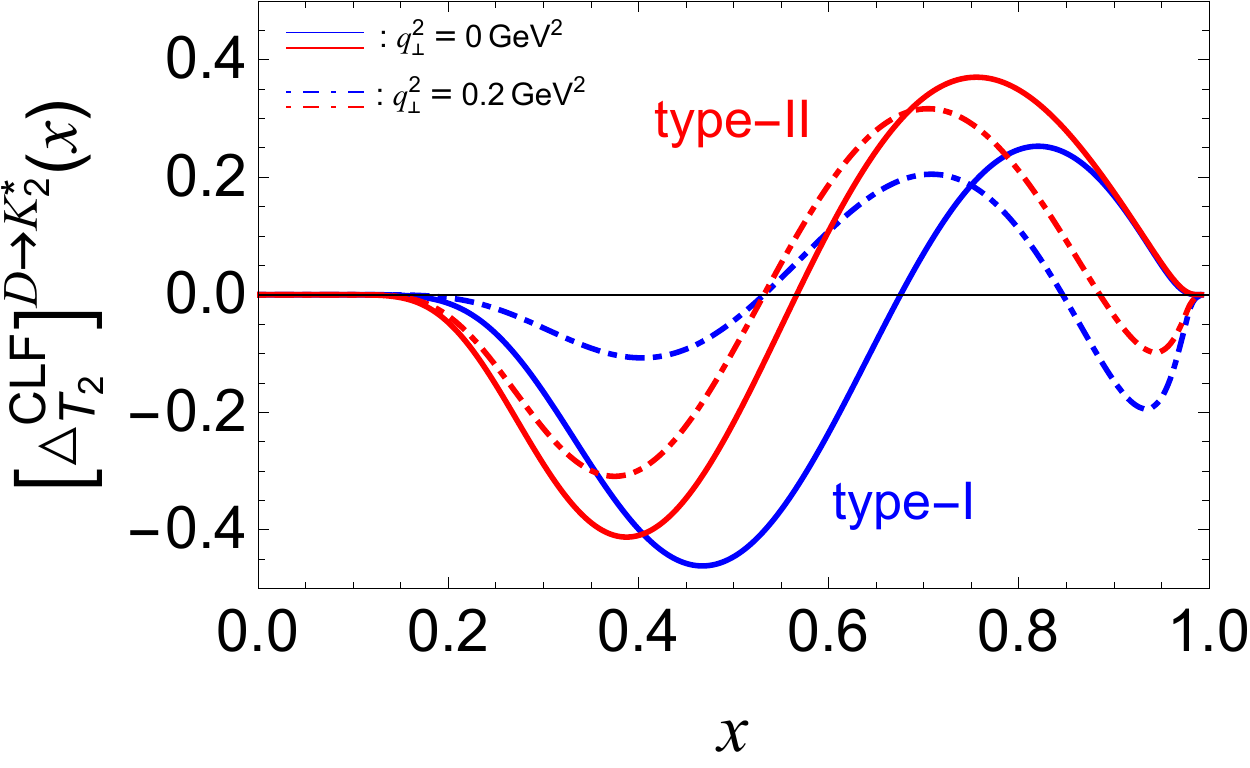}}\quad\quad
\subfigure[]{\includegraphics[scale=0.4]{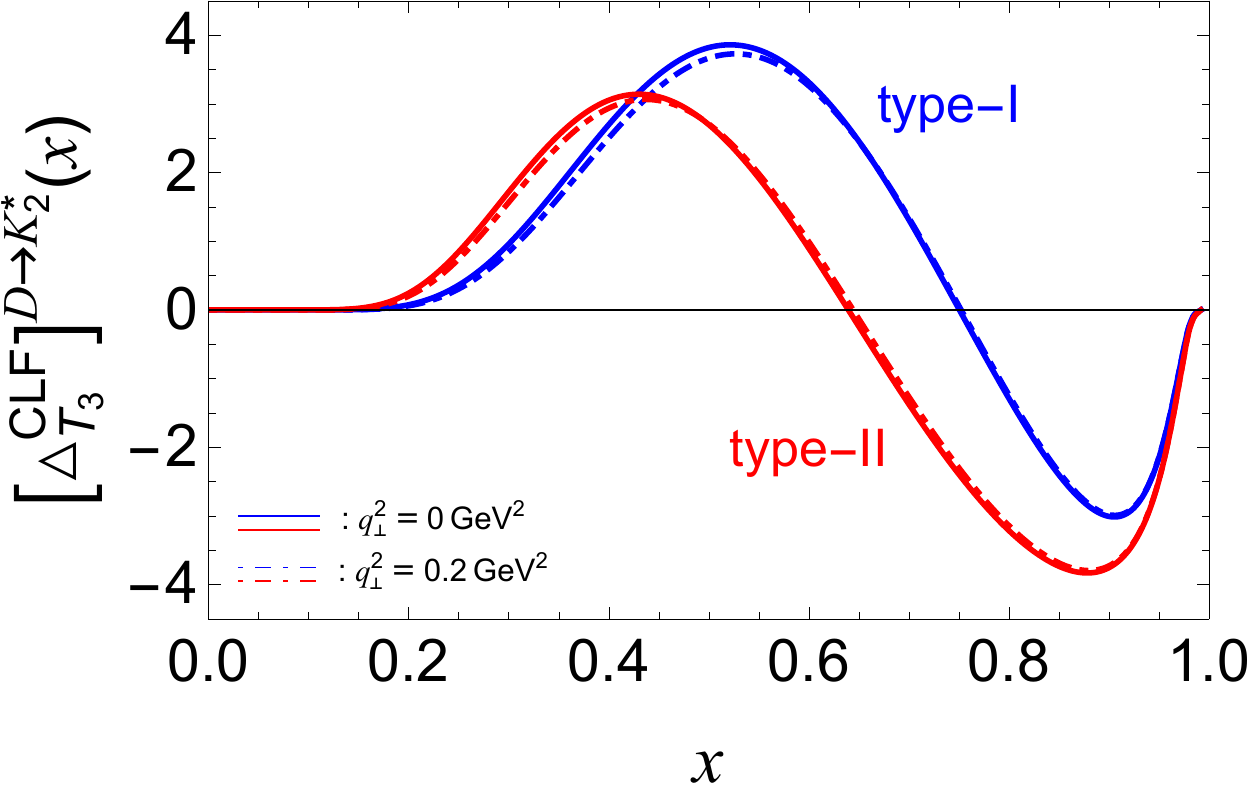}}\\
\subfigure[]{\includegraphics[scale=0.4]{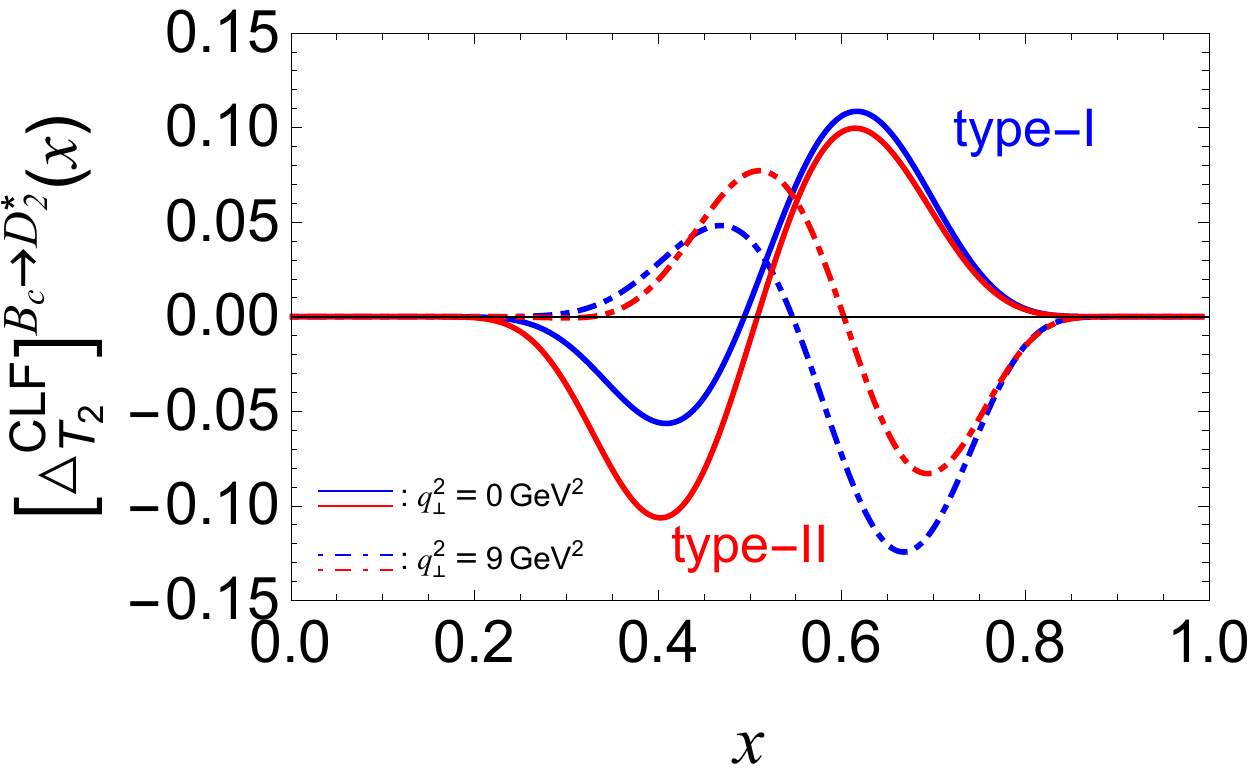}}\quad\quad
\subfigure[]{\includegraphics[scale=0.4]{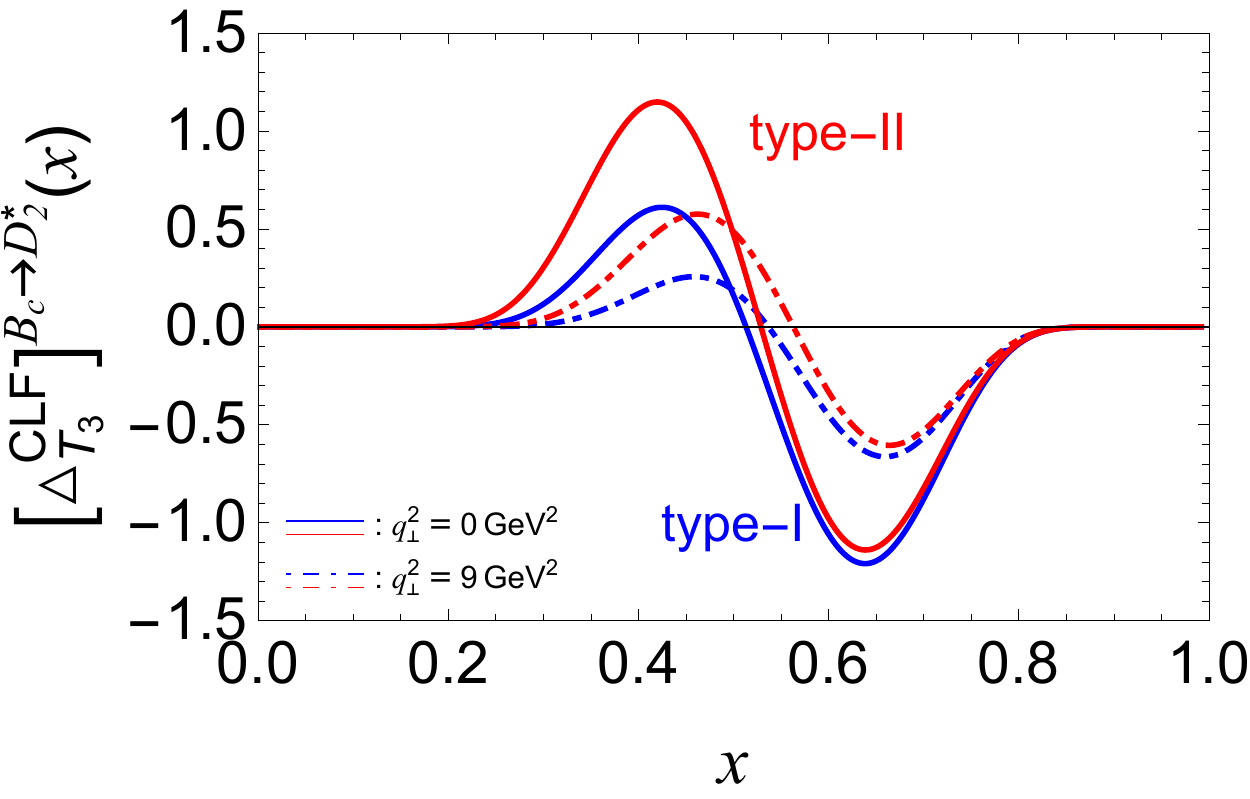}}\\
\end{center}
\label{fig:T23}
\end{figure}

The tensor form factors of  $P \to T$ transition have also been calculated by Cheng and Chua~(CC) ~\cite{Cheng:2004yj} within the traditional CLF QM, their results are given in appendix B. The contributions associated with $B$ functions are not considered in their calculation. Besides, comparing CC's results with ours, it is found that  they are the same  for  $T_1$, but  are obviously different for $T_2$ and $T_3$ in form. In addition, it has been checked that their numerical results for $T_2$ and $T_3$ are different either within type-I scheme. After checking our and CC's calculations, we find  another inconsistent problem caused by the different way for dealing with the trace term $S^{P\to T}_{\u\v\lambda\delta}$ related to the matrix element ${\cal B}^{P\to T}_{\rm CLF}[\Gamma=\sigma_{\mu\nu}\gamma_5]$.
To clarify the origin of  this inconsistent problem, we take the term $2ig_{\nu\lambda}g_{\alpha\mu}g_{\beta\sigma}(P+q)^{\beta}k'^{\sigma}_1k'^{\alpha}_1k'_{1\delta}$ appeared in  $S^{P\to T}_{\u\v\lambda\delta}$ as an example. As has been mentioned in the last section,  some replacements are needed to take the zero-mode contribution into account after integrating out $k'^-_1$. In the CC's calculation, the replacement for $\hat{k}'^{\sigma}_1\hat{k}'^{\alpha}_1\hat{k}'^{\delta}_1$ is used directly though $\sigma$ is a dummy indices, {\it i.e.},
\begin{align}
2ig_{\nu\lambda}g_{\alpha\mu}g_{\beta\sigma}(P+q)^{\beta}\hat{k}'^{\sigma}_1\hat{k}'^{\alpha}_1\hat{k}'_{1\delta}=&2i g_{\nu\lambda}g_{\alpha\mu}g_{\beta\sigma}(P+q)^\b\Big[(g^{\a\sigma}P_\delta+g^\a_\delta P^\sigma+g^{\sigma}_\delta P^\alpha) A^{(3)}_1\nonumber\\
&+(g^{\a\sigma}q_\delta+g^\a_\delta q^\sigma+g^{\sigma}_\delta q^\alpha) A^{(3)}_2+P^\sigma P^\alpha P_\delta A^{(3)}_3\nonumber\\
&+\,...\Big]\nonumber\\
=&2i g_{\nu\lambda}\Big\{g_{\mu\delta}[(P^2+P\cdot q)A^{(3)}_1+(P\cdot q+q^2)A^{(3)}_2]\nonumber\\
&+P_{\mu}[ (2P_\delta+q_\delta)A^{(3)}_1+q_\delta A^{(3)}_2+ P_\delta (P^2+P\cdot q)A^{(3)}_3+\,...]\nonumber\\
&+q_\mu[P_\delta A^{(3)}_1+(P_\delta+2q_\delta) A^{(3)}_2+\,...]\Big\}\,.
\label{eq:s1}
\end{align}
In our calculation, we employ the standard procedure of CLF calculation, and obtain
\begin{align}\
2ig_{\nu\lambda}g_{\alpha\mu}g_{\beta\sigma}(P+q)^{\beta}\hat{k}'^{\sigma}_1\hat{k}'^{\alpha}_1\hat{k}'_{1\delta}=&2i g_{\v\lbd} \hat{k}'_{1\u}\hat{k}'_{1\delta}\,\hat{k}'_1\cdot(P+q)\nonumber\\
=&2i g_{\v\lbd}\Big\{{g_{\mu\delta}\left[(M'^2+m'^2_1-m^2_2+N'_1-Z_2)A_1^{(2)}\right]}\nonumber\\
&+{P_\mu \left[(M'^2+m'^2_1-m^2_2+N'_1)(P_\delta A_2^{(2)}+q_\delta A_3^{(2)})\right]}\nonumber\\
&+q_\mu\Big[(M'^2+m'^2_1-m^2_2+N'_1)(P_\delta A_3^{(2)}+q_\delta A_4^{(2)})\nonumber\\
&{-q_\delta\big( A_4^{(2)}Z_2+2\frac{q\cdot P}{q^2}A_2^{(1)}A_1^{(2)}\big)\Big\}+\,...}\,.
\label{eq:s2}
\end{align}
Comparing Eq.~\eqref{eq:s2} with Eq.~\eqref{eq:s1}, one can easily find that CC's result is different from ours because different replacements are needed. As a result, our and CC's results for $T_{2,3}$ are different in form.

In order to clearly show the divergence between CC's and our results, we define the difference
\begin{align}
\Delta_{\cal F}^{\rm CLF}(x,\qb^2) \equiv \frac{\d [{\cal F}]^{\rm CLF}_{\rm ours}}{ \d x}-\frac{\d [{\cal F}]^{\rm CLF}_{\rm CC}}{ \d x}\,,
\end{align}
where ${\cal F}=T_{2,3}$. Then, taking $D\to K^*_2$ and $B_c \to D^*_2$ as examples, the dependences of  $\Delta_{T_{2,3}}^{\rm CLF}(x,\qb^2)$ on $x$ in type-I and -II schemes are shown in Fig.~\ref{fig:T23}. It can be easily seen from Fig.~\ref{fig:T23} that our and CC's numerical results for $T^{\rm CLF}_{2,3}$ are inconsistent within the type-I scheme; however, such inconsistence problem vanishes  in the type-II scheme due to $\int^1_0 dx \Delta_{{ T}_{2,3}}^{\rm CLF}(x)\dot{=}0 $.

\begin{table}[t]
\begin{center}
\caption{\label{tab:rm} \small The summary of the  resonance masses~(in units of GeV) with different quantum numbers entering  the $z$-series expansions of the $P \to T$ form factors~\cite{Zyla:2020zbs,Detmold:2015aaa,Dowdall:2012ab}.}
\vspace{0.2cm}
\let\oldarraystretch=\arraystretch
\renewcommand*{\arraystretch}{1.1}
\setlength{\tabcolsep}{8.8pt}
\begin{tabular}{lcccccccccc}
\hline\hline
  &${\cal F}(q^2)$      &$J^P$
   &$c\to q$   &$c\to s$
   &$b\to q$   &$b\to s$    &$b\to c$ \\
  \hline
   &$A_0(q^2)$              &$0^-$
   &$1.864$    &$1.968$
   &$5.279$    &$5.367$     &$6.275$ \\
  \hline
  &$V(q^2),T_1(q^2)$        &$1^-$
  &$2.007$     &$2.112$
  &$5.325$     &$5.415$     &$6.329$ \\
  \hline
  &$A_1(q^2),A_{2}(q^2),T_2(q^2),T_{3}(q^2)$   &$1^+$
  &$2.422$     &$2.460$
  &$5.726$     &$5.829$    &$6.767$ \\
  \hline\hline
\end{tabular}
\end{center}
\end{table}

\begin{figure}[t]
\caption{ The $q^2$ dependence of form factors of $c\to (q,s)$ induced $D \to (a_2,K^*_2)$, $D_s \to (K^*_2,f'_{2})$, $\eta_c(1S) \to (D^*_2,D^*_{s2})$, $ B_c \to (B^*_2,B^*_{s2})$ transitions.}
\begin{center}
\subfigure{\includegraphics[scale=0.3]{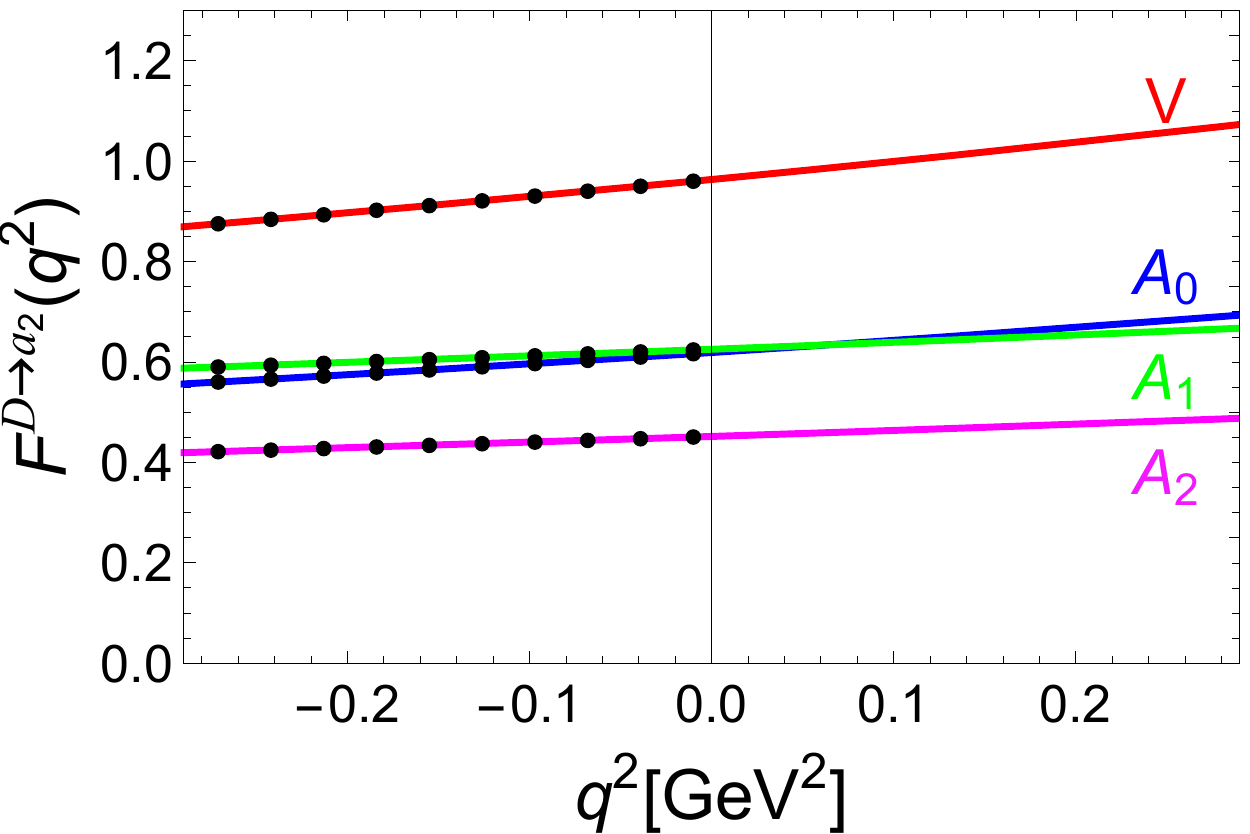}}\,
\subfigure{\includegraphics[scale=0.3]{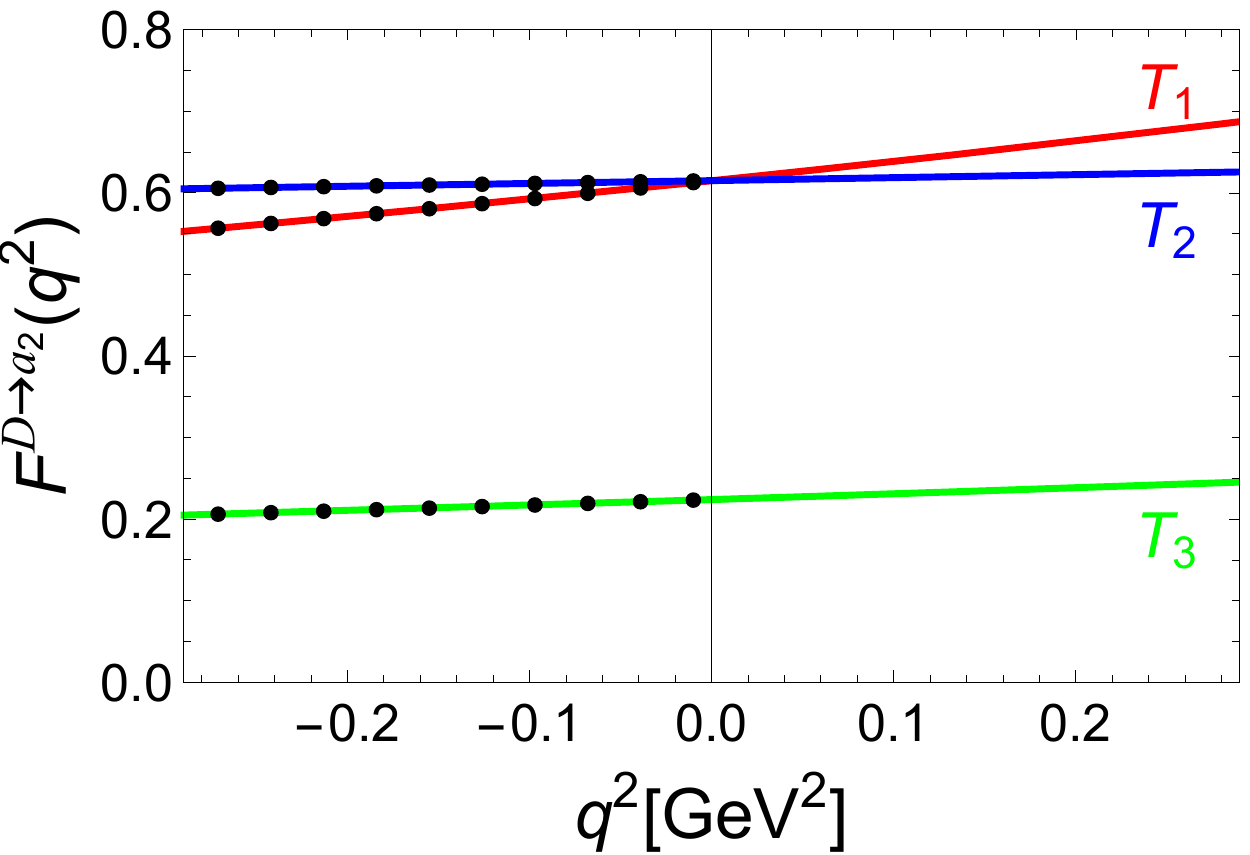}}\,
\subfigure{\includegraphics[scale=0.3]{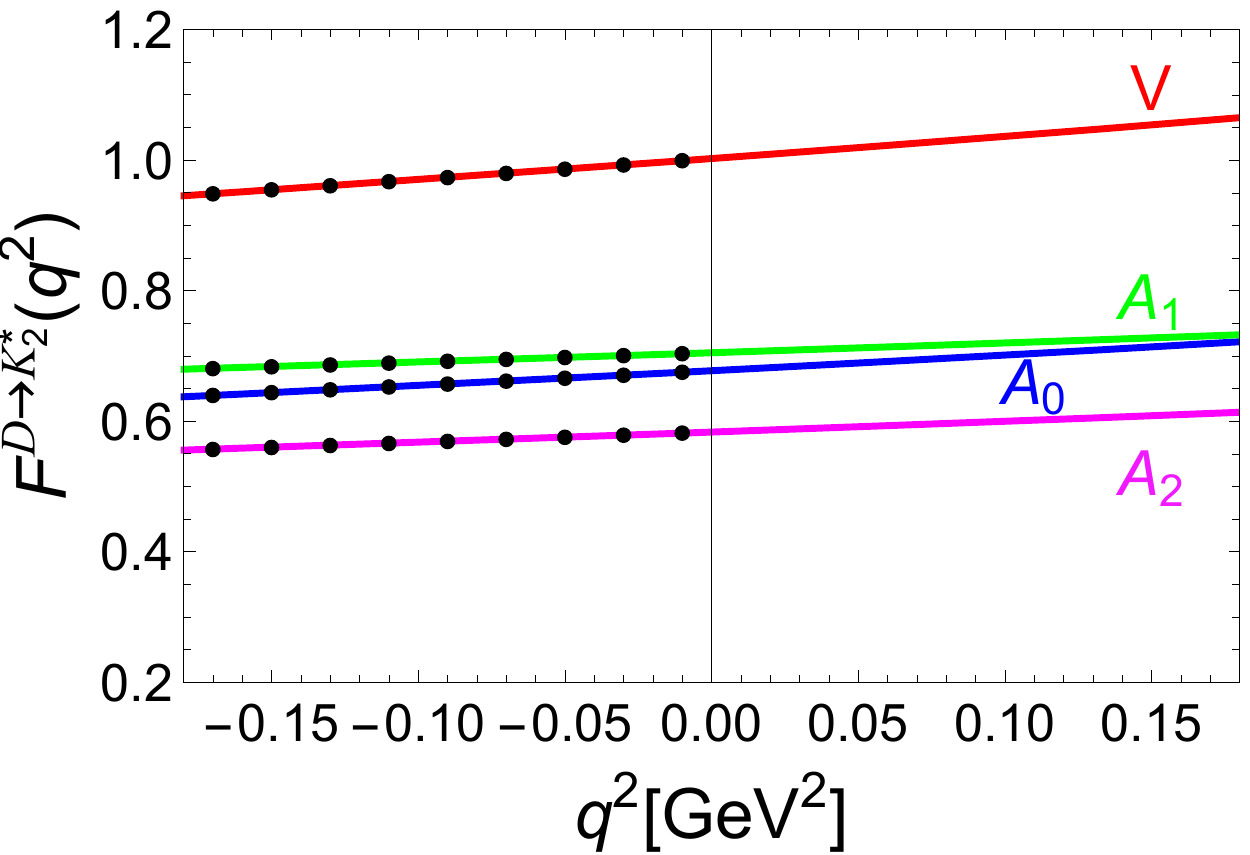}}\,
\subfigure{\includegraphics[scale=0.3]{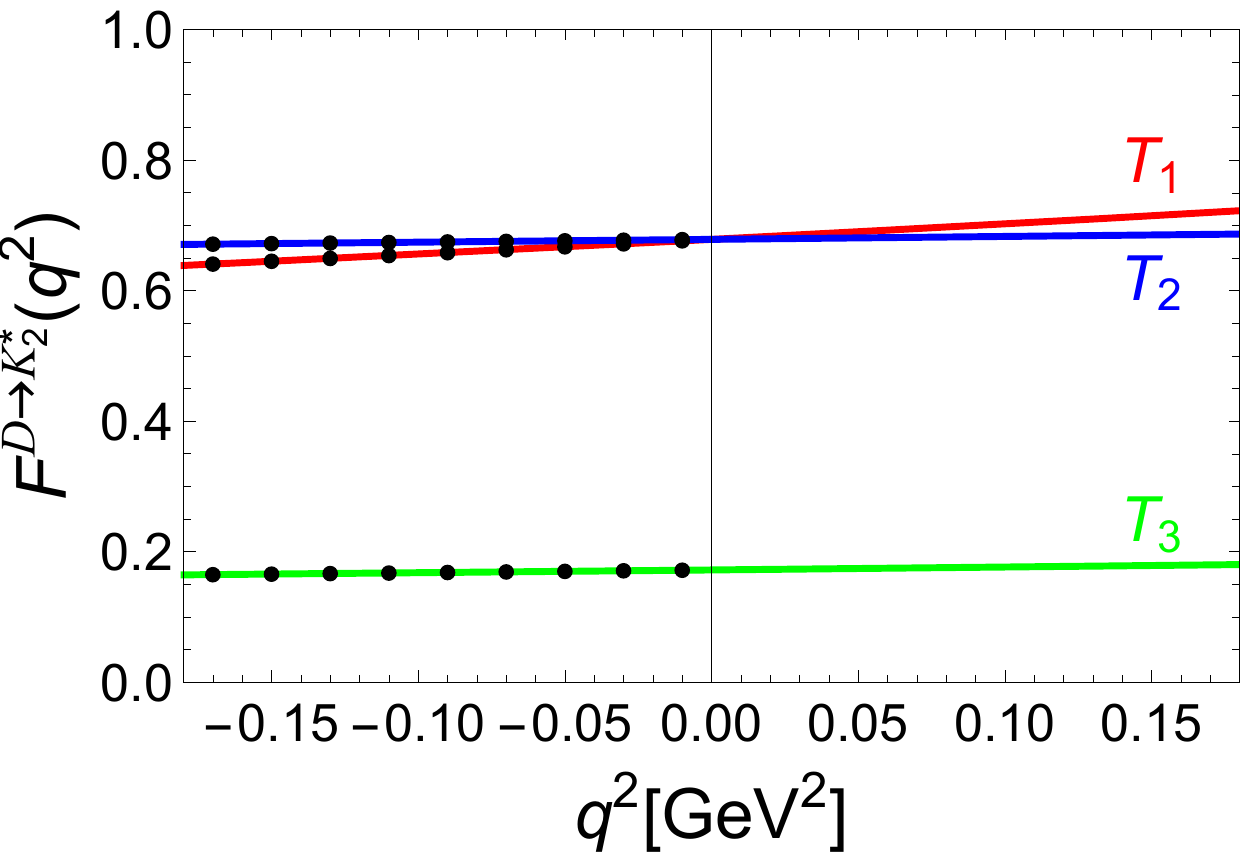}}\,\\
\subfigure{\includegraphics[scale=0.3]{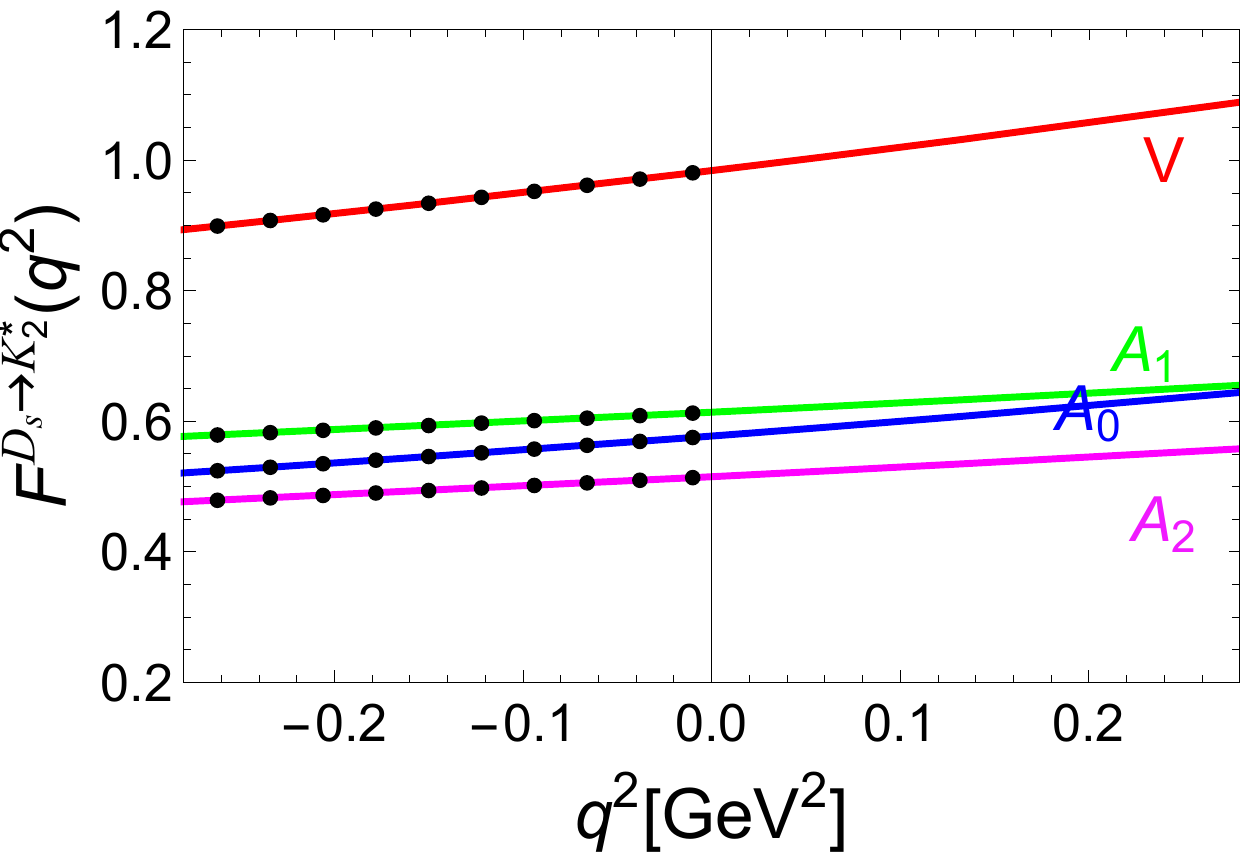}}\,
\subfigure{\includegraphics[scale=0.3]{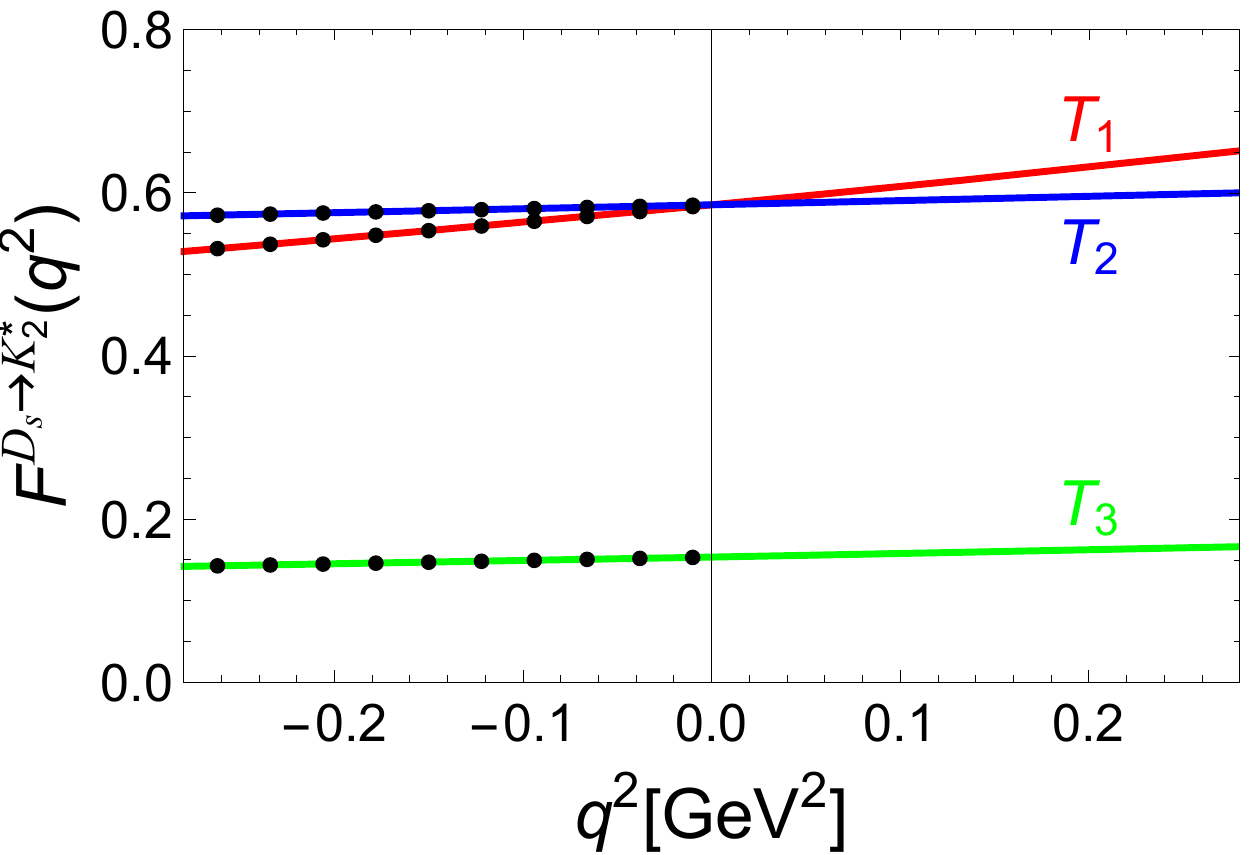}}\,
\subfigure{\includegraphics[scale=0.3]{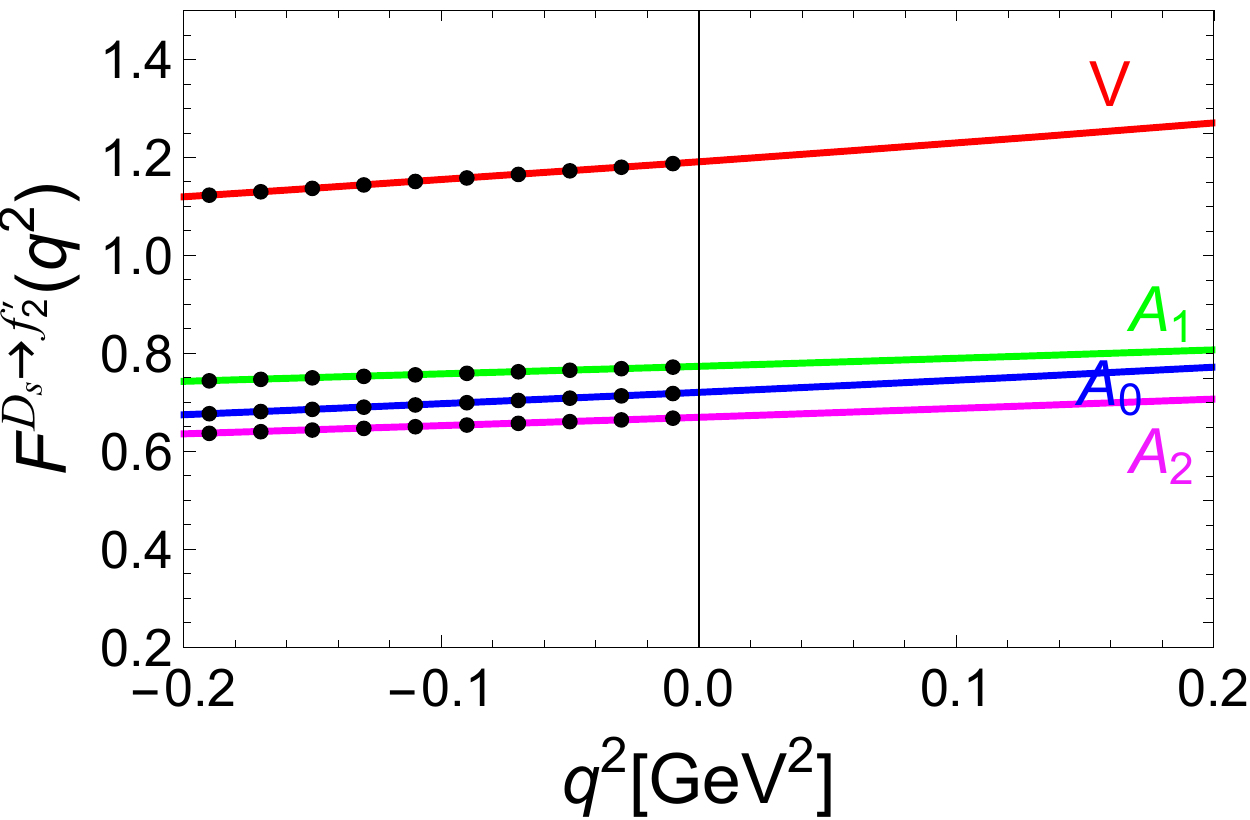}}\,
\subfigure{\includegraphics[scale=0.3]{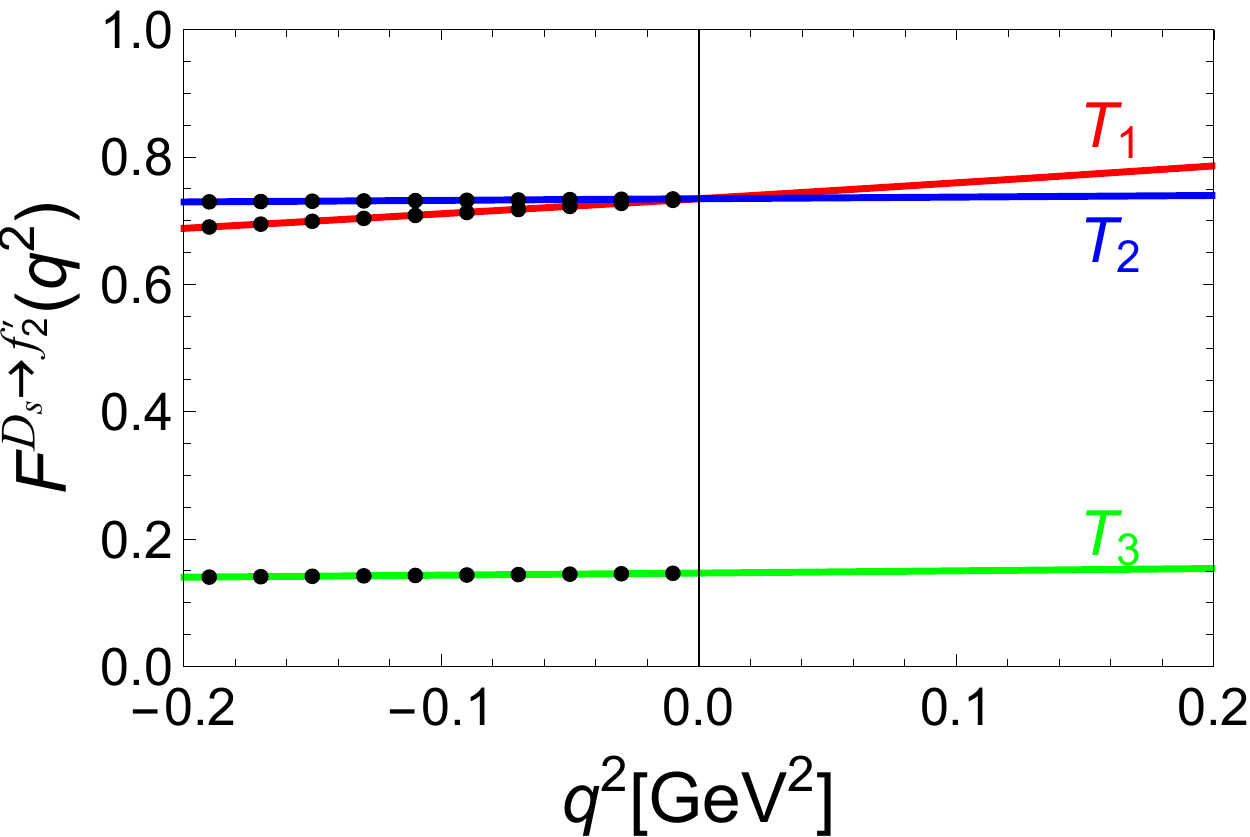}}\\
\subfigure{\includegraphics[scale=0.3]{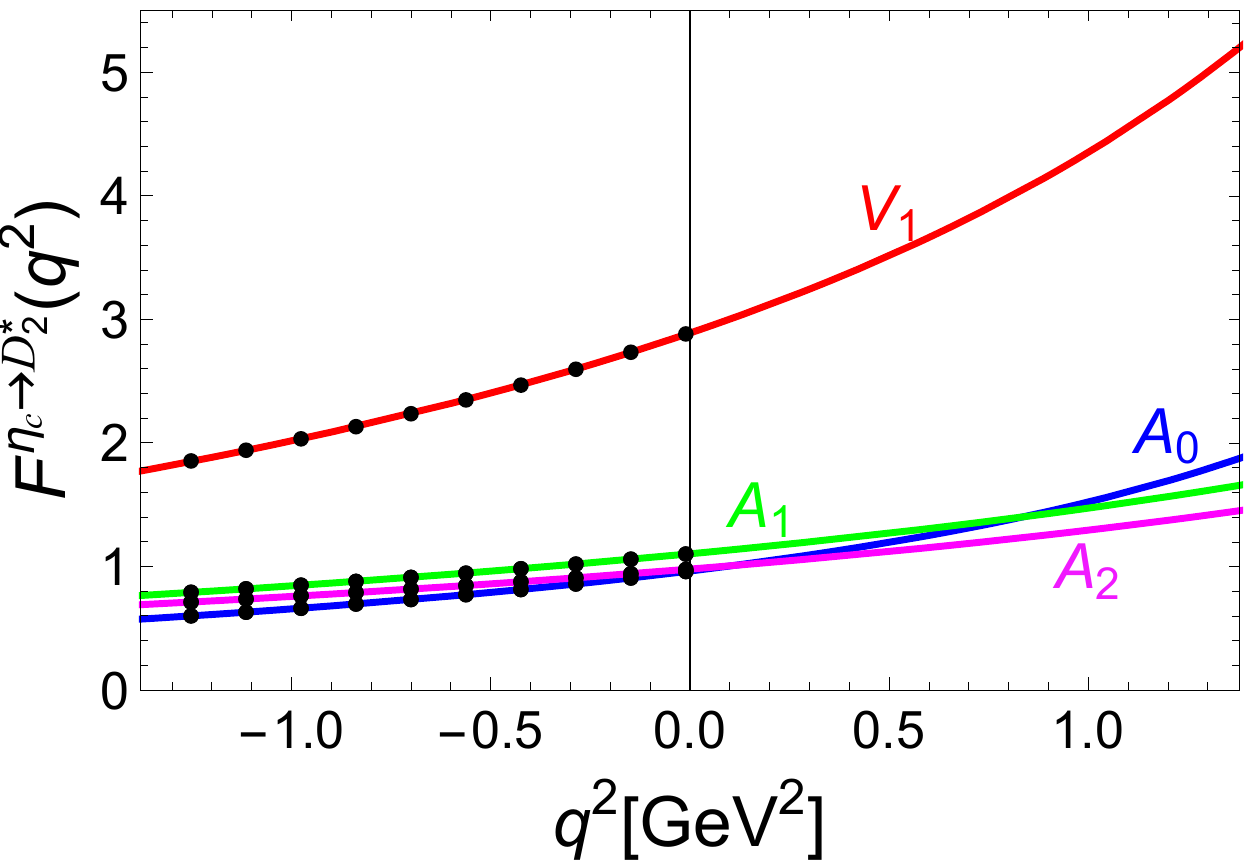}}\,
\subfigure{\includegraphics[scale=0.3]{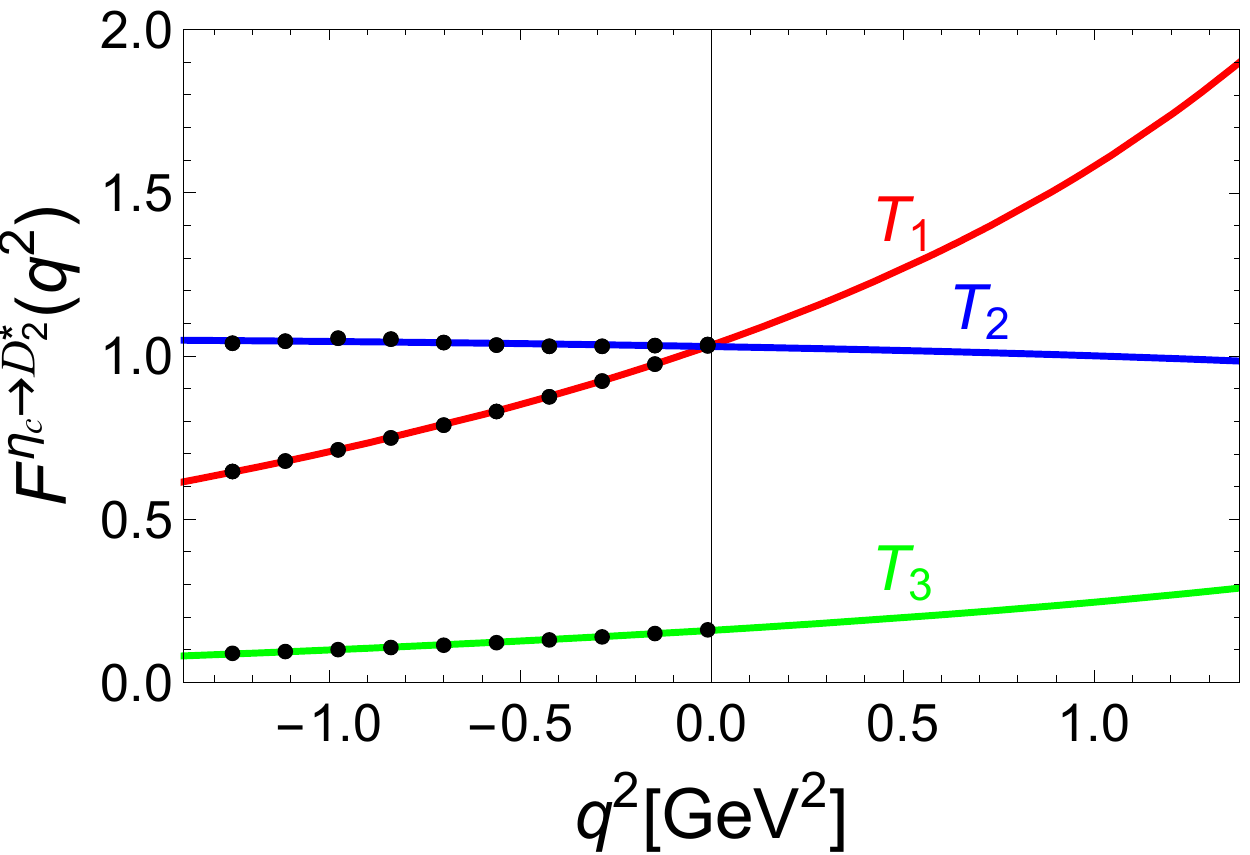}}\,
\subfigure{\includegraphics[scale=0.3]{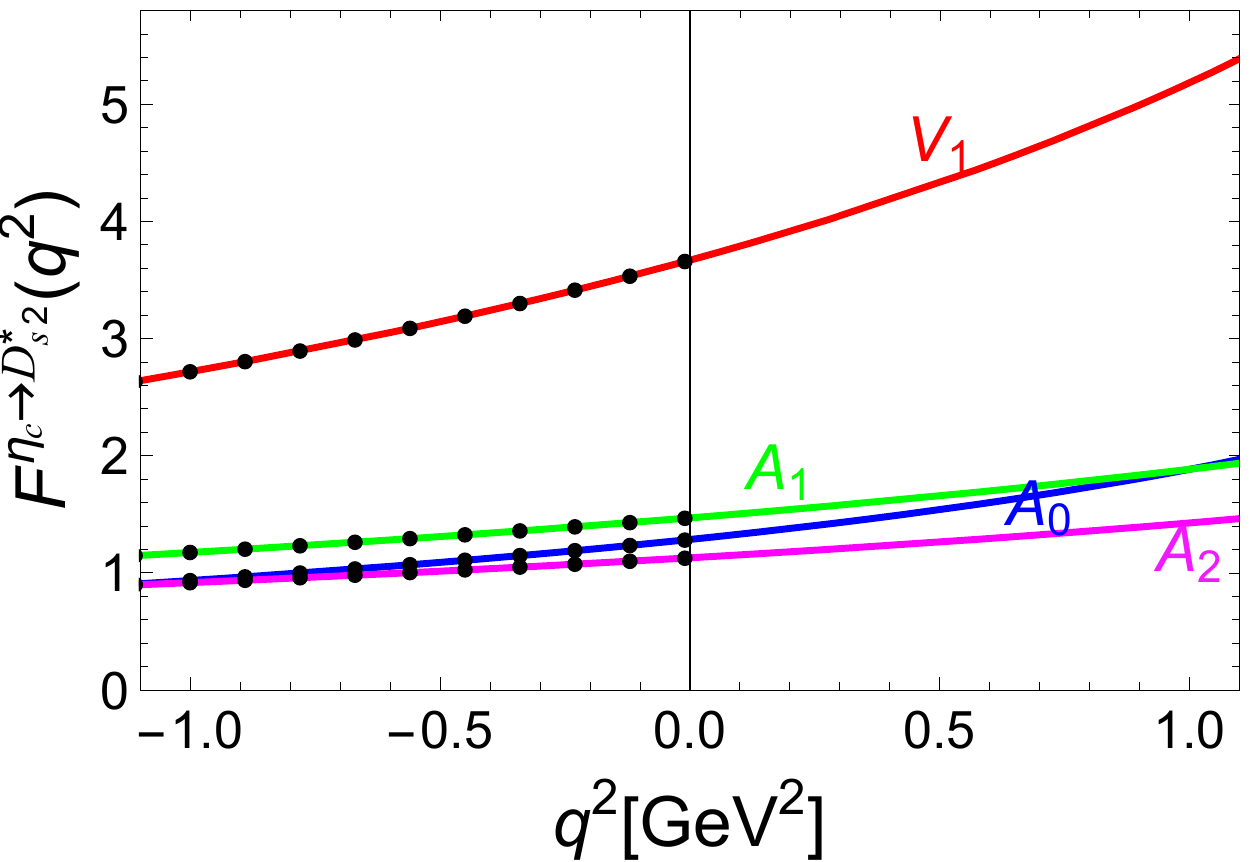}}\,
\subfigure{\includegraphics[scale=0.3]{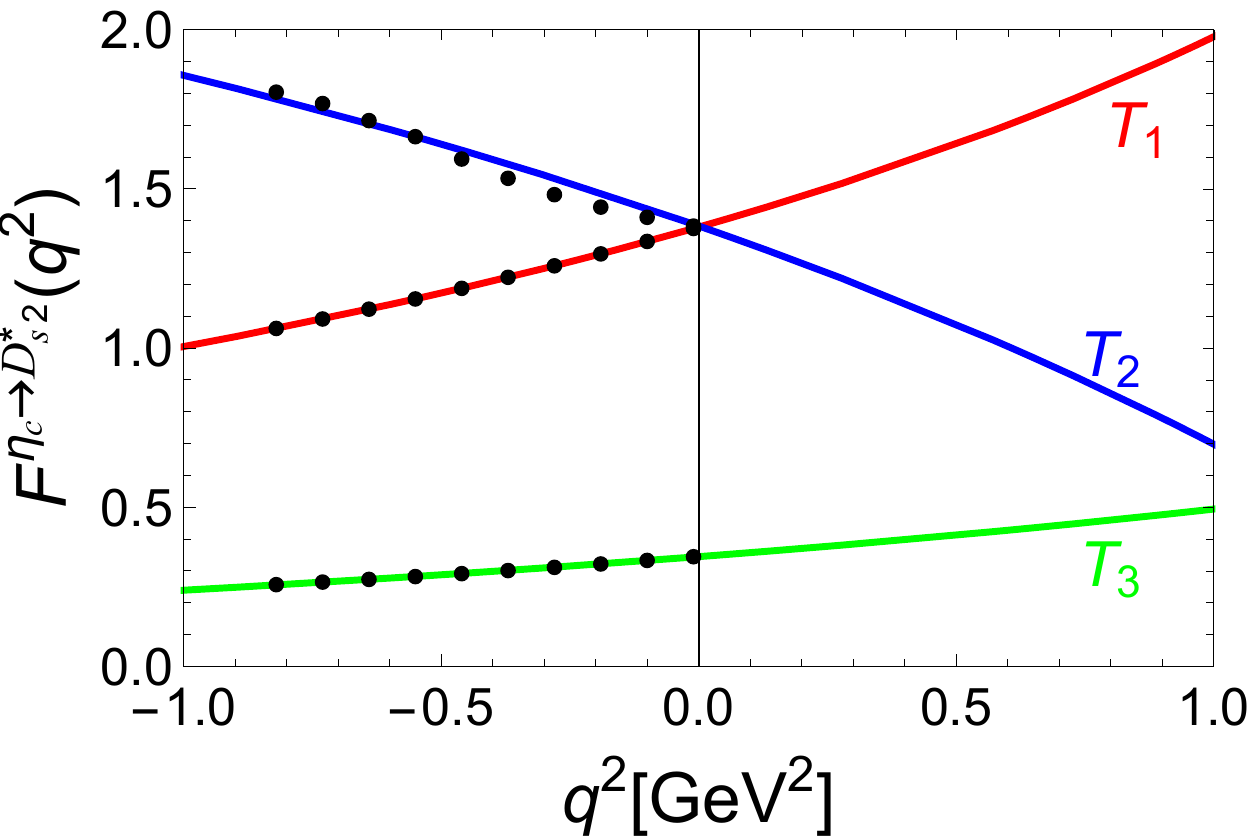}}\\
\subfigure{\includegraphics[scale=0.3]{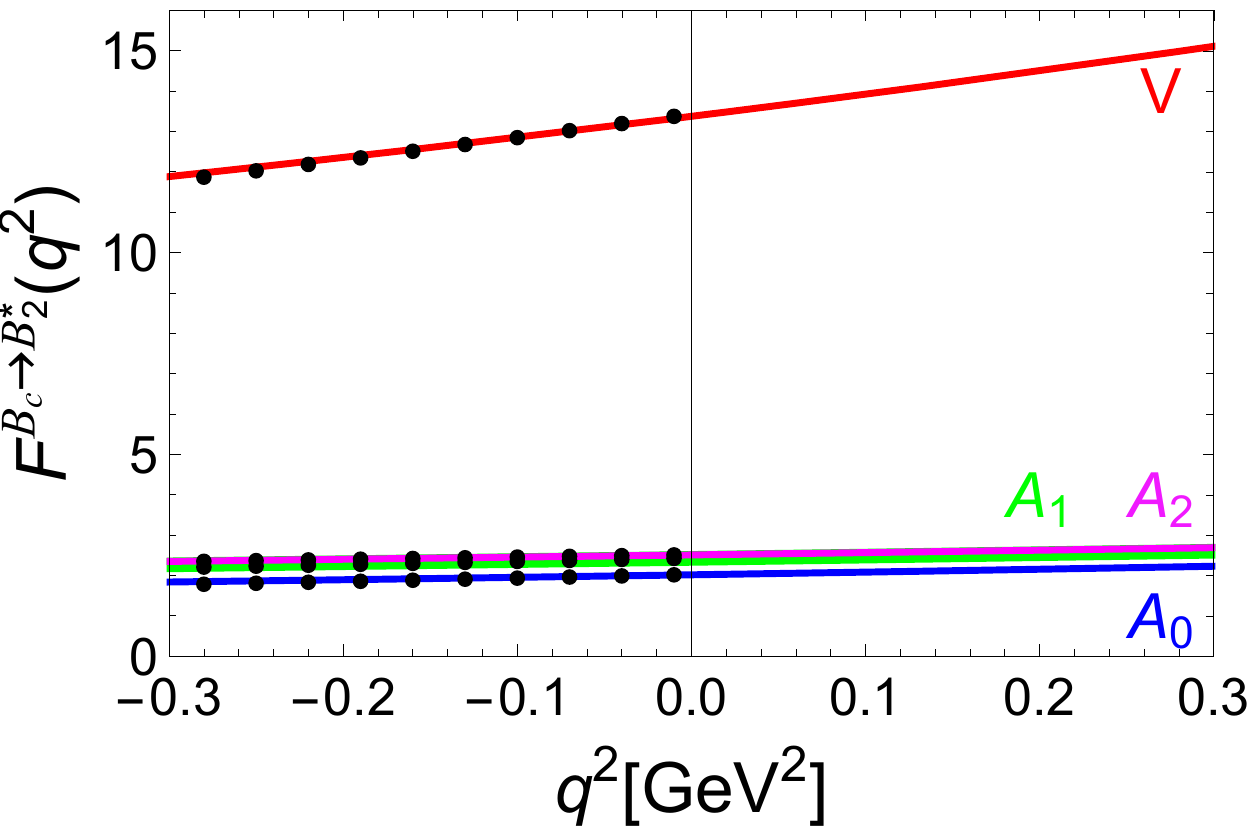}}\,
\subfigure{\includegraphics[scale=0.3]{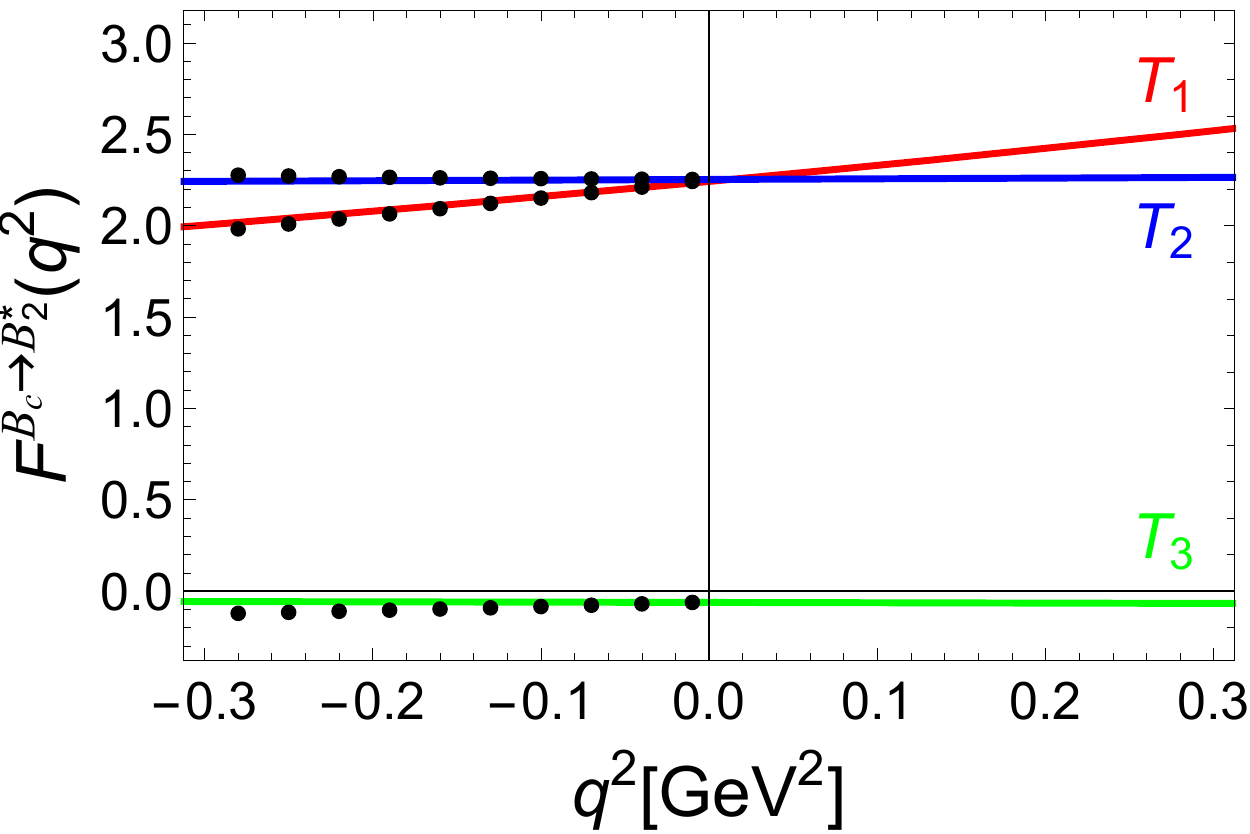}}\,
\subfigure{\includegraphics[scale=0.3]{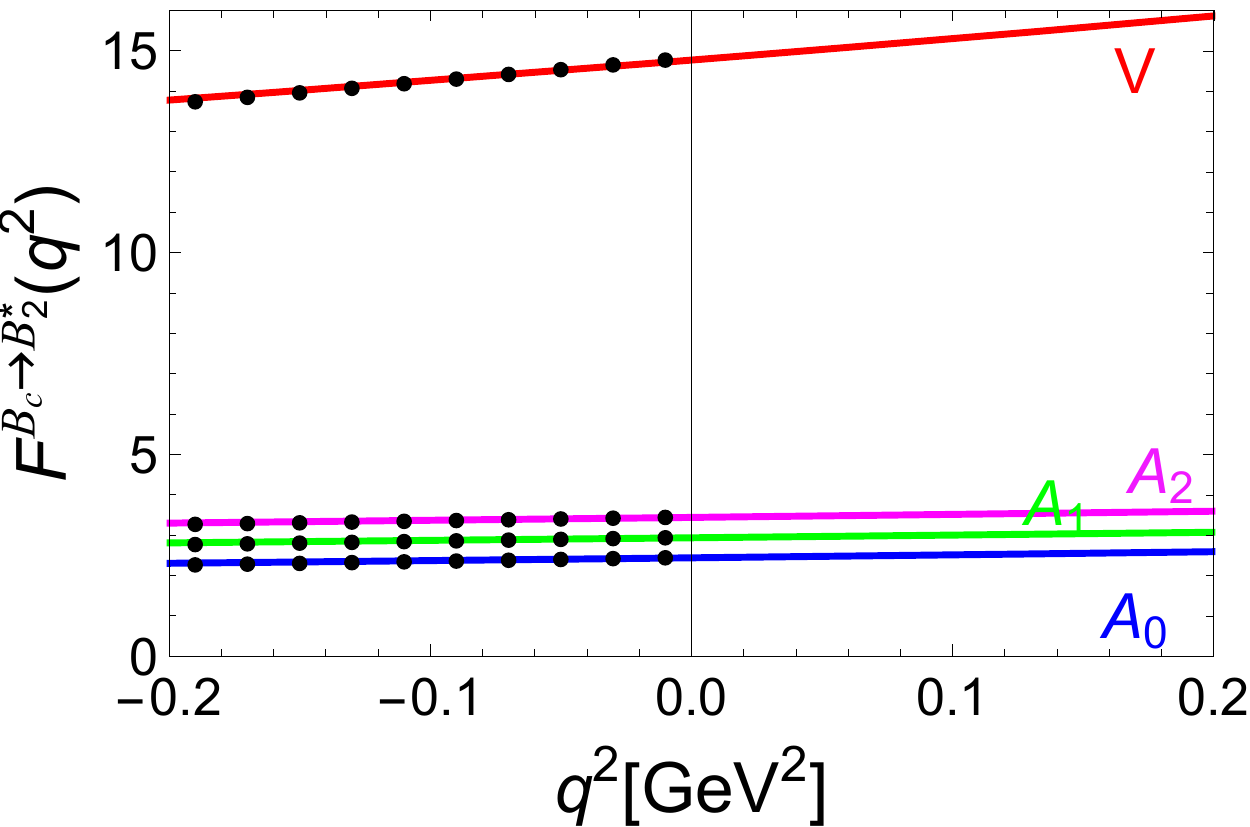}}\,
\subfigure{\includegraphics[scale=0.3]{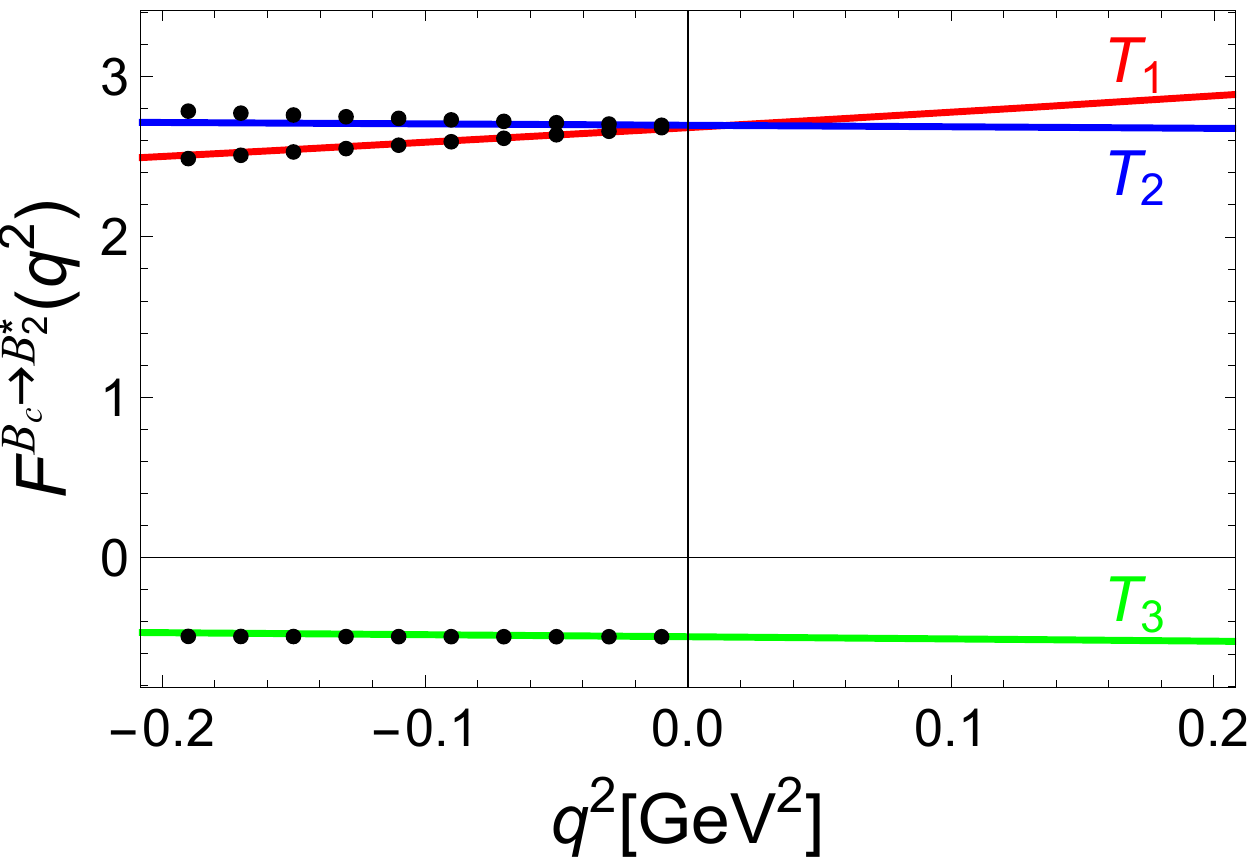}}\\
\end{center}
\label{fig:PTVAc}
\end{figure}

\begin{figure}[t]
\caption{ The $q^2$ dependence of  form factors of $b\to (q,s,c)$ induced $B \to (a_2,K^*_2,D^*_2)$, $B_s \to (K^*_2,f'_2,D^*_{s2})$, $B_c \to (D^*_2,D^*_{s2},\chi_{c2}(1P))$, $\eta_b(1S) \to (B^*_2,B^*_{s2})$   transitions.}
\begin{center}
\subfigure{\includegraphics[scale=0.3]{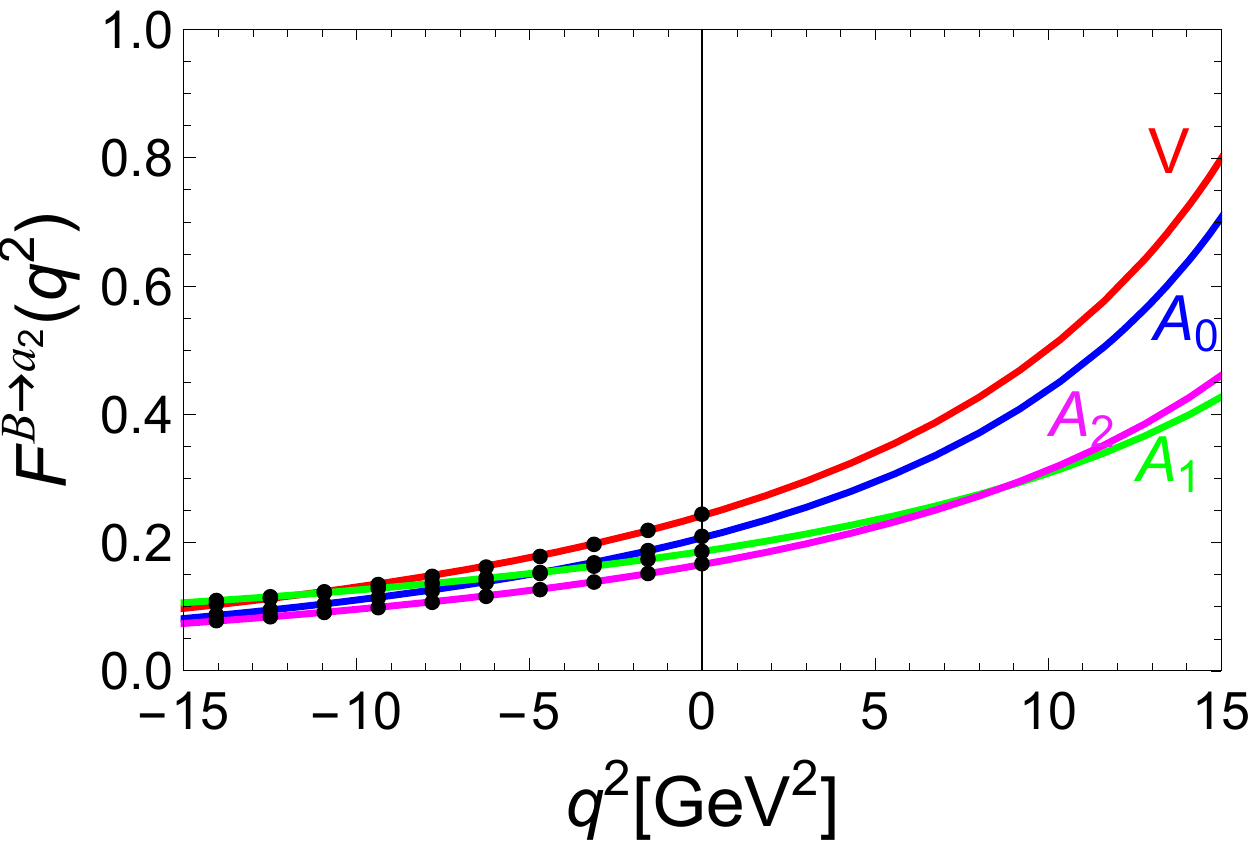}}\,
\subfigure{\includegraphics[scale=0.3]{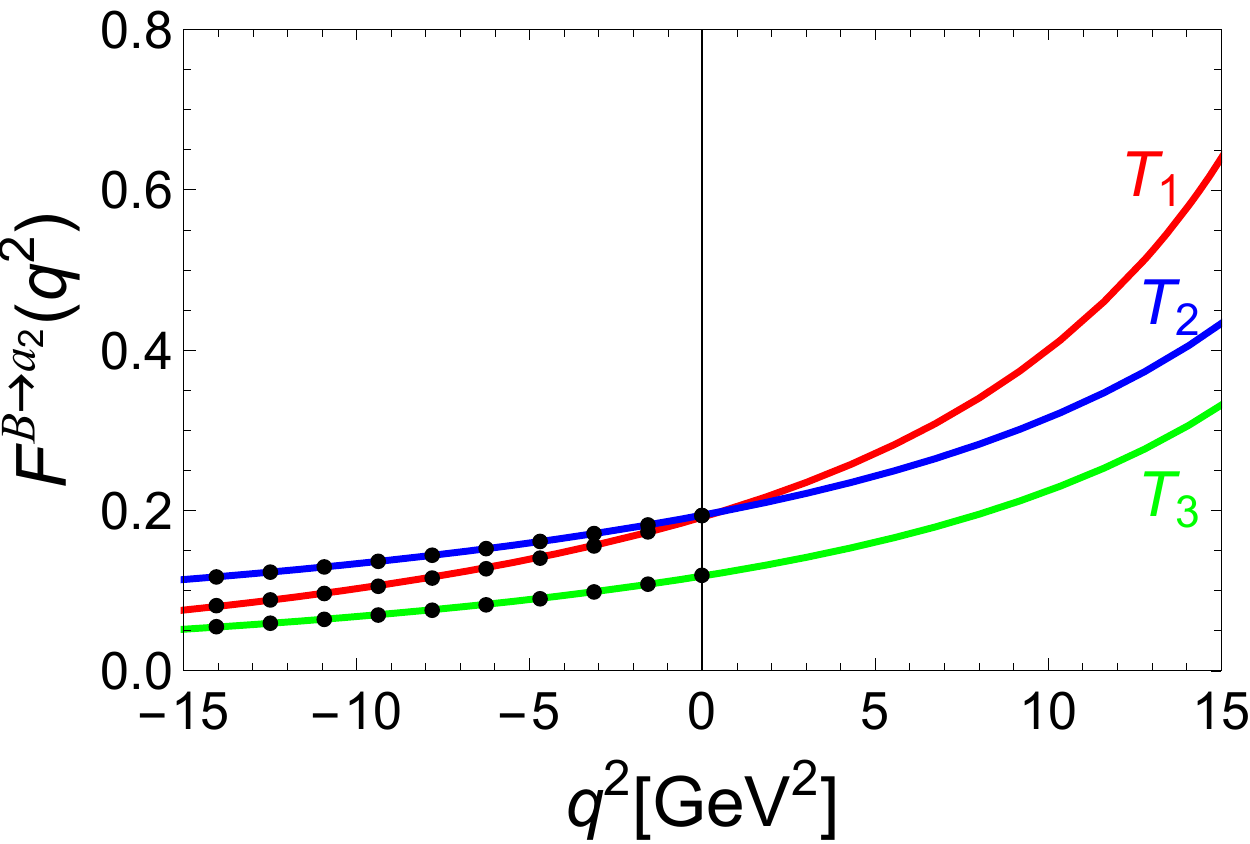}}\,
\subfigure{\includegraphics[scale=0.3]{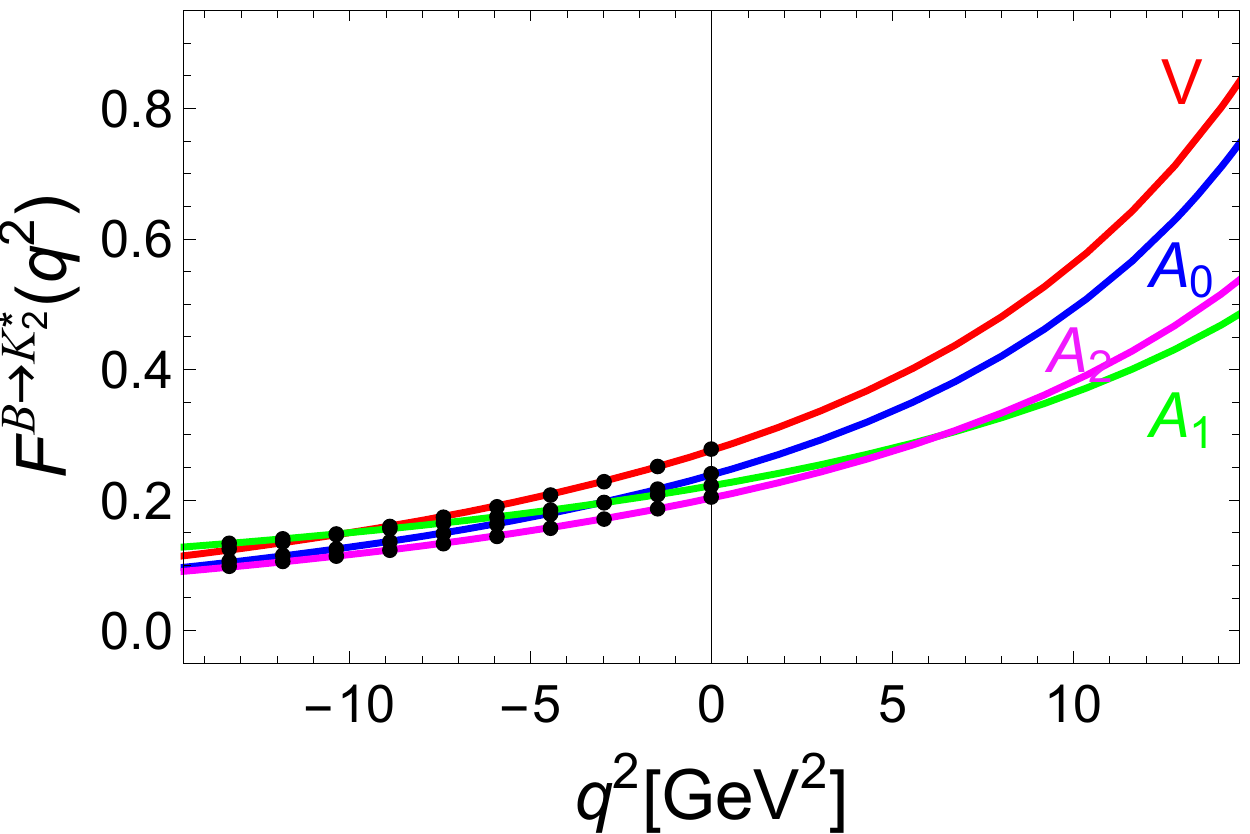}}\,
\subfigure{\includegraphics[scale=0.3]{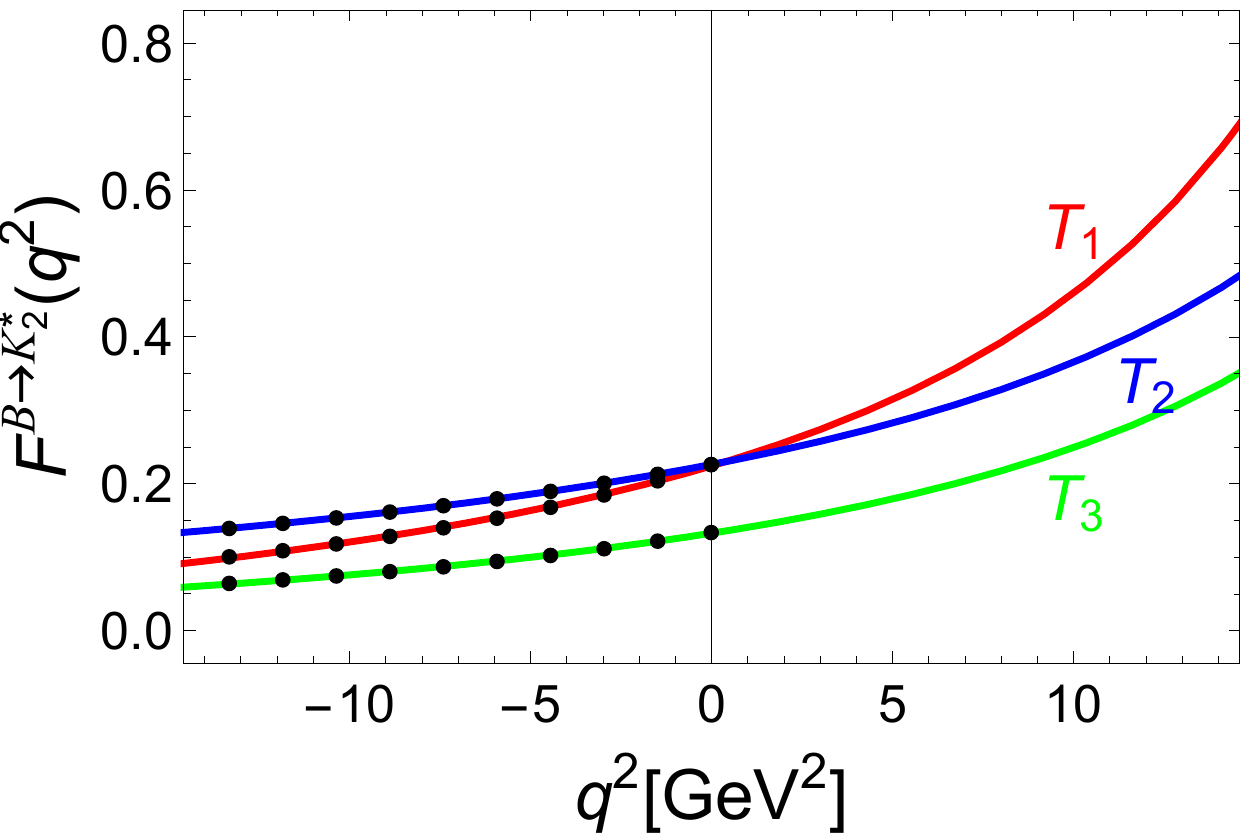}}\\
\subfigure{\includegraphics[scale=0.3]{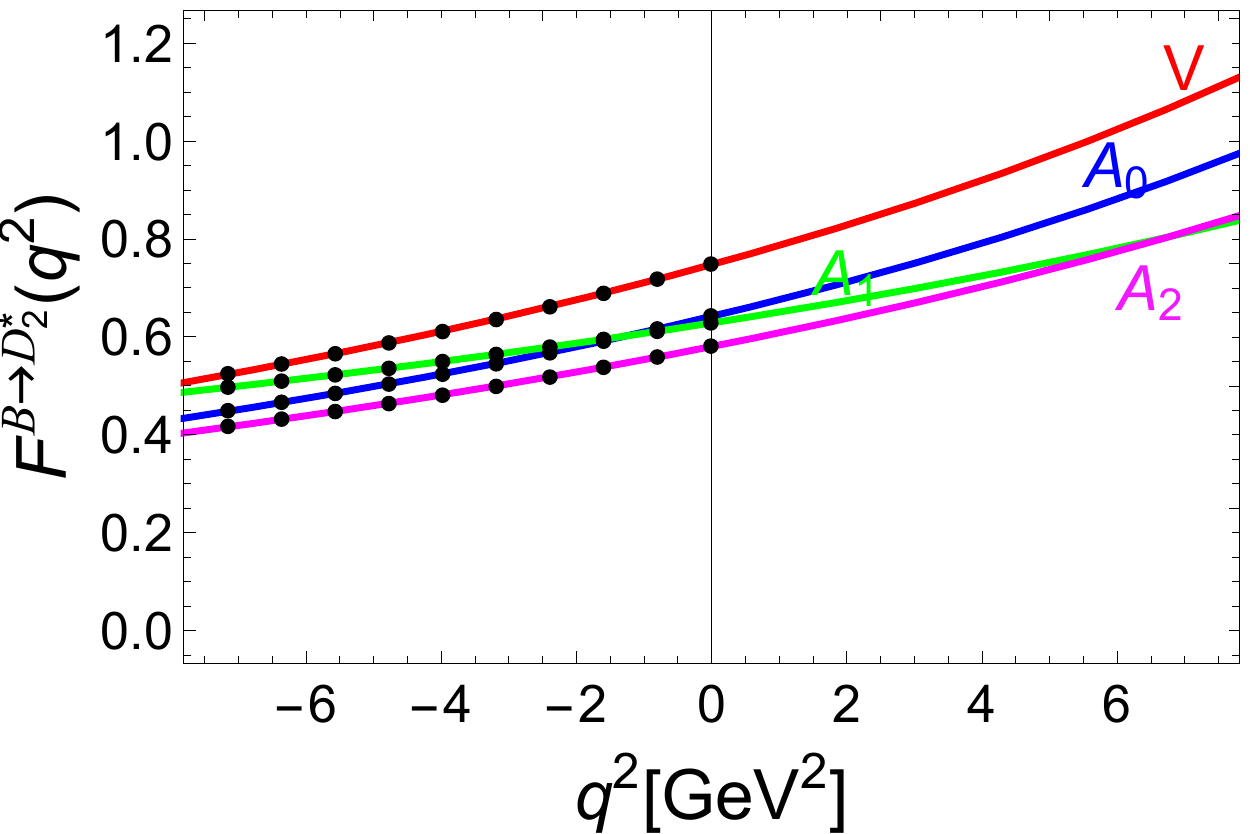}}\,
\subfigure{\includegraphics[scale=0.3]{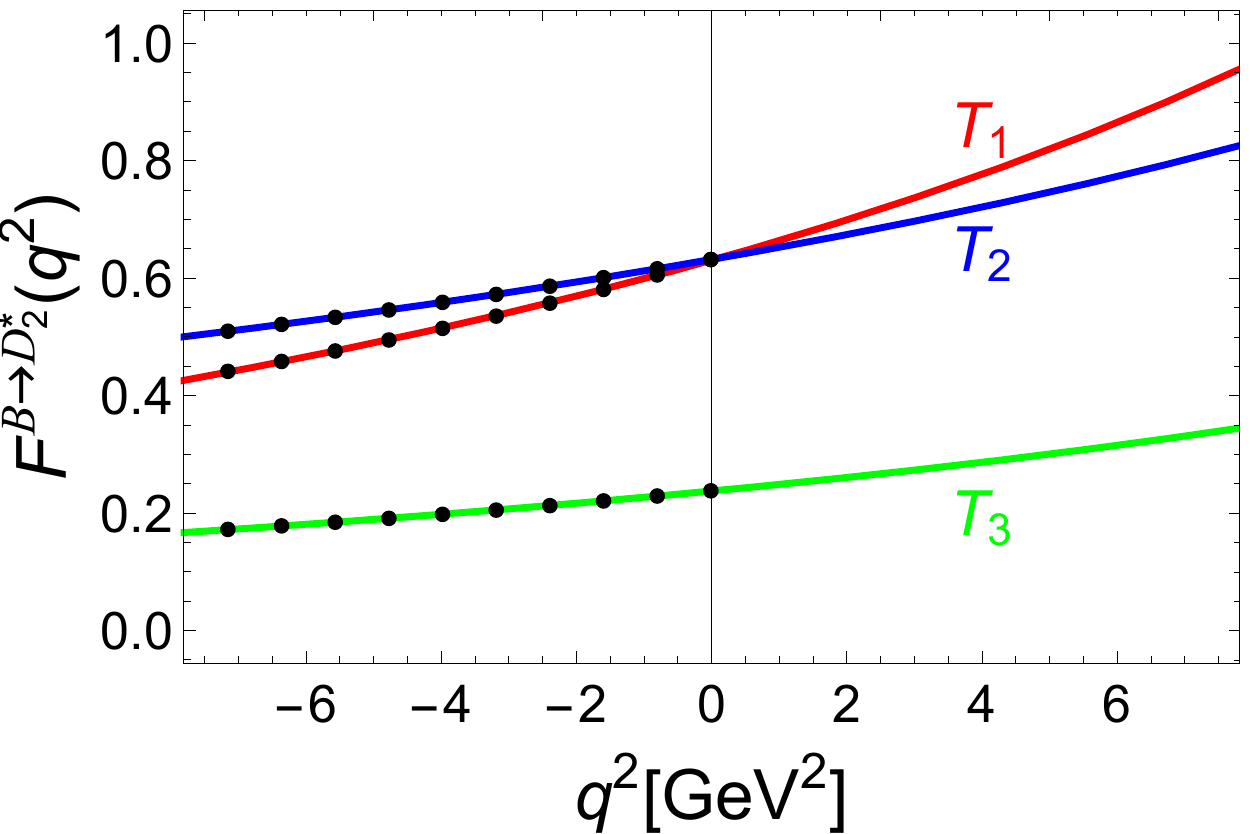}}\
\subfigure{\includegraphics[scale=0.3]{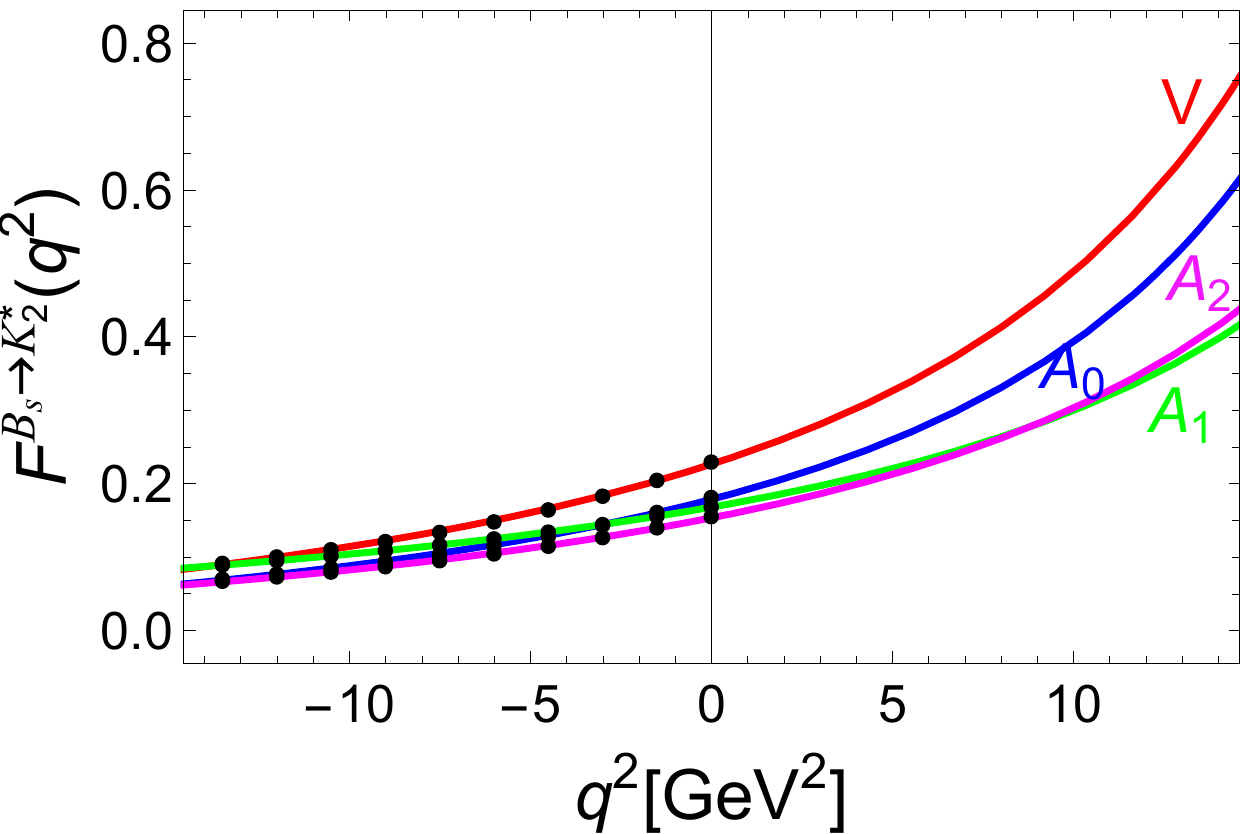}}\,
\subfigure{\includegraphics[scale=0.3]{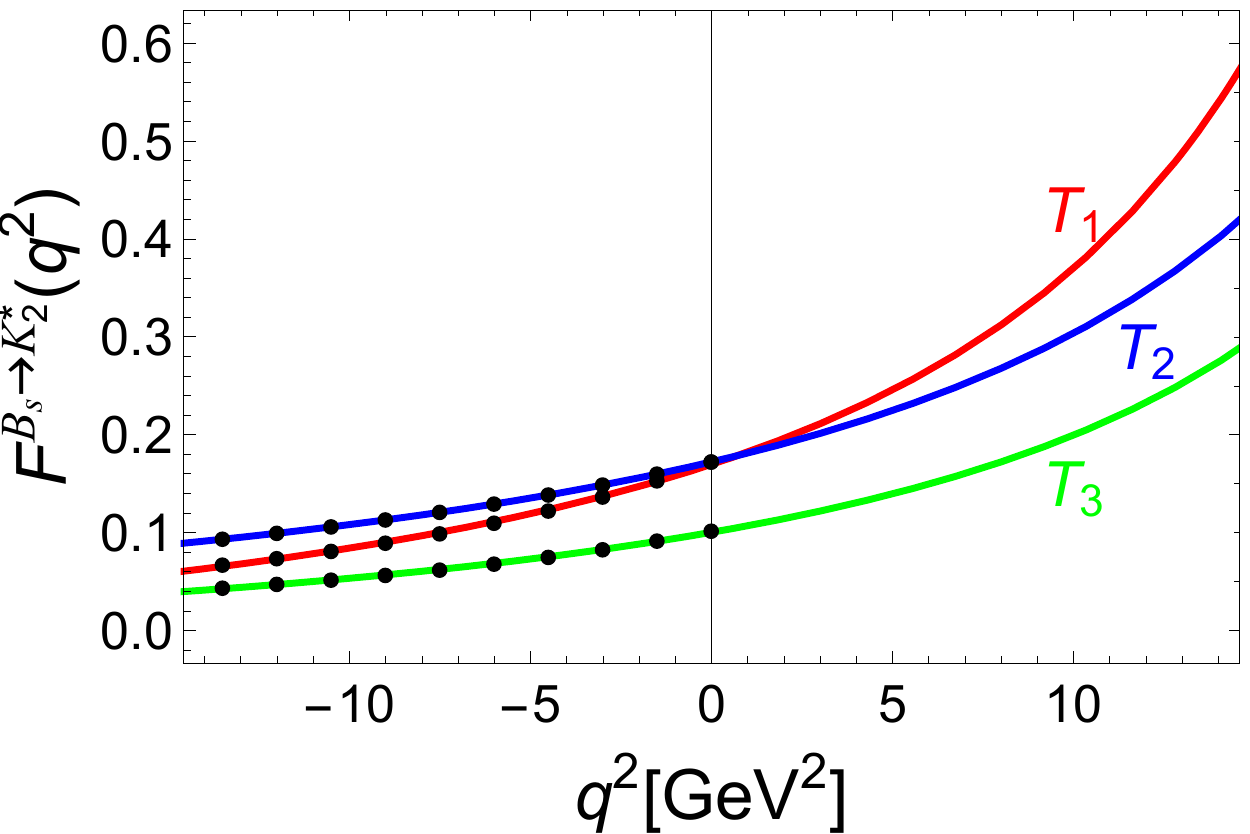}}\\
\subfigure{\includegraphics[scale=0.3]{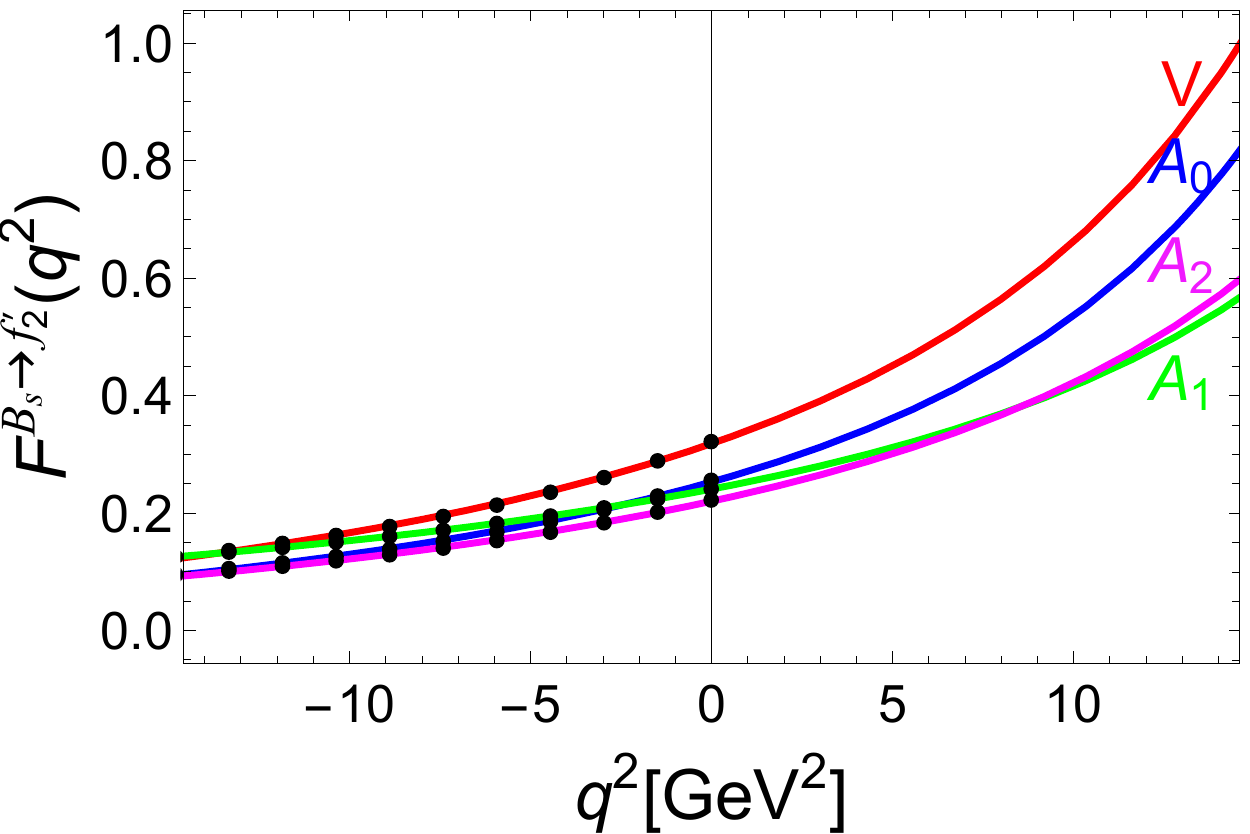}}\,
\subfigure{\includegraphics[scale=0.3]{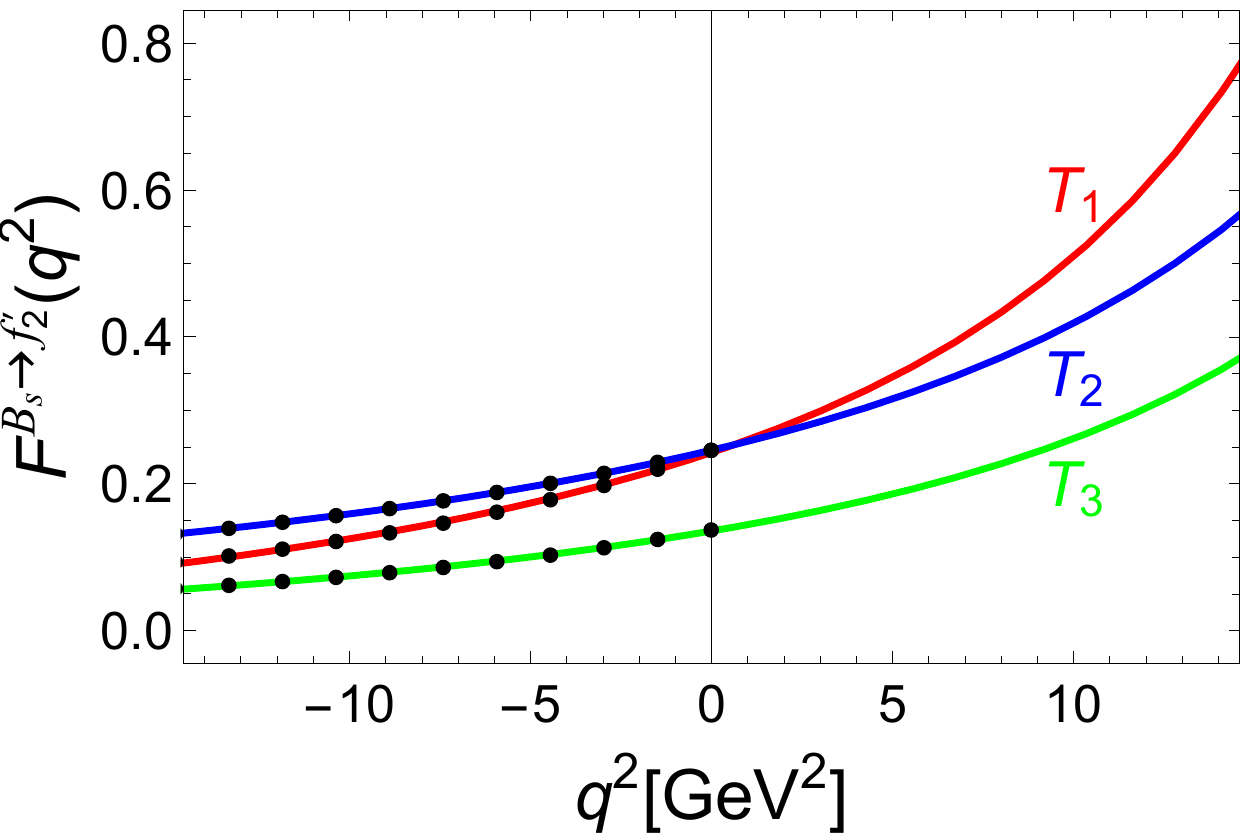}}\
\subfigure{\includegraphics[scale=0.3]{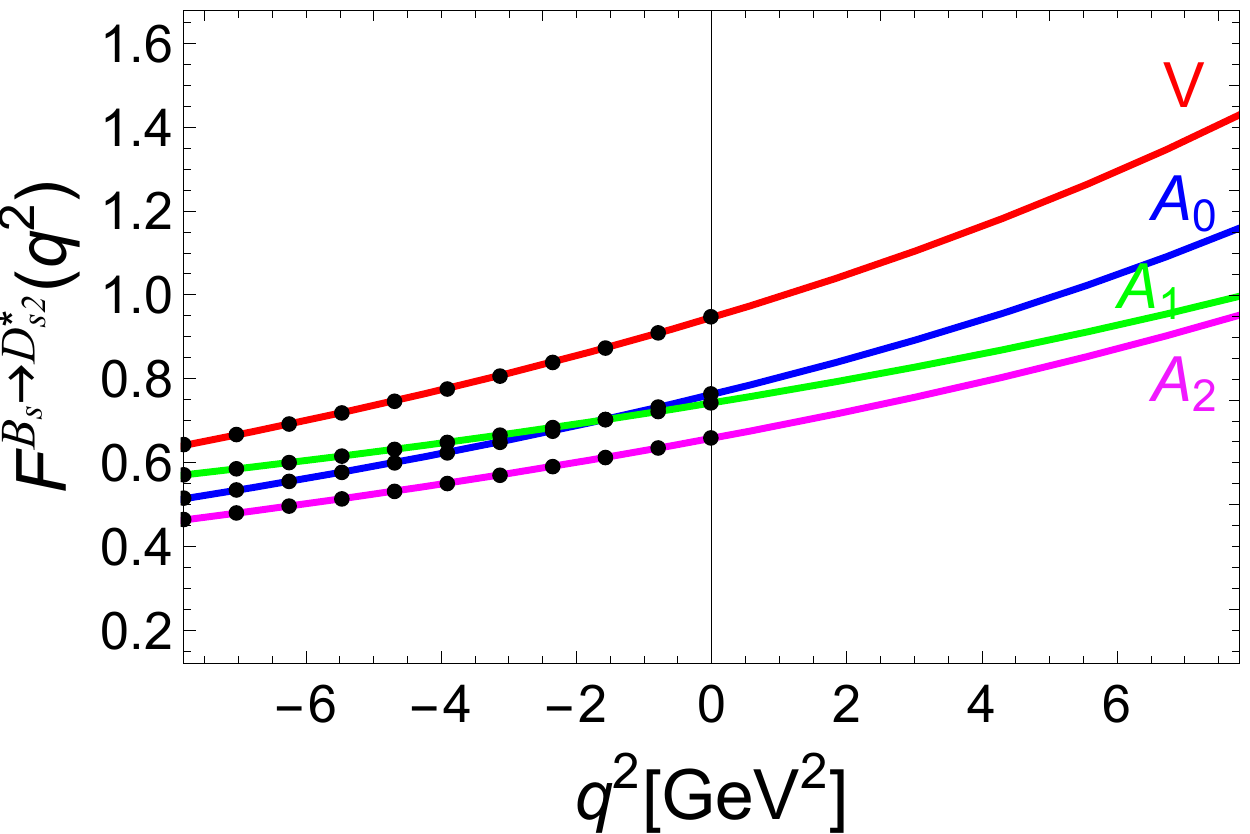}}\,
\subfigure{\includegraphics[scale=0.3]{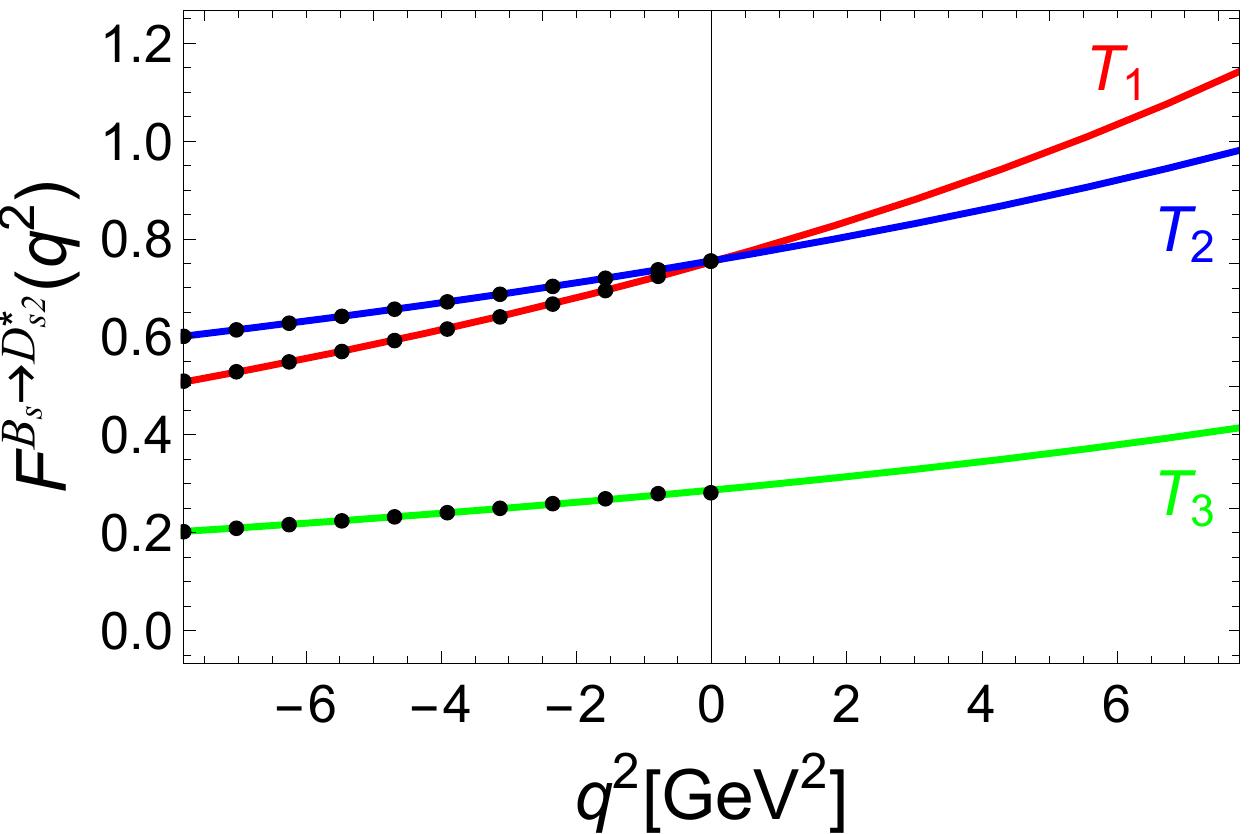}}\\
\subfigure{\includegraphics[scale=0.3]{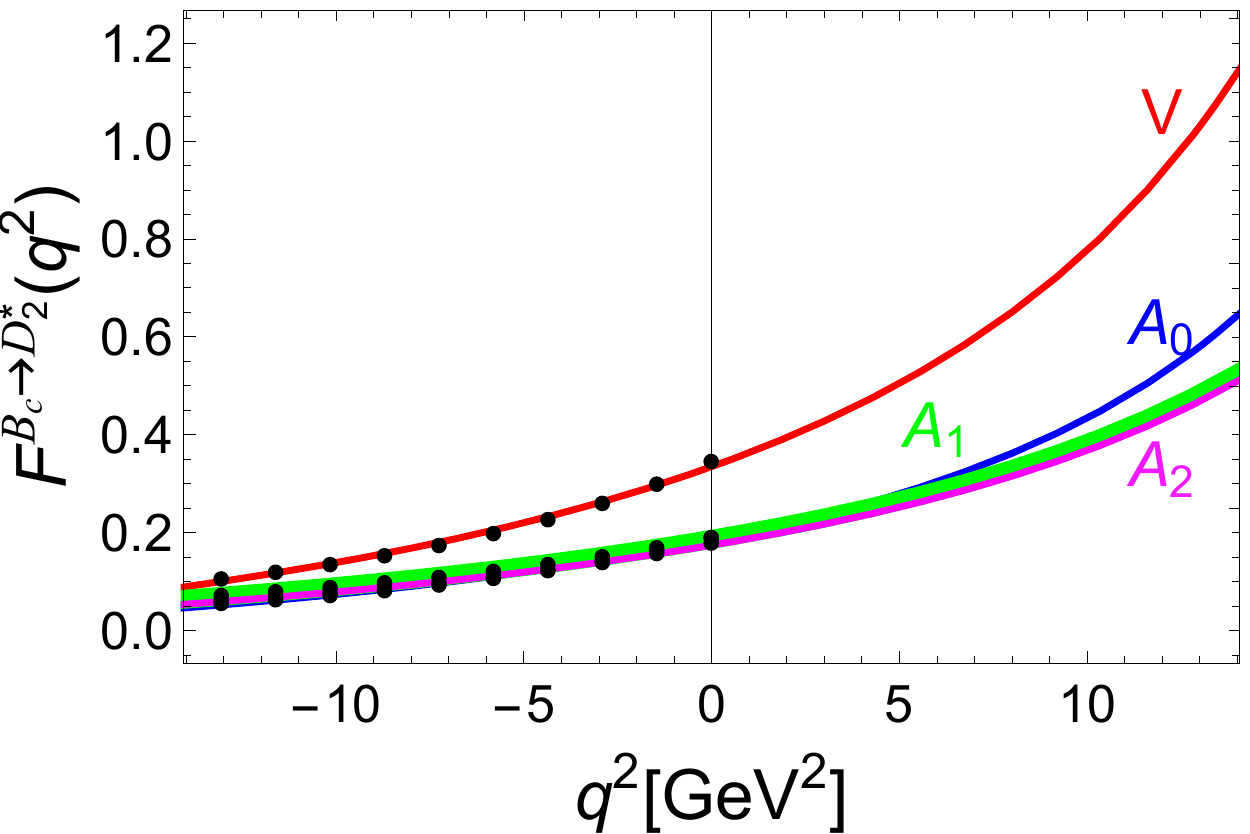}}\,
\subfigure{\includegraphics[scale=0.3]{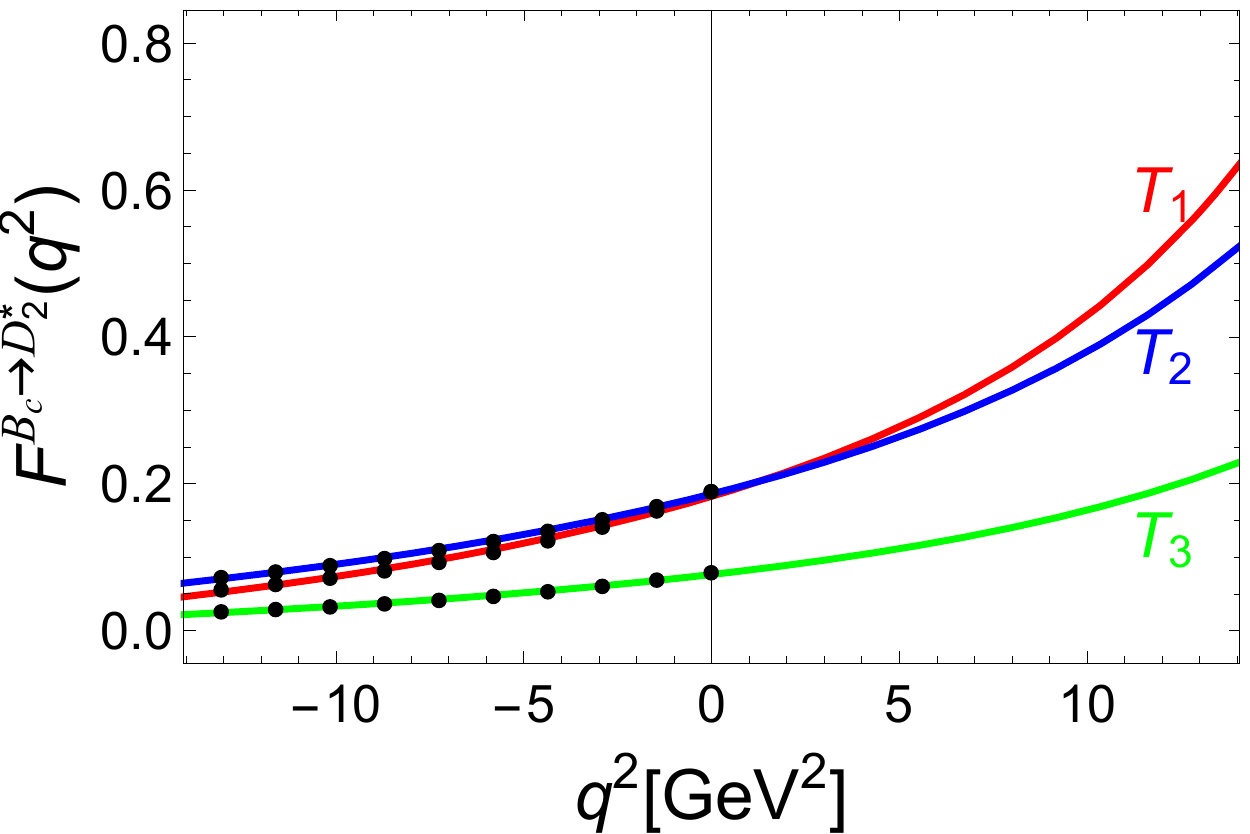}}\,
\subfigure{\includegraphics[scale=0.3]{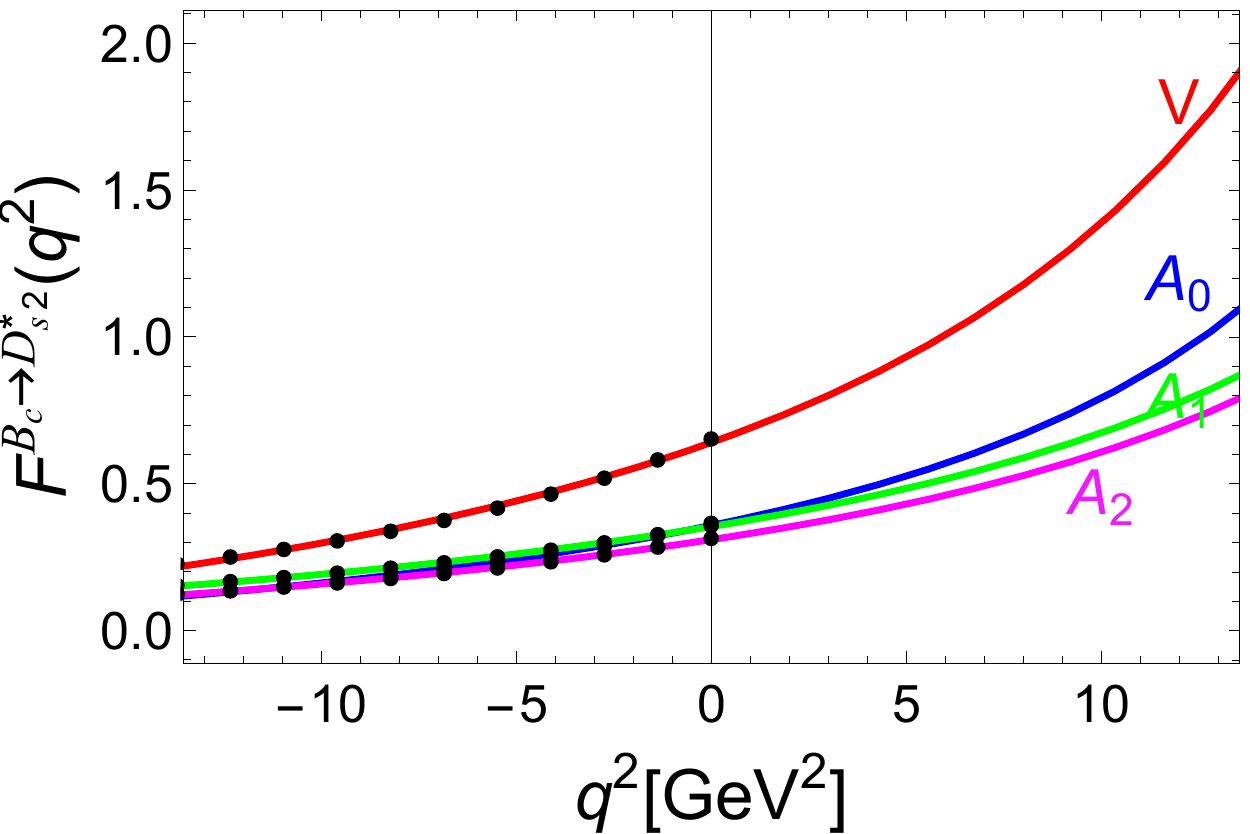}}\,
\subfigure{\includegraphics[scale=0.3]{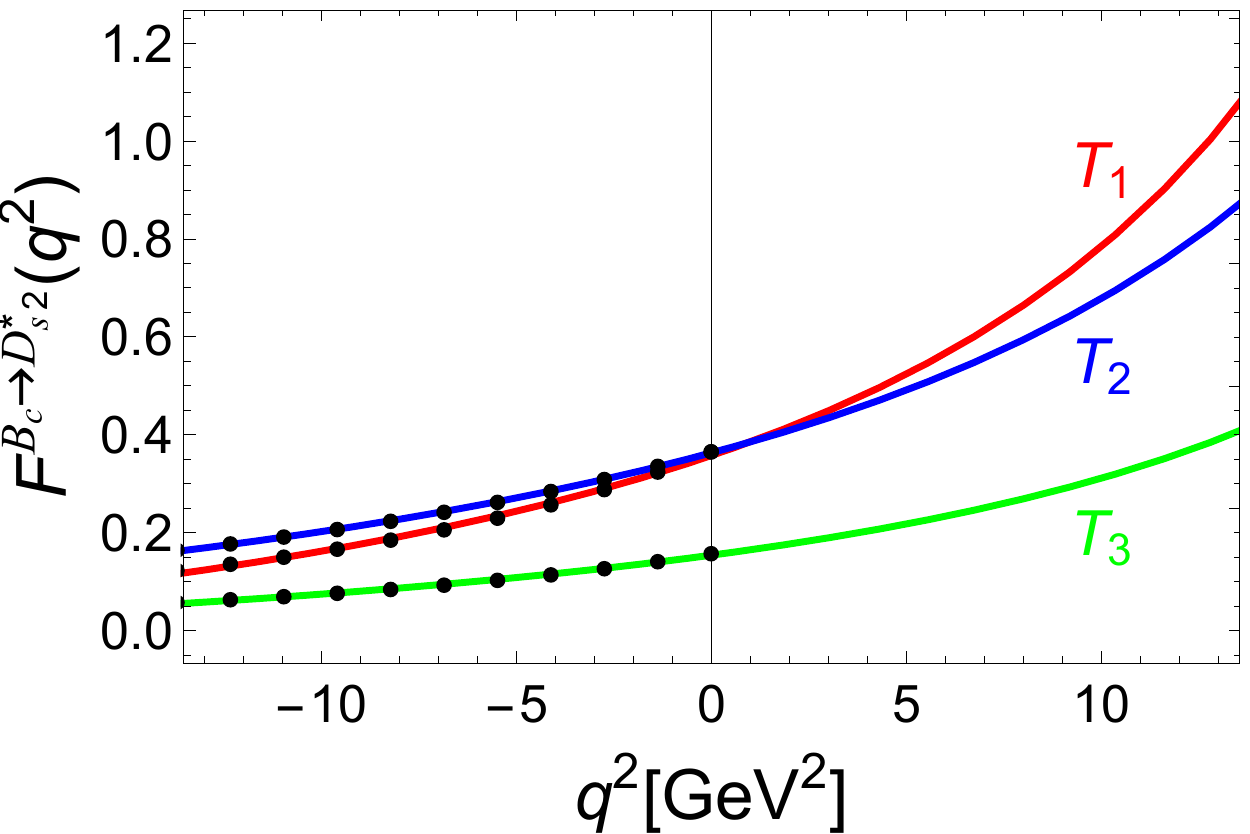}}\\
\subfigure{\includegraphics[scale=0.3]{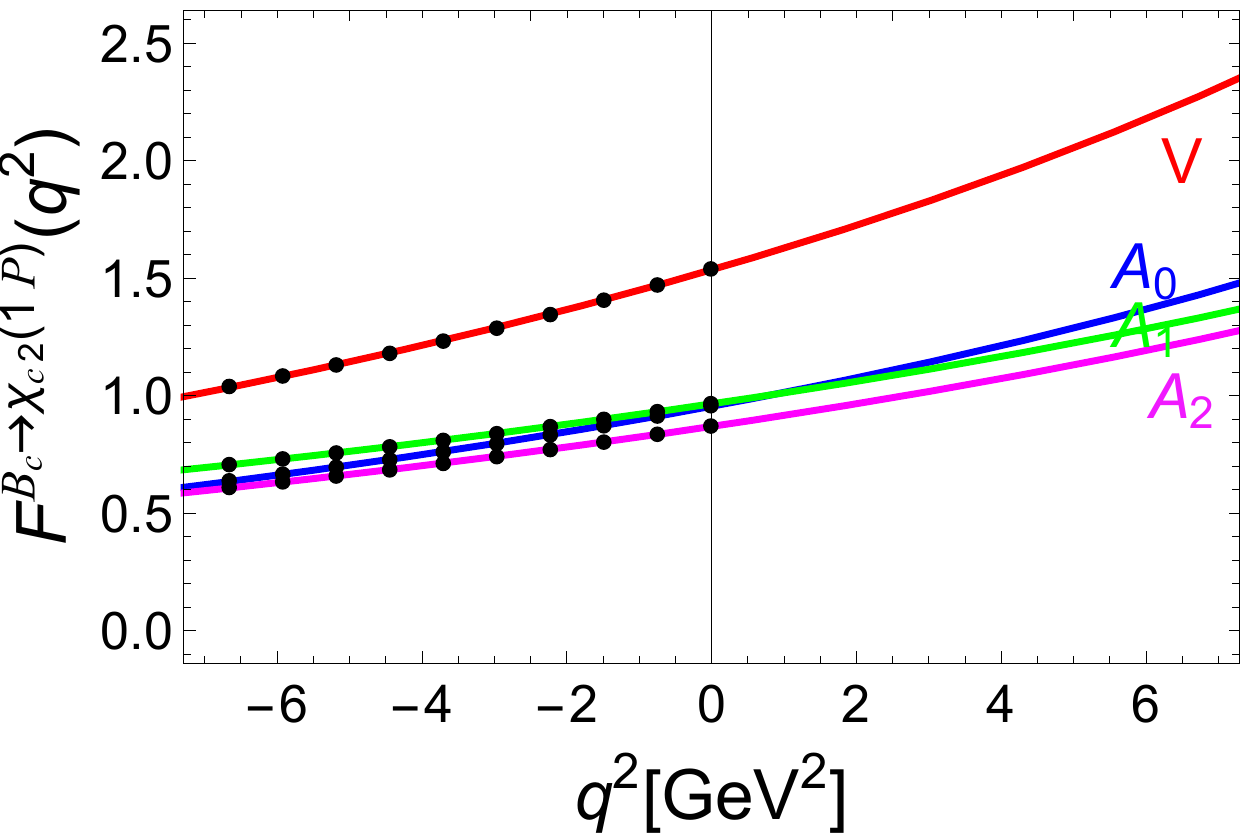}}\,
\subfigure{\includegraphics[scale=0.3]{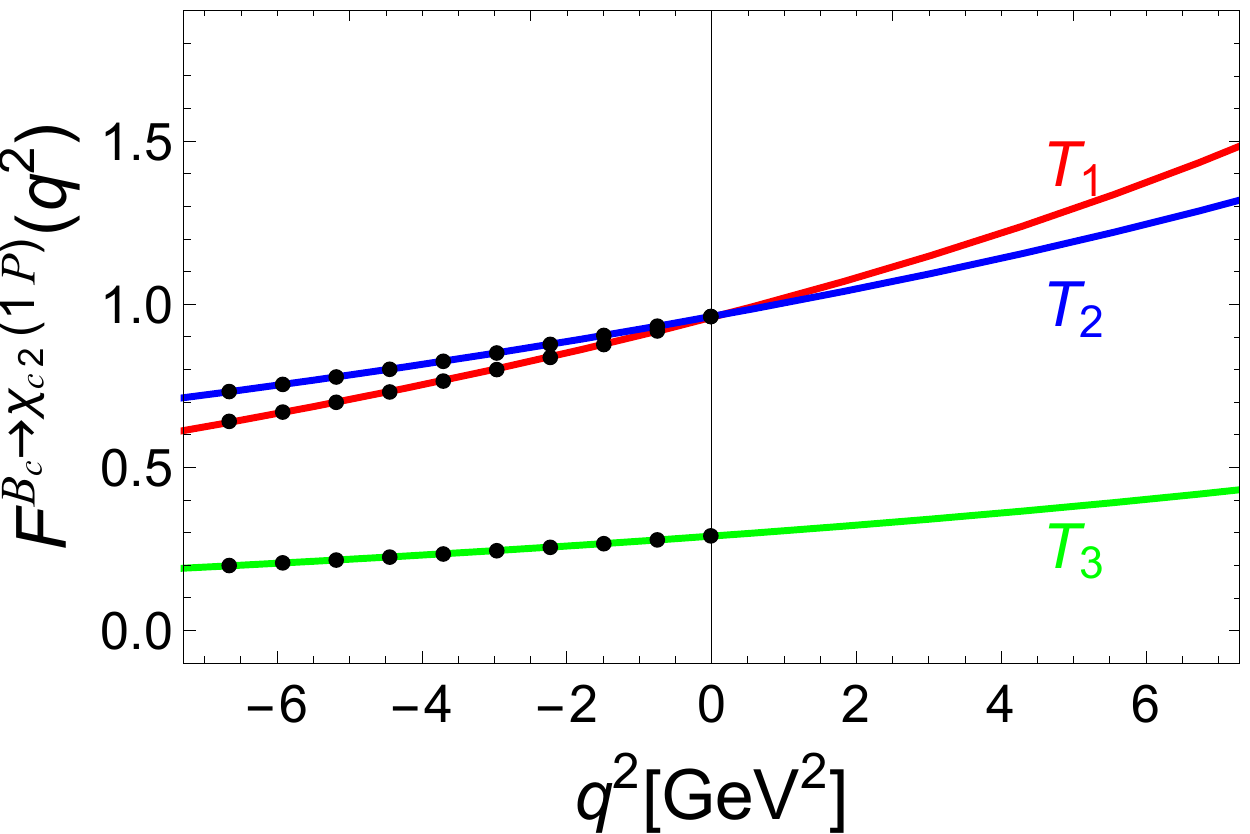}}\,
\subfigure{\includegraphics[scale=0.3]{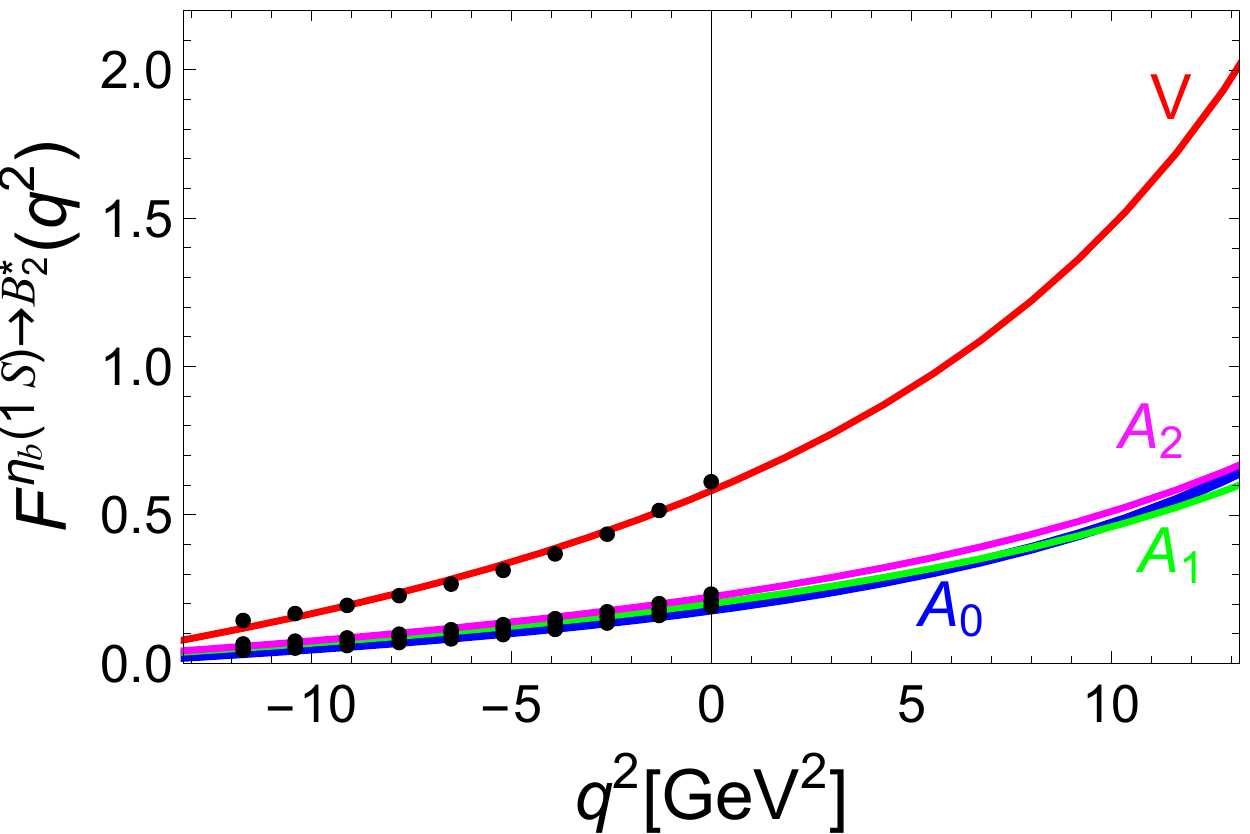}}\,
\subfigure{\includegraphics[scale=0.3]{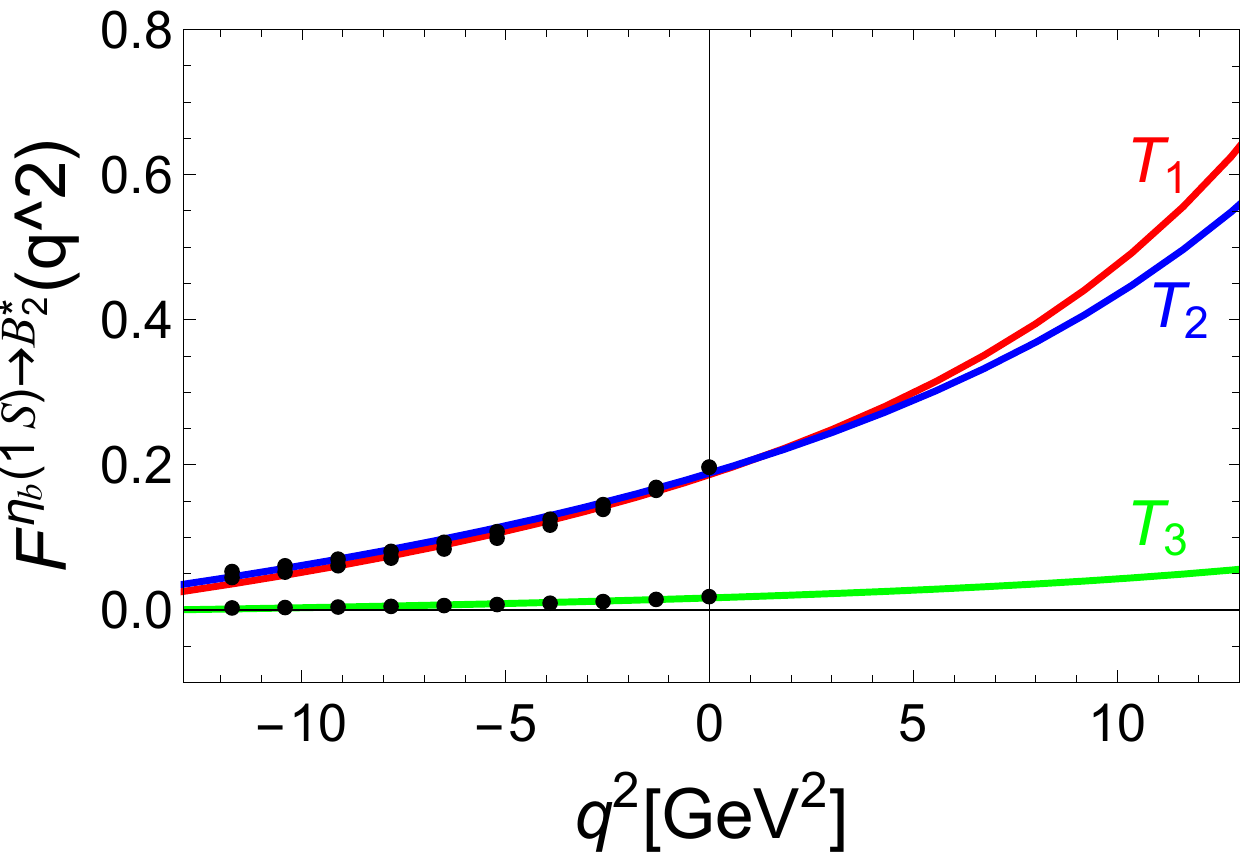}}\\
\subfigure{\includegraphics[scale=0.3]{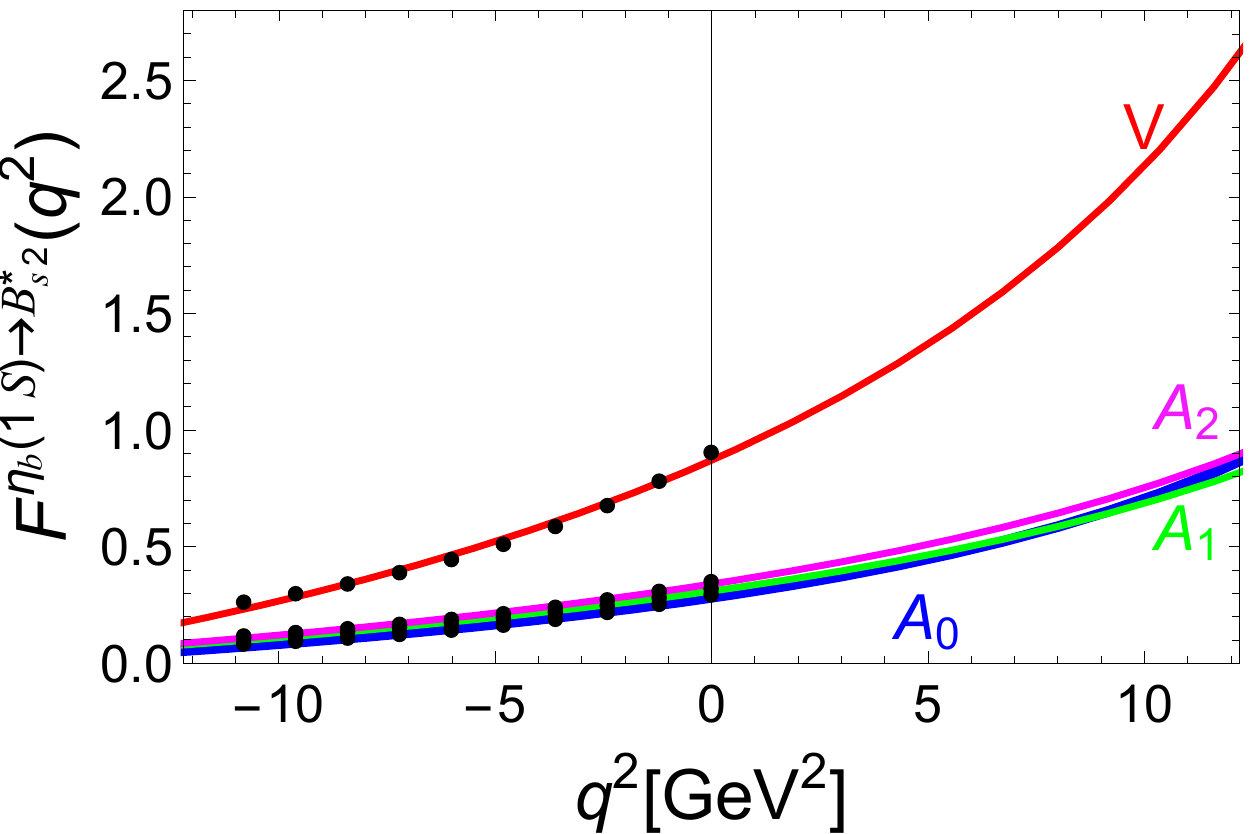}}\,
\subfigure{\includegraphics[scale=0.3]{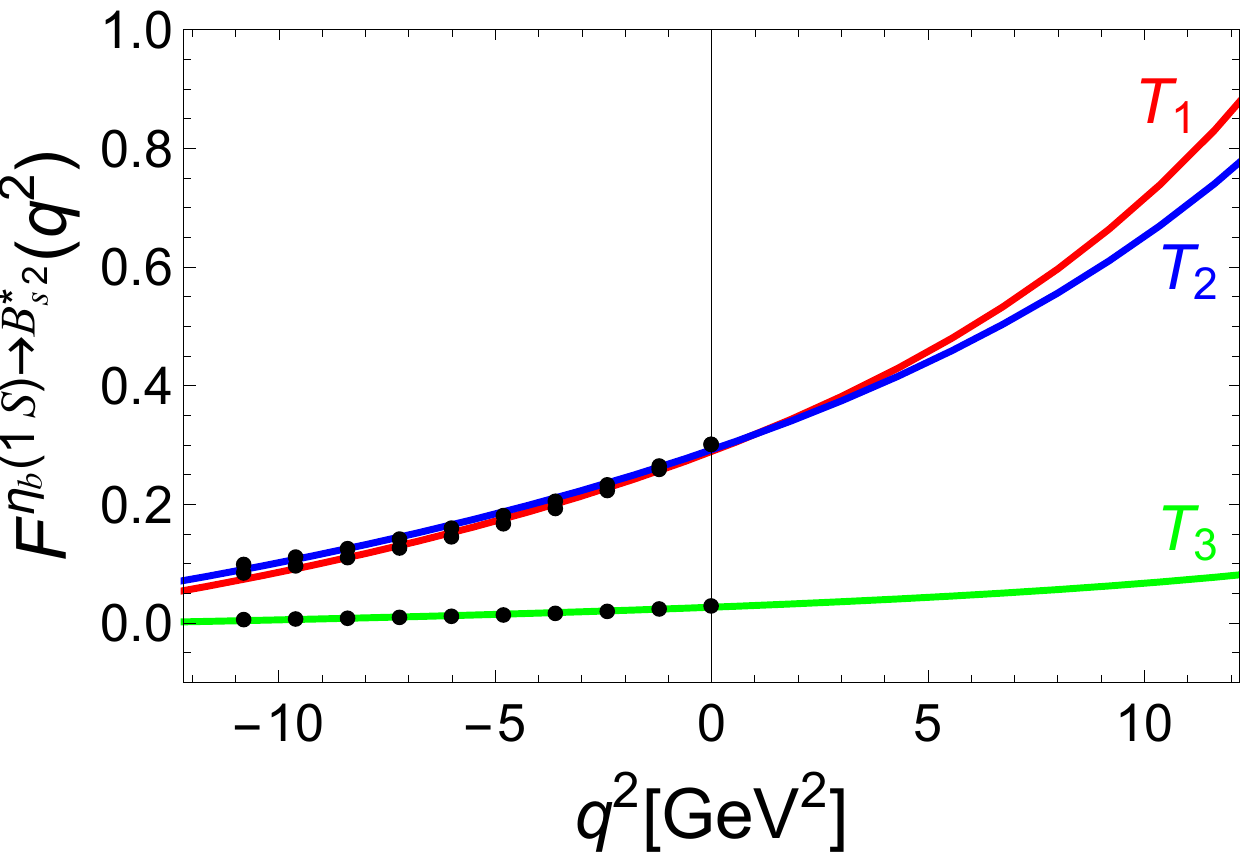}}\\
\end{center}
\label{fig:PTTc}
\end{figure}

\renewcommand{\baselinestretch}{1.2}
\begin{table}[t]
\scriptsize
\begin{center}
\caption{ \label{tab:pvat} Fitting results for form factors of  $c\to (q,s)$ induced $D \to (a_2,K^*_2)$, $D_s \to (K^*_2,f'_{2})$, $\eta_c(1S) \to (D^*_2,D^*_{s2})$, $ B_c \to (B^*_2,B^*_{s2})$  transitions with the parameterization scheme given by Eq.~\eqref{eq:para}.}
\vspace{0.2cm}
\let\oldarraystretch=\arraystretch
\renewcommand*{\arraystretch}{1.4}
\setlength{\tabcolsep}{5.pt}
\begin{tabular}{lcc|ccc|ccc}
\hline\hline
${\cal F}$  &$F(0)$    &$b_1$
&${\cal F}$ &$F(0)$    &$b_1$
&${\cal F}$ &$F(0)$    &$b_1$
\\\hline
$V^{D\to a_2}$       &$0.96_{-0.16}^{+0.18}$     &$-4.60_{-2.70}^{+3.60}$
&$V^{D\to K_2^*}$    &$1.00_{-0.16}^{+0.15}$     &$-4.60_{-2.32}^{+2.84}$
&$V^{D_s\to K_2^*}$  &$0.98_{-0.19}^{+0.22}$     &$-4.80_{-3.10}^{+4.00}$
\\
$A_0^{D\to a_2}$     &$0.62_{-0.07}^{+0.07}$     &$-3.70_{-2.90}^{+3.90}$
&$A_0^{D\to K_2^*}$  &$0.68_{-0.08}^{+0.06}$     &$-3.70_{-2.48}^{+3.10}$
&$A_0^{D_s\to K_2^*}$&$0.58_{-0.08}^{+0.08}$     &$-4.20_{-3.40}^{+4.50}$
\\
$A_1^{D\to a_2}$     &$0.63_{-0.10}^{+0.11}$     &$-2.00_{-1.80}^{+2.60}$
&$A_1^{D\to K_2^*}$  &$0.71_{-0.10}^{+0.09}$     &$-1.90_{-1.63}^{+1.98}$
&$A_1^{D_s\to K_2^*}$&$0.61_{-0.10}^{+0.12}$     &$-2.60_{-2.30}^{+3.00}$
\\
$A_2^{D\to a_2}$     &$0.45_{-0.08}^{+0.10}$     &$-3.70_{-1.90}^{+2.70}$
&$A_2^{D\to K_2^*}$  &$0.58_{-0.10}^{+0.11}$     &$-4.90_{-1.97}^{+2.58}$
&$A_2^{D_s\to K_2^*}$&$0.51_{-0.10}^{+0.14}$     &$-5.10_{-2.50}^{+3.40}$
\\
$T_1^{D\to a_2}$     &$0.61_{-0.08}^{+0.08}$     &$-4.06_{-2.90}^{+3.16}$
&$T_1^{D\to K_2^*}$  &$0.68_{-0.09}^{+0.07}$     &$-3.63_{-2.08}^{+2.27}$
&$T_1^{D_s\to K_2^*}$&$0.58_{-0.09}^{+0.09}$     &$-4.58_{-3.42}^{+3.40}$
\\
$T_2^{D\to a_2}$     &$0.61_{-0.08}^{+0.08}$     &$\quad4.78_{-1.38}^{+2.06}$
&$T_2^{D\to K_2^*}$  &$0.68_{-0.09}^{+0.08}$     &$\quad4.64_{-1.29}^{+1.66}$
&$T_2^{D_s\to K_2^*}$&$0.58_{-0.09}^{+0.09}$     &$\quad4.23_{-1.23}^{+1.87}$
\\
${T_3^{D\to a_2}}$   &$0.22_{-0.06}^{+0.03}$     &$-4.85_{-1.55}^{+1.81}$
&$T_3^{D\to K_2^*}$  &$0.17_{-0.10}^{+0.07}$     &$-2.85_{-1.58}^{+1.35}$
&$T_3^{D_s\to K_2^*}$&$0.15_{-0.08}^{+0.05}$     &$-4.21_{-2.42}^{+1.97}$
\\
\hline
$V^{D_s\to f'_2}$               &$1.19_{-0.20}^{+0.19}$     &$-4.40_{-2.50}^{+3.00}$
&$V^{\eta_c(1S)\to D_2^*}$      &$2.87_{-0.93}^{+1.47}$     &$-19.4_{-10.4}^{+10.8}$
&$V^{\eta_c(1S)\to D_{s2}^*}$   &$3.67_{-1.12}^{+1.53}$     &$-14.6_{-8.34}^{+9.59}$
\\
$A_0^{D_s\to f'_2}$             &$0.72_{-0.08}^{+0.07}$     &$-3.70_{-2.80}^{+3.42}$
&$A_0^{\eta_c(1S)\to D_2^*}$    &$0.96_{-0.20}^{+0.22}$     &$-18.5_{-10.8}^{+11.4}$
&$A_0^{\eta_c(1S)\to D_{s2}^*}$ &$1.28_{-0.24}^{+0.25}$     &$-13.3_{-8.85}^{+9.59}$
\\
$A_1^{D_s\to f'_2}$             &$0.77_{-0.11}^{+0.10}$     &$-2.10_{-1.91}^{+2.21}$
&$A_1^{\eta_c(1S)\to D_2^*}$    &$1.10_{-0.27}^{+0.34}$     &$-15.8_{-8.67}^{+9.23}$
&$A_1^{\eta_c(1S)\to D_{s2}^*}$ &$1.47_{-0.32}^{+0.35}$     &$-10.8_{-7.32}^{+7.86}$
\\
$A_2^{D_s\to f'_2}$             &$0.67_{-0.13}^{+0.15}$     &$-4.90_{-2.36}^{+2.99}$
&$A_2^{\eta_c(1S)\to D_2^*}$    &$0.98_{-0.30}^{+0.55}$     &$-14.2_{-10.1}^{+13.0}$
&$A_2^{\eta_c(1S)\to D_{s2}^*}$ &$1.13_{-0.35}^{+0.62}$     &$-8.39_{-10.9}^{+13.4}$
\\
$T_1^{D_s\to f'_2}$             &$0.73_{-0.10}^{+0.08}$     &$-3.64_{-2.31}^{+2.59}$
&$T_1^{\eta_c(1S)\to D_2^*}$    &$1.04_{-0.23}^{+0.27}$     &$-22.1_{-10.4}^{+10.9}$
&$T_1^{\eta_c(1S)\to D_{s2}^*}$ &$1.38_{-0.28}^{+0.30}$     &$-16.9_{-8.56}^{+9.07}$
\\
$T_2^{D_s\to f'_2}$             &$0.73_{-0.09}^{+0.08}$     &$\quad6.30_{-1.85}^{+1.89}$
&$T_2^{\eta_c(1S)\to D_2^*}$    &$1.04_{-0.23}^{+0.27}$     &$\quad28.4_{-13.8}^{+7.53}$
&$T_2^{\eta_c(1S)\to D_{s2}^*}$ &$1.38_{-0.28}^{+0.30}$     &$\quad88.1_{-21.7}^{+26.1}$

\\
$T_3^{D_s\to f'_2}$             &$0.15_{-0.12}^{+0.07}$     &$-2.27_{-1.83}^{+1.09}$
&$T_3^{\eta_c(1S)\to D_2^*}$    &$0.16_{-0.23}^{+0.18}$     &$-41.1_{-10.6}^{+8.96}$
&$T_3^{\eta_c(1S)\to D_{s2}^*}$ &$0.35_{-0.35}^{+0.28}$     &$-30.0_{-2.50}^{+6.30}$
\\
\hline
$V^{B_c\to B_2^*}$       &$13.4_{-4.11}^{+5.30}$     &$-87.6_{-23.5}^{+26.7}$
&$V^{B_c\to B_{s2}^*}$   &$14.8_{-4.12}^{+5.59}$     &$-74.7_{-14.1}^{+20.2}$
\\
$A_0^{B_c\to B_2^*}$     &$2.01_{-0.37}^{+0.33}$     &$-19.3_{-6.39}^{+3.46}$
&$A_0^{B_c\to B_{s2}^*}$ &$2.45_{-0.14}^{+0.11}$     &$-20.5_{-4.03}^{+5.34}$
\\
$A_1^{B_c\to B_2^*}$     &$2.42_{-0.55}^{+0.60}$     &$-35.4_{-3.74}^{+1.13}$
&$A_1^{B_c\to B_{s2}^*}$ &$2.94_{-0.30}^{+0.10}$     &$-35.4_{-3.63}^{+2.79}$
\\
$A_2^{B_c\to B_2^*}$     &$2.51_{-1.15}^{+2.42}$     &$-31.6_{-14.1}^{+12.0}$
&$A_2^{B_c\to B_{s2}^*}$ &$3.45_{-1.60}^{+3.15}$     &$-29.4_{-7.91}^{+16.5}$
\\
$T_1^{B_c\to B_2^*}$     &$2.24_{-0.44}^{+0.43}$     &$-77.1_{-11.7}^{+10.1}$
&$T_1^{B_c\to B_{s2}^*}$ &$2.69_{-0.49}^{+0.43}$     &$-74.5_{-3.43}^{+4.74}$
\\
$T_2^{B_c\to B_{2}^*}$   &$2.25_{-0.45}^{+0.45}$     &$\quad89.2_{-20.2}^{+22.9}$
&$T_2^{B_c\to B_{s2}^*}$ &$2.69_{-0.18}^{+0.10}$     &$\quad117_{-26.4}^{+23.5}$
\\
$T_3^{B_c\to B_2^*}$     &$-0.06_{-1.43}^{+0.90}\quad$ &$-60.1_{-24.8}^{+13.0}$
&$T_3^{B_c\to B_{s2}^*}$ &$-0.52_{-1.82}^{+1.29}$      &$-61.1_{-26.1}^{+21.3}$
\\
\hline\hline
\end{tabular}
\end{center}
\end{table}
\begin{table}[t]
\scriptsize
\begin{center}
\caption{\label{tab:dc}  Fitting results for form factors of $b\to (q,s,c)$ induced $B \to (a_2,K^*_2,D^*_2)$, $B_s \to (K^*_2,f'_2,D^*_{s2})$, $B_c \to (D^*_2,D^*_{s2},\chi_{c2}(1P))$, $\eta_b(1S) \to (B^*_2,B^*_{s2})$ transitions with the parameterization scheme given by Eq.~\eqref{eq:para}.}
\vspace{0.2cm}
\let\oldarraystretch=\arraystretch
\renewcommand*{\arraystretch}{1.4}
\setlength{\tabcolsep}{5.pt}
\begin{tabular}{lcc|ccc|ccc}
\hline\hline
${\cal F}$  &$F(0)$    &$b_1$
&${\cal F}$ &$F(0)$    &$b_1$
&${\cal F}$ &$F(0)$    &$b_1$
\\\hline
$V^{B\to a_2}$        &$0.24_{-0.04}^{+0.04}$   &$-5.30_{-0.98}^{+1.10}$
&$V^{B\to K_2^*}$     &$0.28_{-0.05}^{+0.05}$   &$-5.50_{-0.99}^{+1.10}$
&$V^{B\to D_2^*}$     &$0.75_{-0.16}^{+0.13}$   &$-6.20_{-0.50}^{+0.52}$
\\
$A_0^{B\to a_2}$      &$0.21_{-0.03}^{+0.03}$   &$-5.40_{-0.61}^{+0.74}$
&$A_0^{B\to K_2^*}$   &$0.24_{-0.04}^{+0.04}$   &$-5.60_{-0.59}^{+0.74}$
&$A_0^{B\to D_2^*}$   &$0.64_{-0.11}^{+0.11}$   &$-6.20_{-0.51}^{+0.53}$
\\
$A_1^{B\to a_2}$      &$0.19_{-0.03}^{+0.04}$   &$-2.30_{-0.37}^{+0.52}$
&$A_1^{B\to K_2^*}$   &$0.22_{-0.01}^{+0.07}$   &$-2.50_{-0.34}^{+0.51}$
&$A_1^{B\to D_2^*}$   &$0.63_{-0.12}^{+0.11}$   &$-3.00_{-0.48}^{+0.51}$
\\
$A_2^{B\to a_2}$      &$0.17_{-0.02}^{+0.03}$   &$-4.80_{-0.90}^{+0.83}$
&$A_2^{B\to K_2^*}$   &$0.20_{-0.03}^{+0.04}$   &$-5.20_{-0.92}^{+1.00}$
&$A_2^{B\to D_2^*}$   &$0.58_{-0.12}^{+0.12}$   &$-6.10_{-0.53}^{+0.52}$
\\
$T_1^{B\to a_2}$      &$0.19_{-0.03}^{+0.03}$   &$-5.40_{-0.61}^{+0.75}$
&$T_1^{B\to K_2^*}$   &$0.23_{-0.04}^{+0.04}$   &$-5.60_{-0.59}^{+0.75}$
&$T_1^{B\to D_2^*}$   &$0.63_{-0.12}^{+0.11}$   &$-6.30_{-0.51}^{+0.52}$
\\
$T_2^{B\to a_2}$      &$0.19_{-0.03}^{+0.03}$   &$-2.00_{-0.30}^{+0.42}$
&$T_2^{B\to K_2^*}$   &$0.23_{-0.04}^{+0.04}$   &$-2.20_{-0.26}^{+0.41}$
&$T_2^{B\to D_2^*}$   &$0.63_{-0.12}^{+0.11}$   &$-2.40_{-0.33}^{+0.20}$
\\
${T_3^{B\to a_2}}$    &$0.12_{-0.05}^{+0.01}$   &$-4.90_{-0.92}^{+1.35}$
&${T_3^{B\to K_2^*}}$ &$0.13_{-0.01}^{+0.03}$   &$-5.20_{-0.62}^{+3.10}$
&$T_3^{B\to D_2^*}$   &$0.24_{-0.06}^{+0.01}$   &$-5.91_{-0.83}^{+0.47}$
\\\hline
$V^{B_s\to K_2^*}$       &$0.23_{-0.05}^{+0.04}$   &$-6.50_{-0.28}^{+0.42}$
&$V^{B_s\to f'_2}$       &$0.32_{-0.06}^{+0.05}$   &$-6.30_{-0.37}^{+0.48}$
&$V^{B_s\to D_{s2}^*}$   &$0.95_{-0.17}^{+0.18}$   &$-6.50_{-0.66}^{+0.66}$
\\
$A_0^{B_s\to K_2^*}$     &$0.18_{-0.04}^{+0.04}$   &$-6.80_{-0.29}^{+0.42}$
&$A_0^{B_s\to f'_2}$     &$0.26_{-0.05}^{+0.05}$   &$-6.50_{-0.37}^{+0.47}$
&$A_0^{B_s\to D_{s2}^*}$ &$0.76_{-0.13}^{+0.13}$   &$-6.60_{-0.67}^{+0.67}$
\\
$A_1^{B_s\to K_2^*}$     &$0.17_{-0.04}^{+0.03}$   &$-4.00_{-0.04}^{+0.08}$
&$A_1^{B_s\to f'_2}$     &$0.24_{-0.05}^{+0.05}$   &$-3.70_{-0.10}^{+0.19}$
&$A_1^{B_s\to D_{s2}^*}$ &$0.74_{-0.14}^{+0.15}$   &$-3.40_{-0.67}^{+0.67}$
\\
$A_2^{B_s\to K_2^*}$     &$0.16_{-0.03}^{+0.03}$   &$-6.10_{-0.18}^{+0.31}$
&$A_2^{B_s\to f'_2}$     &$0.22_{-0.04}^{+0.04}$   &$-6.00_{-0.31}^{+0.40}$
&$A_2^{B_s\to D_{s2}^*}$ &$0.66_{-0.13}^{+0.14}$   &$-6.00_{-0.65}^{+0.64}$
\\
$T_1^{B_s\to K_2^*}$     &$0.17_{-0.03}^{+0.04}$   &$-6.80_{-0.28}^{+0.42}$
&$T_1^{B_s\to f'_2}$     &$0.25_{-0.04}^{+0.05}$   &$-6.50_{-0.37}^{+0.47}$
&$T_1^{B_s\to D^*_{s2}}$ &$0.75_{-0.14}^{+0.14}$   &$-6.70_{-0.67}^{+0.67}$
\\
$T_2^{B_s\to K_2^*}$     &$0.17_{-0.03}^{+0.04}$   &$-3.70_{-0.12}^{+0.02}$
&$T_2^{B_s\to f'_2}$     &$0.25_{-0.04}^{+0.05}$   &$-3.40_{-0.03}^{+0.09}$
&$T_2^{B_s\to D^*_{s2}}$ &$0.75_{-0.14}^{+0.14}$   &$-2.30_{-0.38}^{+0.12}$
\\
${T_3^{B_s\to K_2^*}}$   &$0.10_{-0.02}^{+0.02}$   &$-6.30_{-0.18}^{+0.30}$
&$T_3^{B_s\to f'_2}$     &$0.14_{-0.05}^{+0.01}$   &$-6.10_{-0.61}^{+0.57}$
&$T_3^{B_s\to D^*_{s2}}$ &$0.29_{-0.16}^{+0.32}$   &$-6.00_{-5.80}^{+2.35}$
\\\hline
${V^{B_c\to D_2^*}}$         &${0.35_{-0.10}^{+0.09}}$   &${-14.0_{-0.99}^{+1.10}}$
&${V^{B_c\to D_{s2}^*}}$     &${0.65_{-0.13}^{+0.19}}$   &${-13.0_{-2.60}^{+2.40}}$
&$V^{B_c\to \chi_{c2}(1P)}$  &$1.54_{-0.38}^{+0.51}$     &$-13.0_{-2.80}^{+2.90}$
\\
${A_0^{B_c\to D_2^*}}$       &${0.19_{-0.06}^{+0.06}}$   &${-15.0_{-0.91}^{+0.97}}$
&${A_0^{B_c\to D_{s2}^*}}$   &${0.37_{-0.09}^{+0.08}}$   &${-13.0_{-0.86}^{+0.83}}$
&$A_0^{B_c\to \chi_{c2}(1P)}$&$0.96_{-0.20}^{+0.25}$     &$-13.0_{-2.90}^{+3.00}$
\\
${A_1^{B_c\to D_2^*}}$       &${0.19_{-0.06}^{+0.06}}$   &${-12.0_{-1.70}^{+1.90}}$
&${A_1^{B_c\to D_{s2}^*}}$   &${0.36_{-0.09}^{+0.08}}$   &${-10.0_{-1.40}^{+1.40}}$
&$A_1^{B_c\to \chi_{c2}(1P)}$&$0.97_{-0.22}^{+0.28}$     &$-9.80_{-3.10}^{+3.20}$
\\
${A_2^{B_c\to D_2^*}}$       &${0.18_{-0.06}^{+0.06}}$   &${-13.0_{-1.30}^{+1.40}}$
&${A_2^{B_c\to D_{s2}^*}}$   &${0.31_{-0.08}^{+0.08}}$   &${-11.0_{-1.20}^{+1.20}}$
&$A_2^{B_c\to \chi_{c2}(1P)}$&$0.87_{-0.22}^{+0.32}$     &$-12.0_{-3.00}^{+3.00}$
\\
${T_1^{B_c\to D_2^*}}$       &${0.19_{-0.06}^{+0.06}}$   &${-15.0_{-0.91}^{+0.98}}$
&${T_1^{B_c\to D_{s2}^*}}$   &${0.37_{-0.08}^{+0.08}}$   &${-13.0_{-0.86}^{+0.84}}$
&$T_1^{B_c\to \chi_{2c}(1P)}$     &$0.96_{-0.21}^{+0.26}$     &$-13.0_{-2.90}^{+3.00}$
\\
${T_2^{B_c\to D_2^*}}$       &${0.19_{-0.06}^{+0.06}}$   &${-12.8_{-1.50}^{+1.90}}$
&${T_2^{B_c\to D_{s2}^*}}$   &${0.37_{-0.08}^{+0.08}}$   &${-9.30_{-1.60}^{+1.90}}$
&$T_2^{B_c\to \chi_{2c}(1P)}$     &$0.96_{-0.20}^{+0.26}$     &$-7.70_{-2.80}^{+2.50}$
\\
${T_3^{B_c\to D_2^*}}$       &${0.08_{-0.01}^{+0.03}}$   &${-15.5_{-4.75}^{+6.00}}$
&${T_3^{B_c\to D_{s2}^*}}$   &${0.15_{-0.04}^{+0.03}}$   &${-13.0_{-1.00}^{+1.00}}$
&$T_3^{B_c\to \chi_{2c}(1P)}$     &$0.29_{-0.07}^{+0.03}$     &$-13.0_{-3.10}^{+1.70}$
\\ \hline
${V^{\eta_b(1S)\to B_2^*}}$       &${0.61_{-0.19}^{+0.22}}$   &${-57.4_{-3.43}^{+3.37}}$
&${V^{\eta_b(1S)\to B_{s2}^*}}$   &${0.90_{-0.24}^{+0.27}}$   &${-56.0_{-4.05}^{+4.06}}$
\\
${A_0^{\eta_b(1S)\to B_2^*}}$     &${0.19_{-0.06}^{+0.07}}$   &${-58.5_{-1.27}^{+1.18}}$
&${A_0^{\eta_b(1S)\to B_{s2}^*}}$ &${0.29_{-0.08}^{+0.09}}$   &${-57.2_{-1.22}^{+1.27}}$
\\
${A_1^{\eta_b(1S)\to B_2^*}}$     &${0.21_{-0.07}^{+0.08}}$   &${-53.0_{-1.80}^{+1.81}}$
&${A_1^{\eta_b(1S)\to B_{s2}^*}}$ &${0.32_{-0.09}^{+0.11}}$   &${-51.5_{-1.22}^{+1.26}}$
\\
${A_2^{\eta_b(1S)\to B_2^*}}$     &${0.23_{-0.08}^{+0.10}}$   &${-52.8_{-0.84}^{+0.73}}$
&${A_2^{\eta_b(1S)\to B_{s2}^*}}$ &${0.35_{-0.11}^{+0.14}}$   &${-51.4_{-0.69}^{+0.56}}$
\\
${T_1^{\eta_b(1S)\to B_2^*}}$     &${0.20_{-0.07}^{+0.06}}$   &${-58.7_{-1.06}^{+0.95}}$
&${T_1^{\eta_b(1S)\to B_{s2}^*}}$ &${0.30_{-0.08}^{+0.10}}$   &${-57.4_{-0.21}^{+0.26}}$
\\
${T_2^{\eta_b(1S)\to B_2^*}}$     &${0.20_{-0.07}^{+0.06}}$   &${-54.1_{-1.75}^{+1.78}}$
&${T_2^{\eta_b(1S)\to B_{s2}^*}}$ &${0.30_{-0.08}^{+0.10}}$   &${-52.2_{-0.44}^{+0.52}}$
\\
${T_3^{\eta_b(1S)\to B_2^*}}$     &${0.02_{-0.02}^{+0.01}}$   &${-73.6_{-17.8}^{+25.9}}$
&${T_3^{\eta_b(1S)\to B_{s2}^*}}$ &${0.03_{-0.03}^{+0.02}}$   &${-71.4_{-11.9}^{+22.9}}$
\\
\hline\hline
\label{tab:ptt}
\end{tabular}
\end{center}
\end{table}



From above discussions, one can conclude that the CLF QM with type-II corresponding scheme can make sure the Lorentz covariance of matrix elements and give self-consistent results for the form factors. Using the values of input parameters collected in appendix A and employing the self-consistent type-II scheme, we then present our  numerical predictions for the  form factors of  $c\to (q,s)$~($q=u,d$) induced $D\to (a_2,K^*_2)$, $D_s\to (K^*_2, f'_2)$, $\eta_c(1S)\to (D^*_2,D^*_{s2})$, $B_c\to (B^*_2,B^*_{s2} )$ transitions and  $b\to (q,s,c)$  induced $B\to (a_2,K^*_2, D^*_2), B_s\to (K^*_2, f'_2, D^*_{s2}), B_c\to (D^*_2, D^*_{s2}, X_{c2}(1P))$, $\eta_b(1S)\to (B^*_2,B^*_{s2})$  transitions.

It should be noted that the CLF calculation is made in the $q^+=0$ frame, which implies that the form factors are known only for space-like momentum transfer, $q^2=-\mathbf{q}_\bot^2\leqslant 0$, and the results in the time-like region need  an additional $q^2$ extrapolation. For the phenomenological applications, we adopt the BCL version of the $z$-series expansion~\cite{Bourrely:2008za} in the form adopted in Refs.~\cite{Khodjamirian:2017fxg,Gao:2019lta},
\begin{align}\label{eq:para}
{\cal F}(q^2)=\frac{{\cal F}(0)}{1-q^2/m^{2}_{i,pole}}\left\{1+\sum_{k=1}^N b_k \left[ z(q^2, t_0)^k- z(0, t_0)^k\right]   \right\}\,,
\end{align}
where,  $z(q^2, t_0)=\frac{\sqrt{t_+-q^2}-\sqrt{t_+-t_0}}{\sqrt{t_+-q^2}+\sqrt{t_+-t_0}}$, $t_+=(M'+ M'')^2$, $t_0=(M'+ M'')(\sqrt{M'}-\sqrt{M''})^2$. For the masses of resonances collected in Table~\ref{tab:rm}, we take the values given by  PDG~\cite{Zyla:2020zbs} and lattice QCD~\cite{Detmold:2015aaa,Dowdall:2012ab}.  In the practice, we will truncate the expansion at $N=1$.  The parameter $b_k$ will be obtained by fitting to the results computed  directly by CLF QMs.

\begin{table}[t]
\begin{center}
\caption{\label{tab:Z}{\tb{ Numerical results of $Z_k(q^2)$ at $q^2=(3,5,7)\,{\rm GeV}^2$ for $B\to D^*_2$ transition, and at  $q^2=(6,10,14)\,{\rm GeV}^2$ for $B\to K^*_2$ transition.}}}
\vspace{0.2cm}
\let\oldarraystretch=\arraystretch
\renewcommand*{\arraystretch}{1}
\setlength{\tabcolsep}{4pt}
\begin{tabular}{ccccccccccccc}
\hline\hline
$B\to D^*_2$ &${q}^2=3\,{\rm GeV}^2$ &$q^2=5\,{\rm GeV}^2$ &$q^2=7\,{\rm GeV}^2$\\\hline
$Z_1(q^2)$
&$-1.28\times10^{-2}\quad$&$-2.18\times10^{-2}\quad$&$-3.11\times10^{-2}\quad$\\
$Z_2(q^2)$
&$-2.92\times10^{-4}\quad$&$-3.00\times10^{-4}\quad$&$-1.40\times10^{-4}\quad$\\
\hline\hline
 $B\to K^*_2$ &${q}^2=6\,{\rm GeV}^2$&$q^2=10\,{\rm GeV}^2$ &$q^2=14\,{\rm GeV}^2$\\\hline
$Z_1(q^2)$
 &$-3.57\times10^{-2}\quad$&$-6.27\times10^{-2}\quad$&$-9.30\times10^{-2}\quad$\\
$Z_2(q^2)$
&$-2.28\times10^{-3}\quad$&$-2.31\times10^{-3}\quad$&$-6.05\times10^{-4}\quad$
\\\hline\hline
\end{tabular}
\end{center}
\end{table}

\begin{table}[h]
\begin{center}
\caption{\label{tab:n12}\tb{The values of $b_1$ and $b_2$ for the form factors of  $B\to D^*_2$  and  $B\to K^*_2$ transitions within the truncation-scheme $N=2$. The values given in the parenthesis are the results obtained within the truncation-scheme $N=1$.  }}
\footnotesize 
\vspace{0.2cm}
\let\oldarraystretch=\arraystretch
\renewcommand*{\arraystretch}{1}
\setlength{\tabcolsep}{1.2pt}
\begin{tabular}{ccccccccccccc}
\hline\hline
 $B\to D^*_2$  &$V$&$A_0$&$A_1$&$A_2$&$T_1$&$T_2$&$T_3$\\\hline
$b_1$&$-6.59(-6.20)$&$-6.64(-6.20)$&$-3.19(-3.00)$&$-6.44(-6.10)$&$-6.61(-6.30)$&$-2.47(-2.40)$&$-5.85(-5.91)$
\\
$b_2$&$5.94$&$6.51$&$2.44$&$5.93$&$4.87$&$1.17$&$0.50$
\\\hline\hline
$B\to K^*_2$  &$V$&$A_0$&$A_1$&$A_2$ &$T_1$&$T_2$&$T_3$\\\hline
$b_1$&$-6.49(-5.50)$&$-6.62(-5.60)$&$-2.69(-2.50)$&$-6.20(-5.20)$&$-6.64(-5.60)$&$-2.30(-2.20)$&$-5.78(-5.20)$
   \\
$b_2$&$6.55$&$6.55$&$1.02$&$6.30$&$6.55$&$0.31$&$0.13$
\\\hline\hline
\end{tabular}
\end{center}
\end{table}

Using the parameterization scheme given by Eq.~\eqref{eq:para}, we present our numerical results of ${\cal F}(0)$ and $b_1$ for the $c\to (q,s)$~($q=u,d$) induced $D\to (a_2,K^*_2)$, $D_s\to (K^*_2, f'_2)$, $\eta_c(1S)\to (D^*_2,D^*_{s2})$, $B_c\to (B^*_2,B^*_{s2} )$ transitions and  the $b\to (q,s,c)$  induced $B\to (a_2,K^*_2, D^*_2), B_s\to (K^*_2, f'_2, D^*_{s2}), B_c\to (D^*_2, D^*_{s2}, X_{c2}(1P))$, $\eta_b(1S)\to (B^*_2,B^*_{s2})$  transitions in Tables~\ref{tab:pvat} and \ref{tab:ptt}, respectively. The $q^2$ dependence of form factors are shown in Figs.~\ref{fig:PTVAc} and \ref{fig:PTTc}. 
Some remarks on these results are given in order.
 \begin{itemize}
 \item \tb{Firstly, we would like to test the legality  of the  truncation-scheme $N=1$  employed  in this paper. In the expansion, Eq.~\eqref{eq:para}, the values of  $Z_k(q^2)\equiv z(q^2,t_0)^k-z(0,t_0)^k$ in general  satisfy $Z_{k+1}/Z_k \sim {\cal O}(10^{-1\sim -2})$, which can be found  from the values listed in Table~\ref{tab:Z}~(for convenience of discussion, we mainly study the effect of $k=2$ term, and take $B\to D^*_2$ and  $B\to K^*_2$  transitions as examples). Therefore, the $k=2$ term can be neglected except for $b_{2}\gg b_{1}$.    
From Table~\ref{tab:n12}, it can be found that the values of $b_{1}$ and $b_{2}$ are at the same level. Thus,  the truncation  $N=1$  employed  in this paper is acceptable. It can  be  also clearly seen from Fig.~\ref{fig:n12} that the effect of truncation-scheme $N=2$  on the $q^2$-dependences of form factors are not significant compared with truncation-scheme $N=1$. Such finding can be easily  understood because the CLF result for the form factors can be well reproduced within the truncation-scheme  $N=1$, and thus the higher-order terms are trivial. Some discussions on the effects of higher-order terms have been made in, for instance, Refs.~\cite{Gao:2019lta,Bharucha:2010im}.  }

\begin{figure}[h]
\begin{center}
\caption{\label{fig:n12}\tb{The $q^2$-dependence of form factors of $B\to D^*_2$ and $B\to K^*_2$ transitions within the truncation-schemes $N=1$ and $N=2$.}}
\vspace{0.2cm}
\subfigure{\includegraphics[scale=0.35]{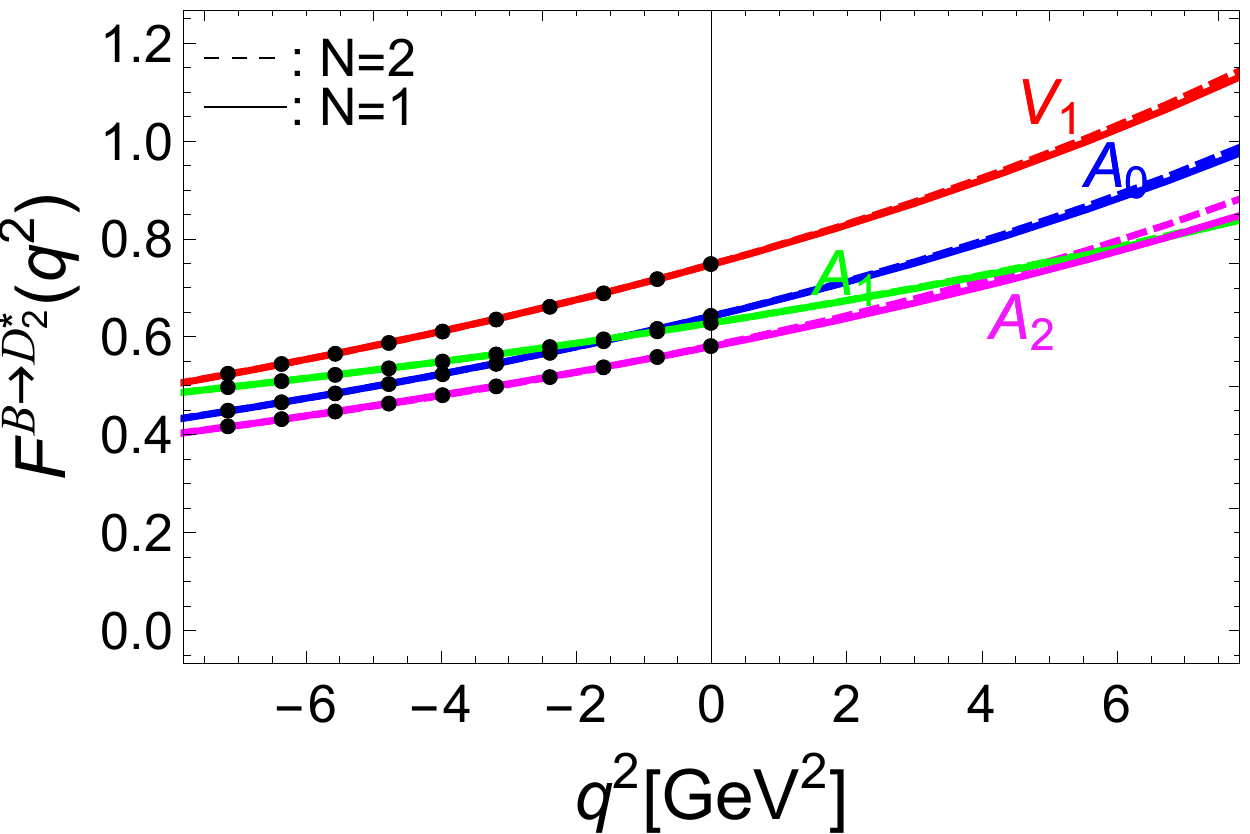}}\qquad
\subfigure{\includegraphics[scale=0.35]{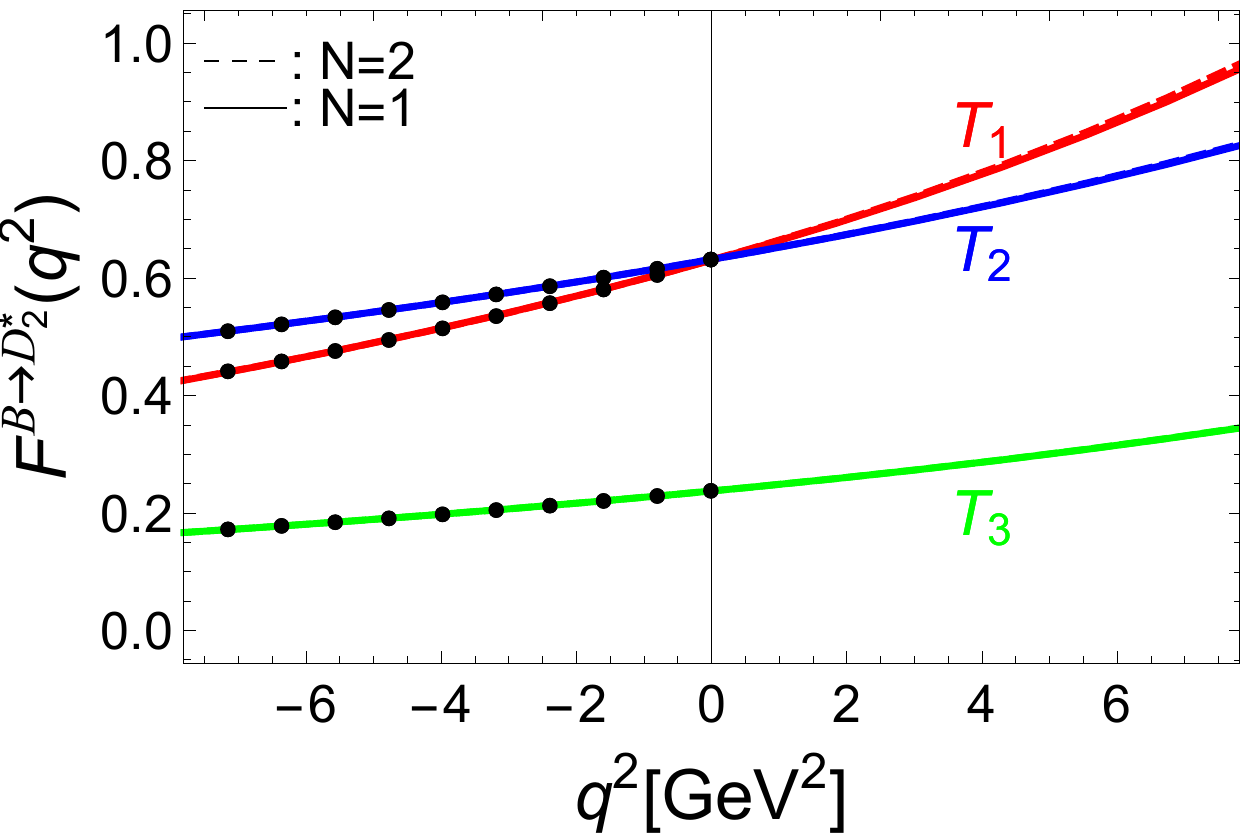}}\\
\subfigure{\includegraphics[scale=0.35]{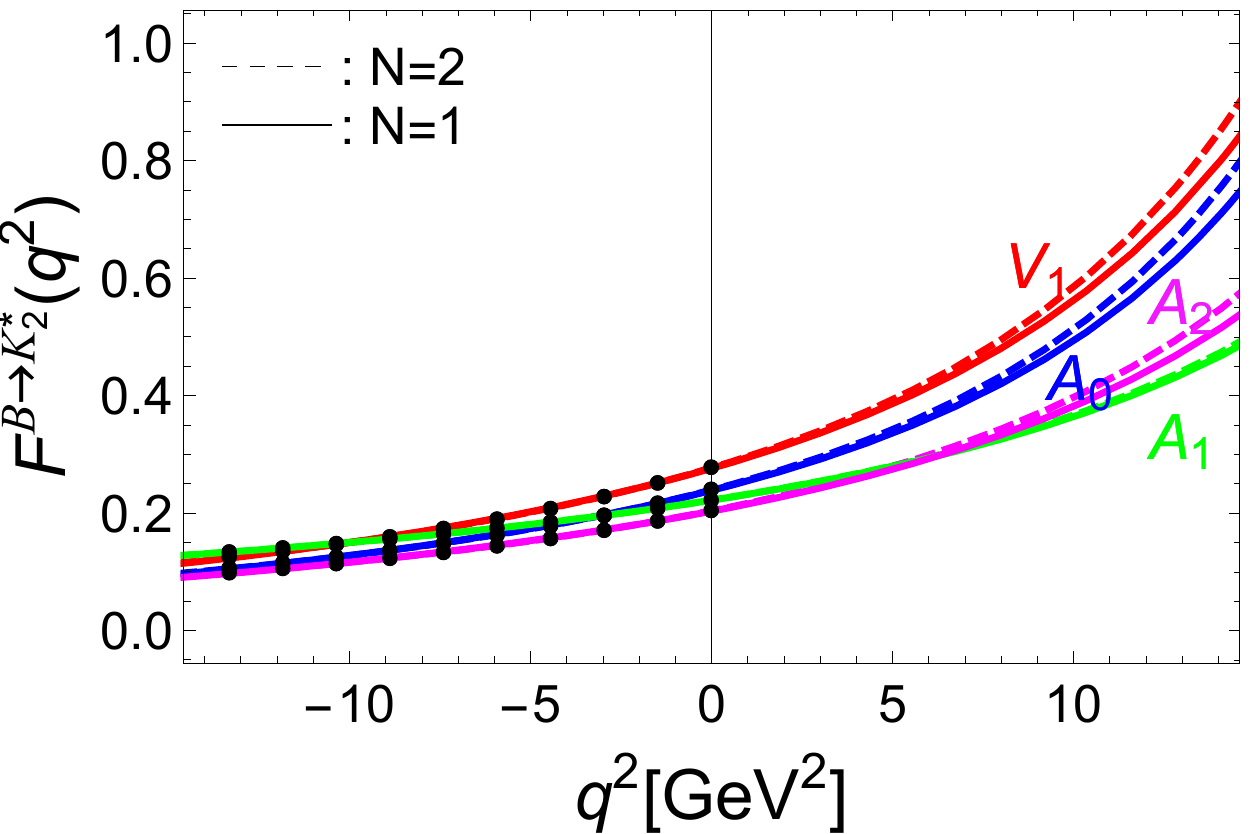}}\qquad
\subfigure{\includegraphics[scale=0.35]{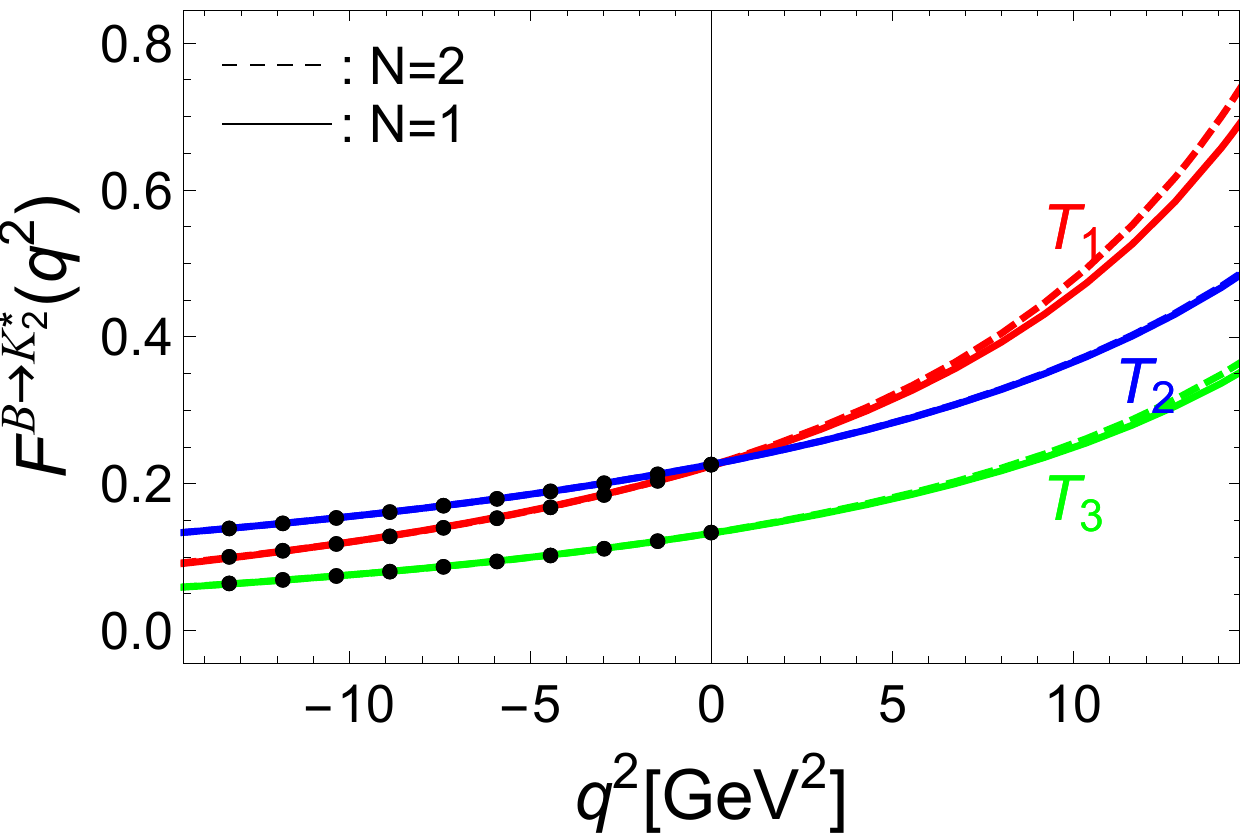}}\\
\end{center}
\end{figure}

 \item
Just like the $\eta-\eta'$ mixing in the pseudoscalar case,  the physical isoscalar tensor states  $f_2(1270)$ and $f'_2(1525)$  also have a mixing. In order to exhibit their flavor components, the mixing relation can be written as
\begin{align}
f_2&\equiv\frac{1}{\sqrt 2}(u\bar u+d\bar d)\,{\rm cos}\,\theta+s\bar s {\,\rm sin} \,\theta\,,\\
f'_2&\equiv\frac{1}{\sqrt 2}(u\bar u+d\bar d)\,{\rm sin}\,\theta-s\bar s {\,\rm cos} \,\theta\,,
\end{align}
where  $\theta$ is the mixing angle. It is obvious that the mixing angle should be small because  $f_2(1270)$ and $f'_2(1525)$ decay  predominantly into $\pi\pi$ and $K\bar{K}$, respectively. Numerically, it is found that $\theta=9^\circ\pm1^\circ$~\cite{Zyla:2020zbs}. Therefore, in our calculation, the  possible mixing effect is neglected, {\it i.e.}, $f_2(1270)$  and $f'_2(1525)$  are assumed to be pure $(u\bar u+d\bar d)$ and $(s\bar{s})$ states, respectively.

\item
From Table~\ref{tab:ptt} and Fig.~\ref{fig:PTTc}, it can be clearly found that all of  transitions respect the relation
\begin{align}
 T_1(0)=T_2(0)\,,
\end{align}
which is essential to assure that the hadronic matrix element of $P\to T$ is divergence free at $q^2 = 0$. However, their dependence on $q^2$ is different, which can be applied further in the relevant phenomenological studies of meson decays.


\renewcommand{\baselinestretch}{1.2}
\begin{table}[!t]
\footnotesize
\begin{center}
\caption{\label{tab:comp} \small  Theoretical predictions for the form factors of $B\to a_2$ and $B\to K^*_2$ transitions at $q^2=0$ in this work, LCSR~\cite{Aliev:2019ojc}, PQCD~\cite{Wang:2010ni} and traditional CLF QM~\cite{Cheng:2010yd,Cheng:2009ms}. }
\vspace{0.2cm}
\let\oldarraystretch=\arraystretch
\renewcommand*{\arraystretch}{1.3}
\setlength{\tabcolsep}{4pt}
\begin{tabular}{l|cccc|cccccccc}
\hline\hline
    &\multicolumn{4}{c|}{$B\to a_2$}  &\multicolumn{4}{c}{$B\to K^*_2$}\\\hline
      &this work  &LCSR~\cite{Aliev:2019ojc}
      &PQCD~\cite{Wang:2010ni}
      &CLF~\cite{Cheng:2010yd}
      &this work  &LCSR~\cite{Aliev:2019ojc}
      &PQCD~\cite{Wang:2010ni}
      &CLF~\cite{Cheng:2010yd,Cheng:2009ms} \\
 \hline
$V(0)$&$0.24^{+0.04}_{-0.04}$&$0.18^{+0.12}_{-0.07}$&$0.18^{+0.05}_{-0.04}$&$0.28$
            &$0.28^{+0.05}_{-0.05}$&$0.22^{+0.11}_{-0.08}$&$0.21^{+0.06}_{-0.05}$&$0.29$\\
$A_0(0)$&$0.21^{+0.03}_{-0.03}$&$0.30^{+0.06}_{-0.05}$&$0.18^{+0.06}_{-0.04}$&$0.24$
            &$0.24^{+0.04}_{-0.04}$&$0.30^{+0.06}_{-0.05}$&$0.18^{+0.05}_{-0.04}$&$0.23$\\
$A_1(0)$&$0.19^{+0.04}_{-0.03}$&$0.16^{+0.09}_{-0.05}$&$0.11^{+0.03}_{-0.03}$&$0.21$
            &$0.22^{+0.07}_{-0.01}$&$0.19^{+0.09}_{-0.07}$&$0.13^{+0.04}_{-0.03}$&$0.22$\\
$A_2(0)$&$0.17^{+0.03}_{-0.02}$&$0.07^{+0.08}_{-0.03}$&$0.06^{+0.02}_{-0.01}$&$0.19$
            &$0.20^{+0.04}_{-0.03}$&$0.11^{+0.05}_{-0.06}$&$0.08^{+0.03}_{-0.02}$&$0.21$\\
            \hline
$T_1(0)$&$0.19^{+0.03}_{-0.03}$&$0.15^{+0.09}_{-0.05}$&$0.15^{+0.04}_{-0.03}$&$$
           &$0.23^{+0.04}_{-0.04}$&$0.19^{+0.09}_{-0.06}$&$0.17^{+0.05}_{-0.04}$&$0.28$\\
$T_2(0)$&$0.19^{+0.03}_{-0.03}$&$0.15^{+0.09}_{-0.05}$&$0.15^{+0.04}_{-0.03}$&$$
           &$0.23^{+0.04}_{-0.04}$&$0.19^{+0.09}_{-0.06}$&$0.17^{+0.05}_{-0.04}$&$0.28$\\
$T_3(0)$&$0.16^{+0.01}_{-0.05}$&$0.07^{+0.06}_{-0.03}$&$0.13^{+0.04}_{-0.03}$&$$
            &$0.12^{+0.03}_{-0.01}$&$0.09^{+0.06}_{-0.04}$&$0.14^{+0.05}_{-0.03}$&$- 0.25\quad$
\\\hline\hline
\end{tabular}
\end{center}
\end{table}

\item
Some  $B\to T$ transitions have been  studied  by employing other approaches.  For instance, the form factors of  $B\to (a_2\,,K^*_2)$ transitions have also been evaluated with the LCSR~\cite{Aliev:2019ojc} and the PQCD approach~\cite{Wang:2010ni}.  These theoretical predictions are collected in Table~\ref{tab:comp},   the traditional CLF QM~(type-I) results given by Cheng~\cite{Cheng:2010yd,Cheng:2009ms} and our results with self-consistent type-II CLF QM are also listed for comparison.
Through comparison of these results listed in Table \ref{tab:comp}, it can be found that our center values are generally larger than the results obtained by the LCSR and the PQCD, but are smaller than the traditional CLF QM results, while they are still  in  consistence  within errors except for $T_3^{B\to K^*_2}(0)$. Our result for $T_3^{B\to K^*_2}(0)$ agrees well with the results obtained by LCSR and  PQCD, however Cheng's CLF result has a different sign. This finding indicates again that the type-II corresponding scheme can improve the CLF predictions.


\item
Compared the numerical results of $P\to T$ transition with the ones of $P\to V$ transition obtained in our previous works~\cite{Chang:2019mmh,Chang:2020wvs} at $q^2=0$,
it is found that: (i)  For the  $c\to q$ and $b \to q$~($q=u,d,s$) induced transition with a light spectator quark, the former are smaller than the later, which is favored by the experimental data of radiative decays. For instance, our result $T_1^{B\to K^*_2}(0)/T_1^{B\to K^*}(0)\simeq 0.72\pm0.18$ agrees well with the result $0.71$ obtained by PQCD~\cite{Wang:2010ni}, both of them are also consistent with the experimental data $0.53\pm0.08$ extracted from the radiative decays of $B$ meson, $B^-\to  K^{*-}_2(1430)\r$  and $B^-\to  K^{*-}\r$~\cite{Zyla:2020zbs}. Such result implies again that the CLF prediction can be improved by  employing the self-consistent   type-II scheme since the traditional CLF results give $T_1^{B\to K^*_2}(0)/T_1^{B\to K^*}(0)= 0.97$~\cite{Cheng:2009ms}, which conflicts with the data.  (ii) For the $b \to (s, c)$  induced transitions with a heavy spectator quark ($ c$ or $b$), the form factors of $P\to T$ transition are generally larger than  the ones of  $P\to V$ transition, for instance,  $T_1^{B_c\to D^*_{s2}}/ T_1^{B_c\to D_s^*}=0.37/0.20$.  It is expected that our results in this work can serve as a useful reference for  relevant studies of meson decays.
\end{itemize}

\tb{The semileptonic decays  play an important role in testing the perditions of form factors, but unfortunately, most of semileptonic decays induced by $(B,D)\to T$ transition have not been measured except for the  decay chains $B\to\bar D_2^{*}(2460)\ell\nu_\ell ( \bar D_2^{*}(2460)\to \bar D\pi)$~($\ell=e,\u$).   The averaged experimental results are~\cite{Zyla:2020zbs} 
\begin{align}
\label{eq:dm}
{\cal B}(B^+\to\bar D_2^{*0}\ell^+\nu_\ell) {\cal B}(\bar D_2^{*0}\to D^{-}\pi^+)&= (1.53\pm0.16)\times 10^{-3} \,, \\
\label{eq:dvm}
{\cal B}(B^+\to\bar D_2^{*0}\ell^+\nu_\ell) {\cal B}(\bar D_2^{*0}\to D^{*-}\pi^+)&= (1.01\pm0.24)\times 10^{-3}\,, \\
 \label{eq:dz}
 {\cal B}(B^0\to D_2^{*-}\ell^+\nu_\ell) {\cal B}( D_2^{*-}\to \bar{D}^{0}\pi^-)&= (1.21\pm0.33)\times 10^{-3}  \,, \\
 \label{eq:dvz}
{\cal B}(B^0\to D_2^{*-}\ell^+\nu_\ell) {\cal B}( D_2^{*-}\to \bar{D}^{*0}\pi^-)&= (0.68\pm0.12)\times 10^{-3}\,,
\end{align}
given by PDG. Theoretically, for the relevant strong decays, the relation  $\Gamma(\bar D_2^{*0}\to D^{(*)-}\pi^+)=\Gamma ( D_2^{*-}\to \bar{D}^{(*)0}\pi^-)$ is required by the isospin symmetry. Further considering the relation $\Gamma(B^+\to\bar D_2^{*0}\ell^+\nu_\ell)\simeq \Gamma(B^0\to D_2^{*-}\ell^+\nu_\ell)$, it is expected that
\begin{align}
\frac{{\cal B}(B^+\to\bar D_2^{*0}\ell^+\nu_\ell) {\cal B}(\bar D_2^{*0}\to D^{(*)-}\pi^+)}{{\cal B}(B^0\to D_2^{*-}\ell^+\nu_\ell) {\cal B}( D_2^{*-}\to \bar{D}^{(*)0}\pi^-)}\simeq \frac{\Gamma(B^0)\Gamma(D_2^{*-})}{\Gamma(B^+)\Gamma(\bar D_2^{*0})}\simeq1.06\, \,,
\end{align}
where, $1.06$ is obtained by using the PDG results for decay widths. This relation is allowed by current experimental data, $1.26\pm0.37~(1.49\pm 0.44)$ obtained from Eqs.~(\ref{eq:dm},\ref{eq:dz})~(Eqs.~(\ref{eq:dvm},\ref{eq:dvz})), within $1\sigma$ error. It is also noted that the SM result obviously  deviates from the central values of data~($1.06\, \text{vs.}\,1.26\,,1.49$), thus the future refined measurements may present a strict test on the SM prediction. }

\tb{
In order to extract the experimental results for ${\cal B}(B\to \bar{D}_2^{*}\ell^+\bar\nu_\ell)$, the decay widths of relevant strong decays are essential. However, these decays have not been measured experimentally for now, and the current theoretical evaluations involve large uncertainties.  Using  the QCD SR predictions $\Gamma(\bar D_2^{*0}\to D^{(*)-}\pi^+)=7.91^{+3.49}_{-3.00}(3.99^{+1.22}_{-1.56})\,{\rm MeV}$~\cite{Wang:2014oca,Li:2015xka} and Eqs.~(\ref{eq:dm},\ref{eq:dvm}), we can obtain the following  experimental results,  
\begin{align}
 {\cal B}(B^+\to\bar D_2^{*0}\ell^+\nu_\ell)=
 \begin{cases}
 (0.92^{+0.10+0.41}_{-0.10-0.35})\times 10^{-2}\,,   \\
(1.20^{+0.29+0.37}_{-0.29-0.47})\times 10^{-2}\,,\\
 \end{cases}
 \label{eq:Smbr}
\end{align}
where, the upper and the lower values are extracted from Eq.~(\ref{eq:dm}) and Eq.~(\ref{eq:dvm}), respectively; for each result, the first and the second errors are caused by Eqs.~(\ref{eq:dm},\ref{eq:dvm}) and QCD SR predictions for $\Gamma(\bar D_2^{*0}\to D^{(*)-}\pi^+)$, respectively.
}

\begin{table}[t]
\caption{\tb{Numerical results for the branching fractions of $B\to \bar{D}_2^*\ell^+\nu_\ell$~($\ell=e\,,\u$) and $B\to \bar{D}_2^*\tau^+\nu_{\tau}$ decays. The errors are caused by the form factors given in Eqs.~(\ref{eq:f0},\ref{eq:b1}).}}
\vspace{-0.1cm}
\begin{center}
\setlength{\tabcolsep}{2pt}
\renewcommand*{\arraystretch}{1.3}
\begin{tabular}{ccccccccc}
\hline\hline
 &type-II &type-I&Ref.~\cite{Azizi:2013aua}&Ref.~\cite{Aliev:2019ojc}\\ \hline
  ${\cal B}(B\to \bar{D}_2^*\ell^+\nu_\ell)$&$(1.23^{+0.36}_{-0.39})\times 10^{-2}$&$(0.63^{+0.39}_{-0.34})\times 10^{-2}$&$(1.01^{+0.30}_{-0.30})\times10^{-3}$&$(3.80^{+0.74}_{-0.74})\times10^{-2}$\\
  ${\cal B}(B\to \bar{D}_2^*\tau^+\nu_{\tau})$&$(0.49^{+0.15}_{-0.16})\times 10^{-3}$&$(0.22^{+0.14}_{-0.13})\times 10^{-3}$&$(0.16^{+0.06}_{-0.06})\times10^{-3}$&$(1.50^{+0.28}_{-0.28})\times10^{-3}$\\
 \hline\hline
\end{tabular}
\end{center}
\label{tab:dw}
\end{table}

\tb{
Theoretically, the differential decay widths of  semileptonic $B\to\bar D_2^{*}l\nu_l$ decays can be written as~\cite{Azizi:2013aua,Wang:2009mi}
\begin{align}
\frac{d\Gamma}{dq^{2}}&=
\frac{\lambda(m_{B}^{2},m_{D_{2}^{*}}^{2},q^{2})}{4m_{D_{2}^{*}}^{2}}\left(\frac{q^{2}-m_{\ell}^{2}}{q^{2}}\right)^{2} \frac{\sqrt{\lambda(m_{B}^{2},m_{D_{2}^{*}}^{2},q^{2})}G_{F}^{2}V_{cb}^{2}}{384m_{B}^{3}\pi^{3}} \Bigg\{\frac{1}{2q^{2}} \Big[ \frac{3m_{\ell}^{2}}{m_{B}^{2}}\lambda(m_{B}^{2},m_{D_{2}^{*}}^{2},q^{2})[{A}_{0}(q^{2})]^{2} \nonumber\\
&+(m_{\ell}^{2}+2q^{2})\Big| -\frac{1}{2m_{D_{2}^{*}}m_{B}}[(m_{B}^{2}-m_{D_{2}^{*}}^{2}-q^{2})( m_{B}+m_{D_{2}^{*}} ){A}_{1}(q^{2})+\frac{\lambda(m_{B}^{2}, m_{D_{2}^{*}}^{2}, q^{2})}{m_{B}+m_{D_{2}^{*}}}{A}_{2}(q^{2})] \Big|^{2} \Big] \nonumber\\
&+\frac{2}{3}(m_{\ell}^{2}+2q^{2})\lambda(m_{B}^{2},m_{D_{2}^{*}}^{2},q^{2})\Big[ \Big| \frac{{V}(q^{2})}{m_{B}(m_{B}+m_{D_{2}^{*}})}-\frac{(m_{B}+m_{D_{2}^{*}}){A}_{1}(q^{2})}{m_{B}\sqrt{ \lambda(m_{B}^{2}, m_{D_{2}^{*}}^{2}, q^{2})}} \Big|^{2} \nonumber\\
&+\Big| \frac{{V}(q^{2})}{m_{B}(m_{B}+m_{D_{2}^{*}})}+\frac{ (m_{B}+m_{D_{2}^{*}}){A}_{1}(q^{2})}{m_{B} \sqrt{\lambda(m_{B}^{2}, m_{D_{2}^{*}}^{2}, q^{2})}} \Big|^{2} \Big]  \Bigg\},
\end{align}
where $\lambda(a,b,c)=a^2+b^2+c^2-2ab-2ac-2bc$ is the $\rm{K\ddot{a}ll\acute{e}}$n function. Using the CLF results for $ \left(V,A_0,A_1,A_2\right)$ in type-I and -II schemes, 
\begin{align}
\label{eq:f0}
F(0)=&
\begin{cases}
\left(0.75^{+0.13}_{-0.16}\,, 0.64^{+0.11}_{-0.11}\,, 0.63^{+0.11}_{-0.12}\,, 0.58^{+0.12}_{-0.12}\,\right)\,,\qquad \text{type-II}\\
\left(0.70^{+0.20}_{-0.18}\,,0.37^{+0.17}_{-0.16}\,,0.43^{+0.17}_{-0.15}\,,0.54^{+0.20}_{-0.14}\,\right)\,,\qquad \text{type-I}\\
\end{cases}\\
b_1=&
\label{eq:b1}
\begin{cases}
\left( -6.20\,,-6.20\,,-3.00\,,-6.10\right)\,,\qquad \text{type-II}\\
\left( -6.42\,,-4.79\,,-0.75\,,-6.22\right)\,,\qquad \text{type-I}\\
 \end{cases}
\end{align}
and the values of the other input parameters given by PDG, we summarize our results for ${\cal B}(B\to \bar{D}_2^*\ell^+\nu_\ell\,,B\to \bar{D}_2^*\tau^+ \nu_{\tau})$ in Table~\ref{tab:dw},  in which  the results obtained in Refs.~\cite{Aliev:2019ojc,Azizi:2013aua} are also listed. It can be found that our results are much larger~(smaller) than the ones given in Ref.~\cite{Azizi:2013aua}~(Ref.~\cite{Aliev:2019ojc}) due to the different form factors.
Comparing our results with experimental data given in Eq.~\eqref{eq:Smbr}, we find that the type-II results are in good consistence with data, while the type-I results can not be excluded due to the large theoretical and experimental errors. More theoretical and experimental efforts are needed to improve the accuracies of results and further test the legality of such two schemes. The errors caused by form factors can be well controlled by evaluating the 
ratio $R_{D_2^*}\equiv \frac{\Gamma(B\to \bar{D}_2^*\tau^+\nu_{\tau})}{\Gamma(B\to \bar{D}_2^*\ell^+\nu_\ell)}$. Our prediction is 
\begin{align}
  R_{D_2^*}=0.040^{+0.002}_{-0.002}\,(0.035^{+0.004}_{-0.003})\,, \qquad \text{type-II~(type-I)}
\label{eq:RD2}
\end{align}
which are consistent with the LCSR prediction $0.041\pm0.002$~\cite{Aliev:2019ojc}, but are different from the result $0.16\pm0.04$~\cite{Azizi:2013aua}.
Such ratio is expected to be measured in the future,  it  will test whether  the $R_{D^{(*)}}$ anomalies in the pseudoscalar~(vector) channels exist also in the tensor channel or not,   and  play a similar role as  $R_{D^{(*)}}$ in testing the lepton flavor universality.}

 \section{Summary}
In this paper,  the matrix elements and  relevant vector, axial-vector and tensor form factors of $P\to T$ transition are  calculated within the CLF approach. The SLF results are also calculated for comparison.  The self-consistency and Lorentz covariance of the CLF QM are analyzed in detail. It is found that the CLF QM with the traditional correspondence scheme~(type-I) between  the  manifest covariant BS  and the LF approaches has  two kinds of  self-consistence problems: one is caused by the non-vanishing $\w$-dependent spurious contributions associated with the $B$ functions, which also violate the strict Lorentz covariance of CLF QM; another one is caused by the different strategies for dealing with the trace term in the calculation of matrix element. The self-consistence and Lorentz covariance problems can be resolved by employing the improved self-consistent  type-II correspondence scheme  which requires an additional replacement $M\to M_0$  relative to type-I scheme. Within the self-consistent type-II scheme, the zero-mode contributions to the form factors exist only in form but vanish numerically, and the valence contributions are exactly the same as the SLF results.
Theses findings confirm again the conclusion obtained via $P\to V$, $P\to A$ and $V'\to V''$ transitions in our previous works.
Finally, we present  our  numerical predictions for the vector, axial-vector and tensor form factors of $c\to (q,s)$ induced $D\to (a_2,K^*_2)$, $D_s\to (K^*_2, f'_2)$, $\eta_c(1S)\to (D^*_2,D^*_{s2})$, $B_c\to (B^*_2,B^*_{s2} )$ transitions and $b\to (q,s,c)$  induced $B\to (a_2,K^*_2, D^*_2), B_s\to (K^*_2, f'_2, D^*_{s2}), B_c\to (D^*_2, D^*_{s2}, X_{c2}(1P))$,  $\eta_b(1S)\to (B^*_2,B^*_{s2})$ transitions by employing a self-consistent CLF approach. These numerical results are collected in Tables \ref{tab:pvat} and \ref{tab:ptt}. Some form factors are first predicted in this work. Our predictions for the form factors of $B\to a_2$ and $B\to K^*_2$ transitions  are generally in consistent with the results obtained by employing LCSR and PQCD approaches, and show that the self-consistent type-II scheme can significantly improve the CLF prediction.  Compared with the form factors of  $P\to V$ transition, it is also found that the form factors of $P\to T $ transition are smaller than the ones of $P\to V $ at $q^2=0$ point when $T$ is a light tensor meson, which is in consistence with the experimental data. \tb{Using the obtained form factors, we also present the predictions for ${B}\to \bar{D}_2^*(2460)\ell^+\nu_\ell$~($\ell=e,\u$) and $\bar{D}_2^*(2460)\tau^+\nu_{\tau}$ decays. }It is expected that our results for the form factors of $P\to T$ transition can be applied further to the relevant phenomenological studies of meson decays.

\section*{Acknowledgements}
This work is supported by the National Natural Science Foundation of China (Grant No. 11875122), the Excellent Youth Foundation of Henan Province (Grant No. 212300410010),  the Youth Talent Support Program of Henan Province (Grant No. ZYQR201912178) and the Program for Innovative Research Team in University of Henan Province (Grant No. 19IRTSTHN018).

\begin{appendix}
\section*{Appendix A: Input parameters}

\begin{table}[ht]
\begin{center}
\caption{\label{tab:input} \small The values of Gaussian parameters $\beta$ (in units of MeV).}
\vspace{0.2cm}
\let\oldarraystretch=\arraystretch
\renewcommand*{\arraystretch}{1.1}
\setlength{\tabcolsep}{8.8pt}
\begin{tabular}{lcccccccccc}
\hline\hline
  &$\beta_{q\bar{q}}$    &$\beta_{s\bar{q}}$   &$\beta_{s\bar{s}}$
  &$\beta_{c\bar{q}}$    &$\beta_{c\bar{s}}$ \\
  \hline
  $P$ &$348\pm1$ &$365\pm2$ &$384\pm3$ &$473\pm12$
    &$543\pm10$ \\ \hline
  $T$ &$312\pm6$ &$313\pm10$ &$348\pm6$ &$429\pm13$
    &$530\pm19$ \\
  \hline\hline
  &$\beta_{c\bar{c}}$    &$\beta_{b\bar{q}}$   &$\beta_{b\bar{s}}$   &$\beta_{b\bar{c}}$   &$\beta_{b\bar{b}}$    \\\hline
  $P$ &$753\pm14$ &$552\pm10$ &$606\pm12$ &$939\pm11$
    &$1394\pm12$\\ \hline
  $T$ &$703\pm7$ &$516\pm15$ &$568\pm10$ &$876\pm20$
    &$1390\pm12$\\
\hline\hline
\end{tabular}
\end{center}
\end{table}

The constituent quark masses and  Gaussian parameters $\beta$ are essential inputs for computing the form factors. The quark masses are model dependent, and their values obtained in the previous works~\cite{Choi:2015ywa,Choi:2009ai,Choi:2007se,Choi:1997iq,Hwang:2010hw,Cheng:2003sm,Verma:2011yw} are different from each other more or less. In this work, we take
\begin{align}
&m_q=230\pm40\, {\rm MeV}\,,\qquad m_s=430\pm60 \, {\rm MeV}\,,\nonumber\\
&m_c=1600\pm300\, {\rm MeV}\,,\quad m_b=4900\pm400 \, {\rm MeV}\,.
\end{align}
which suggested values given in the previous works~\cite{Chang:2019obq}, it covers properly the others values and therefore can reflect roughly the uncertainties induced by the model dependence of quark mass.
Then, the parameters $\beta$ listed  in Table~\ref{tab:input}~\cite{Chang:2020wvs}, in which it have been assumed that $\beta_{q\bar q}$ is same for $V$ and $T $ due to the lack of tensor meson decay constant data .  In addition, the type-II correspondence scheme is employed in the fits, while the fitting results do not affect following comparison between type-I and -II schemes.



\section*{Appendix B: The CLF results for the tensor form factors of $P\to T$ transitions obtained in the previous paper }
The tensor form factors of $P\to T$ transition in the CLF QM have also been calculated  by  Cheng and Chua, the results can also been written as
Eq.~\eqref{eq:FCLF} with the integrands~\cite{Cheng:2004yj},
\begin{align}\label{eq:ptt1c}
\widetilde T_{1}^{\rm CLF}\left(q^{2}\right)=& M^{\prime} \bigg\{2(A_{1}^{(1)}-A_{2}^{(2)}-A_{3}^{(2)}
)\Big[M^{\prime 2}-M^{\prime \prime 2}-2 m_{1}^{\prime 2}-2 \hat{N}_{1}^{\prime}+q^{2}+2(m_{1}^{\prime} m_{2}+m_{1}^{\prime \prime} m_{2}\nonumber\\
&-m_{1}^{\prime} m_{1}^{\prime \prime})\Big]-8(A_{1}^{(2)}-A_{1}^{(3)}-A_{2}^{(3)})+(1-A_{1}^{(1)}-A_{2}^{(1)})\Big[(m'_{1}+m_{1}^{\prime\prime})^{2}+\hat{N}_{1}^{\prime}+\hat{N}_{1}^{\prime \prime}-q^{2}\Big]\nonumber\\
& +4(M^{\prime 2}-M''^{2})(A_{2}^{(2)}-A_{3}^{(2)}-A_{3}^{(3)}+A_{5}^{(3)})+4 q^{2}(-A_{1}^{(1)}+A_{2}^{(1)}+A_{2}^{(2)}+A_{3}^{(2)}-2 A_{4}^{(2)}\nonumber\\
&-A_{4}^{(3)}+A_{6}^{(3)}) -2(A_{1}^{(2)}+2 A_{1}^{(3)}-2 A_{2}^{(3)})-\frac{8}{D_{V}^{\prime \prime}}\Big[(m_{1}^{\prime}+m_{1}^{\prime \prime})(A_{1}^{(2)}-A_{1}^{(3)}-A_{2}^{(3)})\Big]\bigg\}\,,
\\
\widetilde T_{2}^{\rm CLF}\left(q^{2}\right)=& \widetilde T_{1}^{\rm CLF}\left(q^{2}\right)+\frac{ M^{\prime}q^{2}}{\left(M^{\prime 2}-M''^{2}\right)} \bigg\{2(A_{2}^{(1)}-A_{3}^{(2)}-A_{4}^{(2)})\Big[M^{\prime 2}-M^{\prime\prime 2}-2 m_{1}^{\prime 2}-2 \hat{N}_{1}^{\prime}+q^{2}\nonumber\\
&+2(m_{1}^{\prime} m_{2}+m_{1}^{\prime \prime} m_{2}-m_{1}^{\prime} m_{1}^{\prime \prime})\Big]-8(A_{1}^{(2)}-A_{1}^{(3)}-A_{2}^{(3)}) +(1-A_{1}^{(1)}-A_{2}^{(1)})\Big[-2 M^{\prime 2}\nonumber\\
&+2 m_{1}^{\prime 2}+(m_{1}^{\prime}+m_{1}^{\prime \prime})^{2}+2(m_{2}-2 m_{1}^{\prime}) m_{2}+3 \hat{N}_{1}^{\prime}+\hat{N}_{1}^{\prime \prime}-q^{2}\Big] +2\Big[Z_{2}(1-A_{2}^{(1)})\nonumber\\
&-\frac{P \cdot q}{q^{2}} A_{1}^{(2)}\Big]+4(q^{2}-2 M^{\prime 2}-2 M^{\prime\prime 2})(A_{2}^{(2)}-A_{3}^{(2)}-A_{3}^{(3)}+A_{5}^{(3)}) -4(M^{\prime 2}\nonumber\\
&-M^{\prime \prime 2})(-A_{1}^{(1)}+A_{2}^{(1)}+A_{2}^{(2)}+A_{3}^{(2)}-2 A_{4}^{(2)}-A_{4}^{(3)}+A_{6}^{(3)})+2(A_{1}^{(2)}+2 A_{1}^{(3)}\nonumber\\
&-2 A_{2}^{(3)})-\frac{8}{D_{V}^{\prime \prime}}\left[(m''_{1}-m_{1}^{\prime}+2 m_{2})(A_{1}^{(2)}-A_{1}^{(3)}-A_{2}^{(3)})\right]
\bigg\}\,,
\\
\label{eq:ptt3c}
\widetilde T_{3}^{\rm CLF}\left(q^{2}\right)=& M^{\prime} \bigg\{-2(A_{2}^{(1)}-A_{3}^{(2)}-A_{4}^{(2)})\Big[M^{\prime 2}-M''^{2}-2 m_{1}^{\prime 2}-2 \hat{N}_{1}^{\prime}+q^{2}+2(m_{1}^{\prime} m_{2}+m_{1}^{\prime \prime} m_{2}\nonumber\\
&-m_{1}^{\prime} m_{1}^{\prime \prime})\Big]+8(A_{1}^{(2)}-A_{1}^{(3)}-A_{2}^{(3)})-(1-A_{1}^{(1)}-A_{2}^{(1)})\Big[-2 M^{\prime 2}+2 m_{1}^{\prime 2}+(m_{1}^{\prime}\nonumber\\
&+m_{1}^{\prime \prime})^{2}+2(m_{2}-2 m_{1}^{\prime}) m_{2}+3 \hat{N}_{1}^{\prime}+\hat{N}_{1}^{\prime \prime}-q^{2}\Big]-2\Big[Z_{2}(1-A_{2}^{(1)})-\frac{P \cdot q}{q^{2}} A_{1}^{(2)}\Big]\nonumber\\
&-4(q^{2}-M^{\prime 2}-3 M^{\prime\prime 2})(A_{2}^{(2)}-A_{3}^{(2)}-A_{3}^{(3)}+A_{5}^{(3)})-2(A_{1}^{(2)}+2 A_{1}^{(3)}-2 A_{2}^{(3)})\nonumber \\
&+\frac{4}{D_{V}^{\prime \prime}}\Big\{(m''_{1}-m'_{1}+2 m_{2})\Big[2(A_{1}^{(2)}-A_{1}^{(3)}-A_{2}^{(3)})+(M^{\prime 2}-M^{\prime\prime 2})(-A_{1}^{(1)}+2 A_{2}^{(2)}\nonumber\\
&+2 A_{3}^{(2)}-A_{3}^{(3)}-2 A_{4}^{(3)}-A_{5}^{(3)})\Big]+(m_{1}^{\prime}+m_{1}^{\prime \prime})(M^{\prime 2}-M^{\prime\prime2})(A_{2}^{(1)}-2 A_{3}^{(2)}-2 A_{4}^{(2)}\nonumber\\
&+A_{4}^{(3)}+2 A_{5}^{(3)}+A_{6}^{(3)})+m_{1}^{\prime}(M^{\prime 2}-M^{\prime\prime 2})(-1+2 A_{1}^{(1)}+2 A_{2}^{(1)}-A_{2}^{(2)}-2 A_{3}^{(2)}-A_{4}^{(2)})\Big\}\bigg\}.
\end{align}

\end{appendix}

\end{document}